\documentclass[fleqn,usenatbib]{mnras} 

\usepackage{newtxtext,newtxmath}
\usepackage[T1]{fontenc}

\DeclareRobustCommand{\VAN}[3]{#2}
\let\VANthebibliography\thebibliography
\def\thebibliography{\DeclareRobustCommand{\VAN}[3]{##3}\VANthebibliography}
\usepackage{graphicx}	% Including Fig. files
\usepackage{amsmath}	% Advanced maths commands
\usepackage[dvipsnames]{xcolor}
\usepackage{units}
\usepackage{bm}
\usepackage{threeparttable}
\usepackage{subcaption}
\title[Multiphase Study of Cepheids]{A multiphase study of theoretical and observed light curves of classical Cepheids in the Magellanic Clouds}

\author[Kurbah et al.]{
Kerdaris Kurbah,$^{1}$\thanks{E-mail: kerdarisbahkur@gmail.com}
Sukanta Deb,$^{1,2}$\thanks{E-mail:sukanta.deb@cottonuniversity.ac.in}
Shashi M. Kanbur,$^{3}$\thanks{E-mail:shashi.kanbur@oswego.edu}
Susmita Das$^{4}$, Mami Deka$^{1}$, Anupam Bhardwaj$^{5}$, 
\newauthor{Hugh Riley Randall $^{3}$, Selim Kalici $^{3}$} \\
$^{1}$Department of Physics, Cotton University, Guwahati 781001,
Assam, India \\
$^{2}$Space and Astronomy Research Center, Cotton University, Guwahati 781001, Assam, India \\
$^{3}$ Department of Physics, State University of New York, Oswego, NY 23126, USA \\
$^{4}$Konkoly Observatory, Research Centre for Astronomy and Earth Sciences, Eötvös Loránd Research Network (ELKH), Konkoly-Thege Miklós út 15-17, H-1121,\\ Budapest, Hungary\\
$^{5}$INAF-Osservatorio Astronomico di Capodimonte, Via Moiariello 16, I-801301, Napoli, Italy
}
\begin{document}
\label{firstpage}
\pagerange{\pageref{firstpage}--\pageref{lastpage}}
\maketitle
%Abstract of the paper
\begin{abstract}
We present an analysis of the theoretical and observed light curve parameters of the fundamental mode (FU) classical Cepheids in the Magellanic Clouds in $V$- and $I$- photometric bands. The state-of-the-art 1D non-linear radial stellar pulsation (RSP) code in MESA (\textsc{mesa-rsp}) has been utilized to generate the  theoretical light curves using four sets of convection parameters. Theoretical light curves with two chemical compositions: $Z=0.008$ and $Z=0.004$  appropriate for the Large Magellanic Cloud (LMC) and Small Magellanic Cloud (SMC), respectively, covered a wide range of periods ($3<P (\rm{d})<32$). The observed light curves are taken from the OGLE-IV database. We compare theoretical and observed Fourier parameters (FPs), and investigate the period-luminosity (PL), period-colour (PC), and amplitude-colour (AC) relations as a function of pulsation phase for short ($\log{P}<1$), long ($\log{P}>1$) and all periods. The multiphase relations  obtained from theoretical and observed light curves in the PL/PC/AC plane are found to be dynamic in nature, with the effect more pronounced at $\Phi \sim 0.75-0.85$. 
Furthermore, a contrasting behaviour of the theoretical/observed multiphase PL and PC relations between the short and long periods has been found for both LMC and SMC. The analysis shows that multiphase PL relations are more stringent to test the models with observations over the FPs. Distances to the LMC/SMC determined using long period Cepheids are found to be in good agreement with the literature values when the term $R_{21}$ is added to the PL relation.  
\end{abstract}

\section{Introduction}
Classical Cepheids (hereafter, Cepheids) serve as proxies in determining the galactic and extra-galactic distances. They belong to a radially pulsating class of variable stars which pulsate in fundamental (FU) and
other overtone modes. A vast majority of them have single periods \citep{sosz15}. They are relatively young stars  $\sim 10$ to $300$ Myr with masses $\sim3$ to $11$ M$_{\odot}$ located in the instability strip of the Hertzsprung-Russel (HR) diagram. Cepheids in the Magellanic Clouds
were discovered by Henrietta Leavitt \citep{leav08} who first showed that their luminosities vary with periods in a very regular fashion: the longer the period of the Cepheid, the higher is its luminosity \citep{leav12}. The Leavitt Law has been the key factor in Galactic as well as extra-galactic distance determinations. Numerous observational and theoretical investigations  carried out in the literature have shown the sensitivity of Cepheid PL relation with metallicity, \citep{gier18,beat18,ripe20,bhar20,ripe21,breu21} as well as  the linearity and non-linearity of the relation over a period range \citep{ngeo05,ngeo06b,subr15, ripe16,bhar16,ripe22}. The non-linearity of the mean light  PL relation in the MCs was also studied by other authors: \citet{subr15} and \citet{ripe16} investigated the non-linearity in the SMC, in the optical and near-infrared, both finding a break for the FU Cepheids at period $\sim~3$~d. Recently, \citet{ripe22} reported a break at $P=0.58$ d in the mean light LMC FO Cepheid PL relation. These studies revealed that the Cepheid PL relation is non-universal. 

The PL relation is closely related to the PC relation via the PLC relation; a change in the PL relation can be explained by the changes
in the PC relation. Moreover, changes in the PC relation is also correlated to changes in AC relations.
Extensive investigations in the literature were carried out on the PL, PC and AC relations of the Galactic- and the Magellanic Clouds- Cepheids at maximum, minimum light and mean light \citep{simo93,kanb04,kanb04b,kanb06,kanb07,kanb10,bhar14,das20}. Changes in the PC and AC relations using pulsating stars at maximum and minimum light were explained satisfactorily using the 
theory of interaction between  hydrogen ionization front (HIF)  and the stellar photosphere \citep[and references therein]{simo93, das18, das20, deka22}. 

A comparison between the observed and theoretical PC,  AC and PL relations at maximum and minimum light was carried out
for the Galactic/LMC/SMC Cepheids by \citet{kanb04,kanb06,kanb07}. In those studies, a 1D non-linear radiation hydrodynamical model was used with the  mass-luminosity (ML) relation adopted from \citet{chio89} and \citet{bono00} to generate the theoretical Cepheid light curves. The input parameters of the Cepheid models used by the authors are summarized in their respective papers. The computational codes and numerical methods as described in \citet{yeck98} and \citet{koll02} were used in those studies
to  carry out the pulsation modelings. Recently, using \textsc{mesa-rsp}, \citet{das20} presented  theoretical explanations of the observed PC and AC relations at maximum and minimum light for the pulsating stars such as RR Lyraes, type I and type II Cepheids, 
whereas the same was provided for the $\delta$ Scuti stars by \citet{deka22}.

The first investigation of the observed multiphase  PC and AC relations of Cepheids was carried out by \citet{ngeo06} using OGLE-II data. Using multiphase light curve analysis, they found a strong evidence for the linearity of Galactic/SMC PC/PL and non-linearity of LMC PC/PL relations at mean light. Studies by \citet{ngeo06,ngeo10} and \citet{ngeo12a} have shown that Cepheid PC,  AC and PL relations as a function of phase are highly dynamic in nature, with more pronounced effect occurring at phases corresponding to minimum light. Investigations of observed Cepheid PC/AC/PL relations at multiphase and comparing these relations with those from models can be used to constrain theoretical Cepheid pulsation and evolutionary models.

Comparison between the observed and theoretical Cepheid PC,  AC and PL relations in the literature were mostly done at maximum and minimum light. In addition, multiphase study of these relations was carried out by \citet{kanb10}. In that study, Cepheid models were computed using the code based on \citet{stel82, bono94, bono99b} and the ML relation given by \citet{bono00b} was adopted.
The study by \citet{kanb10} considered Cepheid models with masses: $3.5~M_{\odot}, 4~M_{\odot}, 4.5~M_{\odot}, 5~M_{\odot}, 7~M_{\odot}, 9~M_{\odot}, 11~M_{\odot}$ consisting of different metallicities.

In this work, we investigate the theoretical light curves of Cepheid variables at multiphase using Modules for Experiments in Stellar Astrophysics (\textsc{mesa}, \citet{paxt11,paxt13,paxt15,paxt18,paxt19}. Recently, non-linear radial stellar pulsation (RSP) code \citep{smol08} has been incorporated into \textsc{mesa} which can be utilized to generate the theoretical light curves of Cepheids. Moreover, \textsc{mesa-rsp} also supplies the means to incorporate different prescriptions for modelling time dependent turbulent convection through seven user defined parameters. This provides an opportunity to explore the effect of convection on the theoretical PC/AC/PL relations. The present study deals with the analysis of the  theoretical Cepheid light curves appropriate for the LMC and SMC using the four sets of values outlined in \citet{paxt19}. We further compare our results with observations. The study comprises one of the largest set of models carried out in the literature.

The work done in estimating distances using multiphase approach is very limited. Using 48 Mira variables in the $JHK_{s}$  bands, \citet{kanb97} carried out the first study of PL and PLC relations at minimum, maximum and mean light in the LMC. It was found that in the $JH$-bands, the PL relations of Miras at maximum light exhibit a significant smaller dispersion than those at mean light. In an extension to this work, using a much larger sample of Mira variables belonging to the LMC and SMC based on OGLE-III and Gaia DR2, \citet{bhar19} found that the scatter in the PL- and PLC-relations at maximum-light reduces to $\sim 30\%$ more as compared to the mean-light. This serves as a strong motivation to make use of multiphase relations of Cepheid variables for distance determination and is a subject of our future research. As opposed to PL relations at mean light which may exhibit greater scatter, the existence of a PL relation with minimum scatter at a particular phase would prove to be crucial in the precise estimation of extra-galactic distances.

The remaining paper is organized as follows. Section~\ref{sec:data} gives a description of the observed data, method utilized to carry out this study, and models related to \textsc{mesa-rsp}. We investigate the  theoretical and observed FPs, theoretical/observed multiphase PL, PC and AC relations for the LMC and SMC in Sections \ref{sec:pl_lmc}. Distance determinations to the LMC and SMC using the theoretically obtained PL relation are discussed in Section~\ref{sec:distance}. The summary and conclusion of the present study are discussed in Section~\ref{sec:diss}. 

\section {Data and Methodology}
\label{sec:data}
\subsection{Observed data}
\label{subsec:obs_data}
The observed Cepheid light curves are taken from the Optical Gravitational Lensing Experiment (OGLE-IV) database, which are available in $V$- and $I$- photometric bands. This database contains the apparent mean magnitude, period, epoch, the Fourier parameters for the $I$-band light curves obtained from Fourier cosine decomposition \citep[and reference therein]{sosz15}. The stars listed in 'remarks.text' file provided by OGLE-IV are removed to avoid any contamination to the analysis, for both LMC and SMC. Complementary light curves with more than $30$ data points are selected from both the photometric bands leaving $2055$ and $2419$ number of stars in the LMC and SMC, respectively.

The light curves of these stars in the $V$- and $I$-band are decomposed with cosine Fourier decomposition method
\citep[and reference therein]{deb09}
\begin{equation}
m(\Phi)= A_{0} + \sum_{i=1}^{N} A_{i}\cos\left(2\pi i\Phi + \phi_{i}\right),
\label{equa:n01}
\end{equation}
using the fifth- and seventh-order Fourier fits for $V$- and $I$-band, respectively. $m(\Phi)$ is the observed
apparent magnitude, $A_{0}$ is the apparent mean magnitude, $\Phi$ is the pulsation phase given by 
\begin{equation}
\Phi = \frac{(t-t_{0})}{P} - {\rm Int} \left[\frac{(t-t_{0})}{P}\right],
\label{equa:n02}
\end{equation}
where $t$ is the time of observation in days, $t_{0}$ is the epoch of maximum light and $P$ denotes the period in days. 
The value of $\Phi$ ranges from $0$ to $1$ which represents one pulsation cycle. The light curves in both $V$- and $I$- bands are shifted such that  $\Phi=0$ corresponds to maximum light, following the steps from  \citet{ngeo06}. The 
phase difference is given by $\phi_{i1}=\phi_{i} - i\phi_{1}$ and the amplitude ratio by  $R_{i1} = \frac{A_{i}}{A_{1}}$. Errors in the parameters are calculated using the formulae of \citet{deb09,deb15}. The mean light ($A_{0}$) obtained from Fourier decomposition in $V$- and $I$-bands are corrected for extinction using the relation:
\begin{equation}
A_{0,\lambda}=A_{0,\lambda}-R_{\lambda} E(B-V),
\end{equation} 

where $\lambda \equiv \left(V,I\right)$. $R_{\lambda}$  is the ratio of the total to selective absorption in $\lambda$-photometric band. $E(B-V)$ is the interstellar reddening along the line of sight. To correct for interstellar extinction, \citet[][hereafter SM]{skow21} reddening map is utilized. This map is based on the latest OGLE-IV data plus dust reddening maps and covers the entire 
Magellanic system.  $E(V-I)$ values obtained from the map are converted into $E(B-V)$  using the relation, $E(V-I)=1.26 E(B-V)$
following the standard Cardelli Law \citep{card89} keeping $(R_{V}, R_{I})=(3.23,2.05)$ fixed \citep{deb18,deb19}.

 \subsection{Theoretical models}
 \label{sec:theoretical_models}
In order to generate the theoretical light curves of Cepheids  with different sets of convection parameters: A, B, C and D,
\textsc{mesa-rsp} version ``\textsc{mesa} r15140'' is used. Set A corresponds to the simplest convection model without turbulent pressure, turbulent flux and radiative cooling ($\alpha_{p} = \alpha_{t} = \gamma_{r} = 0$); set B adds radiative cooling without turbulent pressure and turbulent flux ($\alpha_{p} = \alpha_{t} = 0$); set C adds turbulent pressure and turbulent flux without radiative cooling ($\gamma_{r} = 0$); and set D include these effects simultaneously. For a brief explanation of the
$\alpha$ parameters used in each convection sets, the readers are referred to  \citet[and references therein]{das20}. The free parameters for each of these convection sets are taken as provided in Tables 3 and 4 of \citet{paxt19}. 
 The other input parameters required are metal abundance ($Z$),  hydrogen mass fraction ($X$), stellar mass ($M/ \rm M_{\odot}$), stellar luminosity ($L/\rm {L_{\odot}}$) and effective temperature ($T_{\rm eff}$).
Chemical compositions of $Z=0.008, X=0.742$ and $Z=0.004, X=0.746$, appropriate for the LMC and SMC, respectively, are adopted from \citet{bono98}. The Cepheid mass is chosen in the range $3.6-6.8~M_{\odot}$. It is worth mentioning here  that a mass range of $5.4-6.8~M_{\odot}$ was used by \citet{bhar17a} for generating the theoretical light curves of Cepheids. Extension of mass range to a lower mass of $3.6~M_{\odot}$ has been done in the present study to include a few more masses lower than $5.4~M_{\odot}$ for a fixed luminosity following period-mean density relation. This takes into account the better coverage of the finite width of the instability strip for Cepheids. 

For each chemical composition and mass, the luminosity is calculated using the rotation-averaged ML relation of \citet[their Equation (5)]{ande14}. From the luminosity thus obtained, the upper/lower luminosity limit is roughly set by increasing/decreasing the luminosity level by $\Delta \log L/\rm L_{\odot} \approx 0.25$ dex. Luminosity values lying  within these upper/lower limits are taken in steps of $500~L_{\odot}$. Effective temperatures in the range $\sim 4000$ to $8000$ K in steps of $50$ K are taken for each combination of chemical composition, mass, and luminosity. 

\begin{figure*}
%\scalebox{0.95}{
\begin{tabular}{cc}
\resizebox{0.48\linewidth}{!}{\includegraphics*{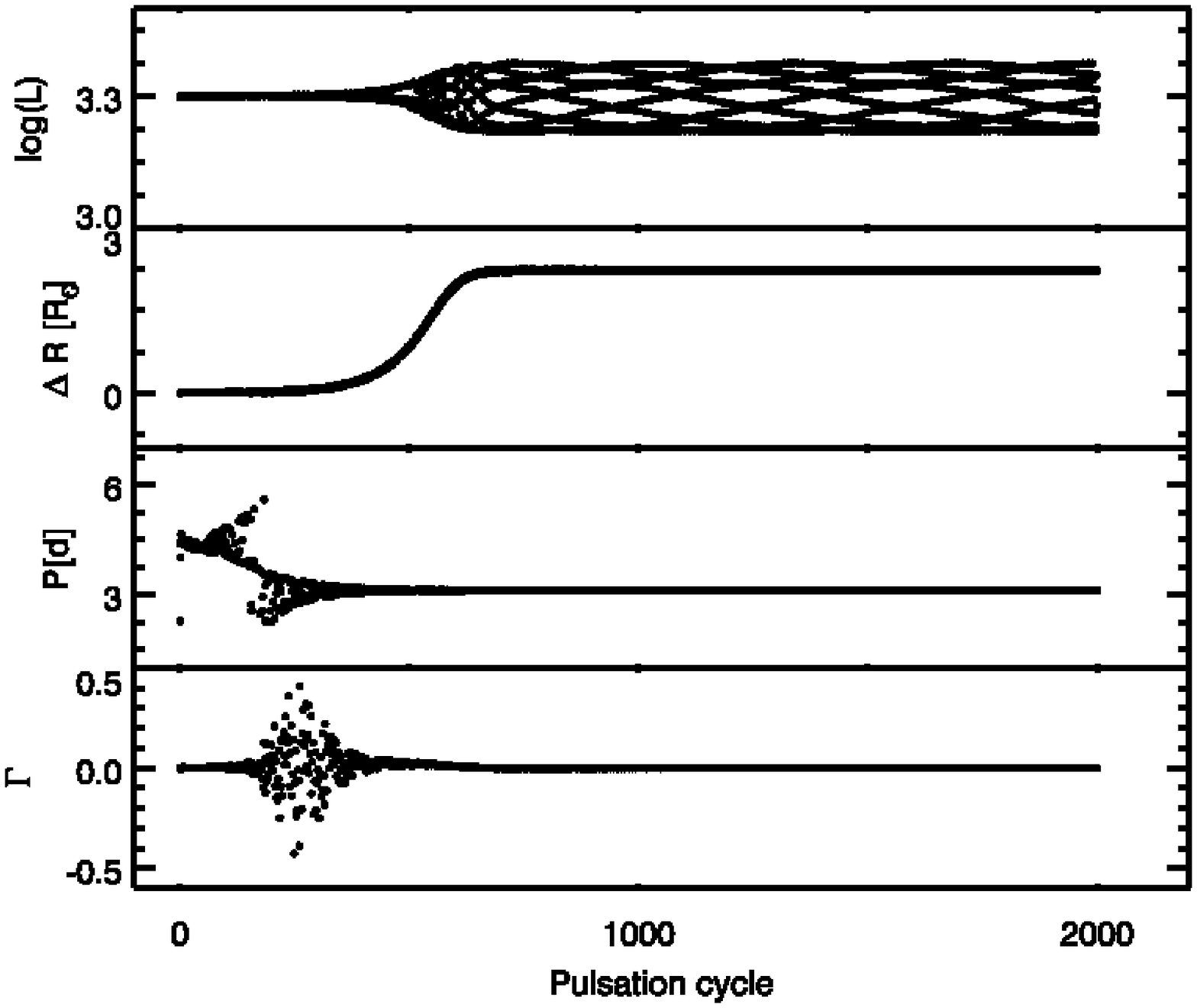}}& 
\resizebox{0.48\linewidth}{!}{\includegraphics*{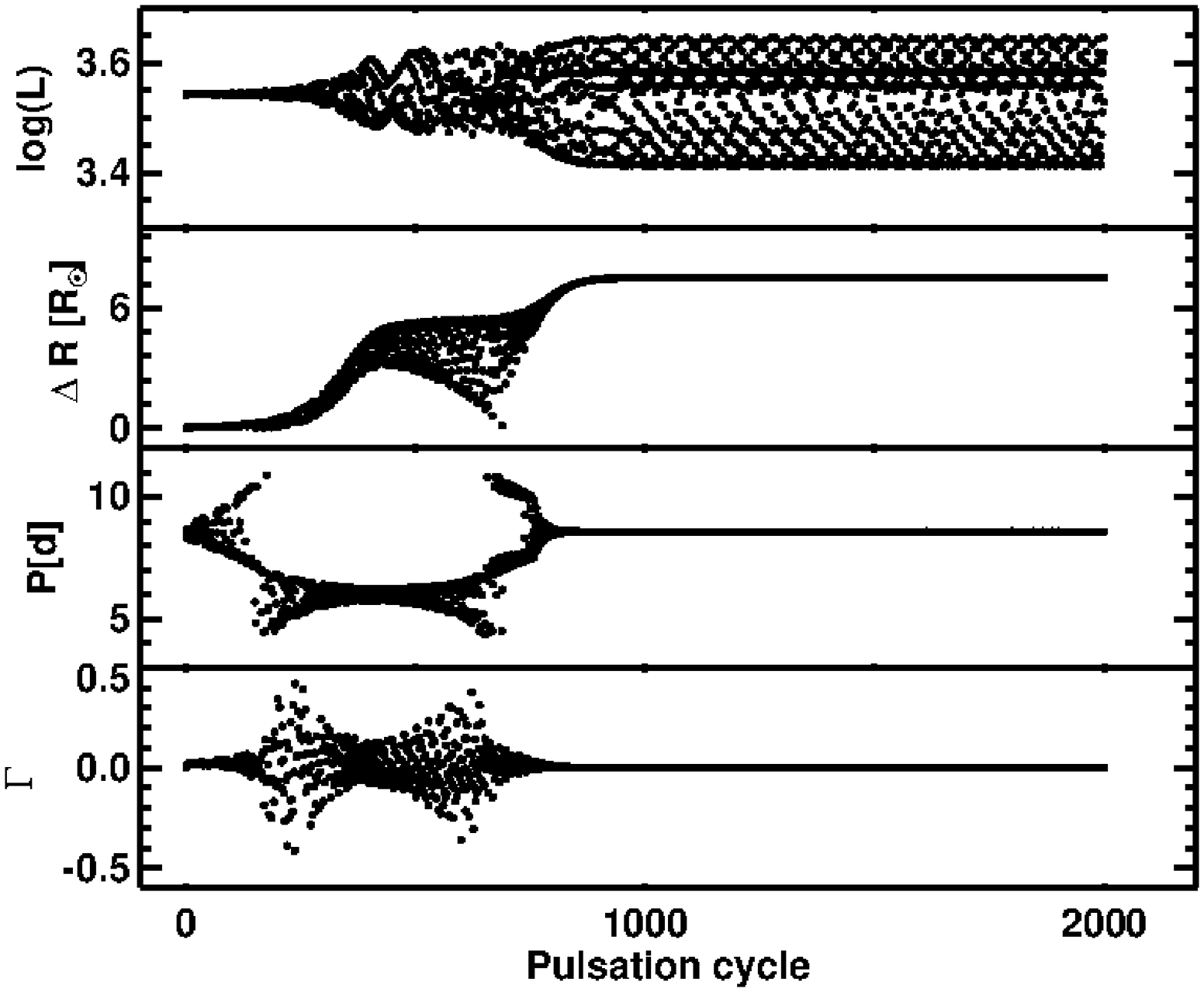}}\\
\end{tabular}
\caption{Plots of converging parameters of the model: (a) (\textit{left panel:}$M=6.0~\rm M_{\odot}$, $L=2000~\rm L_{\odot}$, $T_{\rm eff}=5950$~K). Here the model changes mode and converges in the FO mode (second row from bottom); (b)   
(\textit{right panel:}$M=6.6\rm M_{\odot}$, $L=3500 \rm L_{\odot}$, $T_{\rm eff}=5550$K) depicting that after a cycle of beating between FU and FO mode, the model returns to the FU mode and finally converges there (second row from the bottom).}
\label{fig:param_lmc}
\end{figure*}

The `inlist' used for computing the models in the present study is provided in Appendix~\ref{app:inlist}. Linear computations of the LMC and SMC grids as described in \citet[and references therein]{das21} are performed, from where the linear period and growth rate of the models are obtained. To obtain the pulsation state of the models and properties of light/radial velocity curves, non-linear computation has to be carried out \citep{smol08}. The non-linear integration of the models is carried out for $2000$ pulsation cycles. However, non-linear computation of models is time-consuming. We have selected $350$ LMC and $320$ SMC models covering a wide range of periods ($3<P \rm(d) <32$) for carrying out the non-linear analysis. While computing the models, even though the calculation was started in the F-mode, some of the models converged to stable pulsation in the FO mode (left panel of Fig.~\ref{fig:param_lmc}). Also for few other models, after a cycle of beating between FU and FO modes, the model again converged in the FU mode (right panel of Fig.~\ref{fig:param_lmc}). These cases of mode switching of the models are largely seen in sets B and D as compared to sets A and C. The presence of radiative cooling in sets B and D seems to have played a major role in mode switching of the models. Due to this, the number of models which pulsate in the FU mode for the chosen LMC/SMC subset  gets reduced in the analysis that follows.

Full-amplitude and stable pulsation state of the models are checked before generating their light curves. The condition of full-amplitude and stable pulsation state of the model is satisfied when the parameters; amplitude of radius variation $\Delta R$, the pulsation period computed on a cycle to cycle basis ($P$), fractional growth of the kinetic energy per pulsation period ($\Gamma$) vary by
less than $0.01$ in the last $100$ pulsation cycles of the total pulsation cycles computed.   
Based on this condition, light curves are selected for further analysis. The details on the transformation of bolometric light curves into the optical $(V,I)$-bands are given in \citet{paxt18} and have been briefly summarized in Section 2.3 of \citet[and references therein]{das21}. MESA-RSP provides pre-processed tables of bolometric corrections based on stellar model atmospheres of \citet{leje98}. We follow \citet{das21} to transform the theoretical light curves into the observed ones
based on the pre-processed table from \citet{leje98}. Further details on the use of different model atmospheres to transform the bolometric light curves into the observational bands can be found in \citet{das21}.
The total number of light curves generated are $1014$ and $938$ for  the LMC and SMC, respectively. 
These theoretical light curves are then decomposed with a cosine Fourier decomposition method, with the order of the fit set  to $N=20$ for both the $V$- and $I$-bands. The number of models, stellar mass, stellar luminosity and effective temperature in each convection set for the LMC and SMC subsets are summarized in Tables \ref{tab:model_abcd_lmc}.  

\section{Light Curve Analysis}
\label{sec:pl_lmc}

\begin{table}
\begin{threeparttable}[b]
\caption{Summary of LMC and SMC Cepheid models using sets A, B, C, D.}
\begin{tabular}{c c c c c c c c} \\ \hline \hline
 &  &$N$   & $M/\rm M_{\odot}$ &$L/\rm L_{\odot}$& $T_{\rm eff}$(K) \\ \hline
LMC & Set A &$307$ & $3.6-6.8$ & $1000-8500$ & $4750-6050$\\
    & Set B &$237$ & $3.6-6.8$ & $1000-8500$ & $4850-6150$\\
    & Set C &$257$ & $3.6-6.8$ & $1000-8500$ & $4700-5800$    \\  
    & Set D &$213$ & $3.6-6.8$ & $1000-8500$ & $4700-5750$       \\ \hline
SMC & Set A &$293$& $3.6-6.8$ & $1000-8500$ & $4850-6200$  \\
    & Set B &$234$& $3.6-6.8$ & $1000-8500$ & $4900-6200$  \\
    & Set C &$240$& $3.6-6.8$ & $1000-8500$ & $4850-6200$  \\  
    & Set D &$171$& $3.6-6.8$ & $1000-8500$ & $4900-6250$        \\ \hline
\end{tabular}
\label{tab:model_abcd_lmc}
\begin{tablenotes}
\item $N$- Number of full amplitude stable models
\end{tablenotes}
\end{threeparttable}
\end{table}

\subsection{Comparison of Theoretical and Observed Fourier Parameters}
\label{subsec:lmc_fourier}

\begin{figure*}
%\scalebox{0.95}{
\begin{tabular}{ccc}
\resizebox{0.48\linewidth}{!}{\includegraphics*{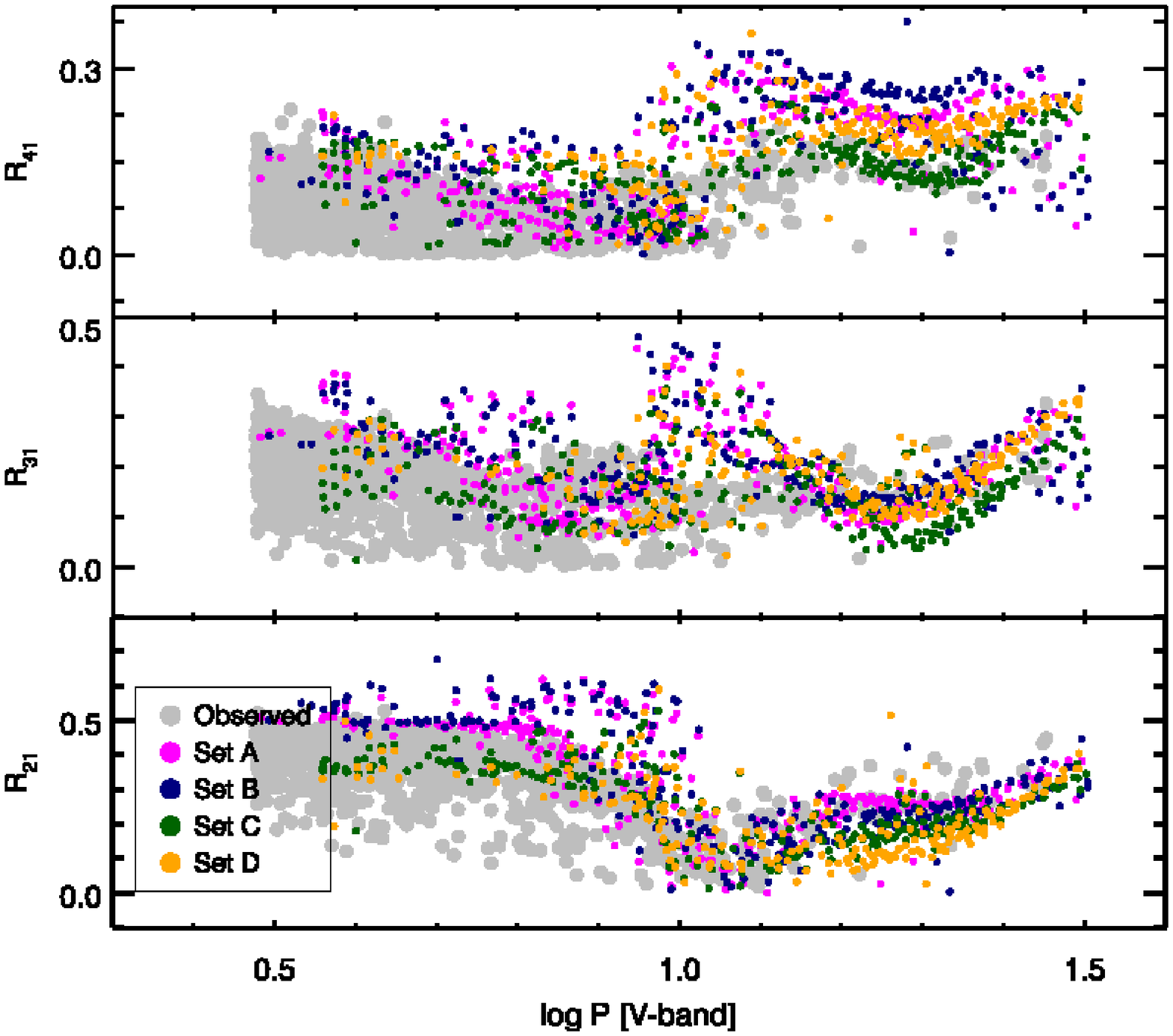}}&
\resizebox{0.48\linewidth}{!}{\includegraphics*{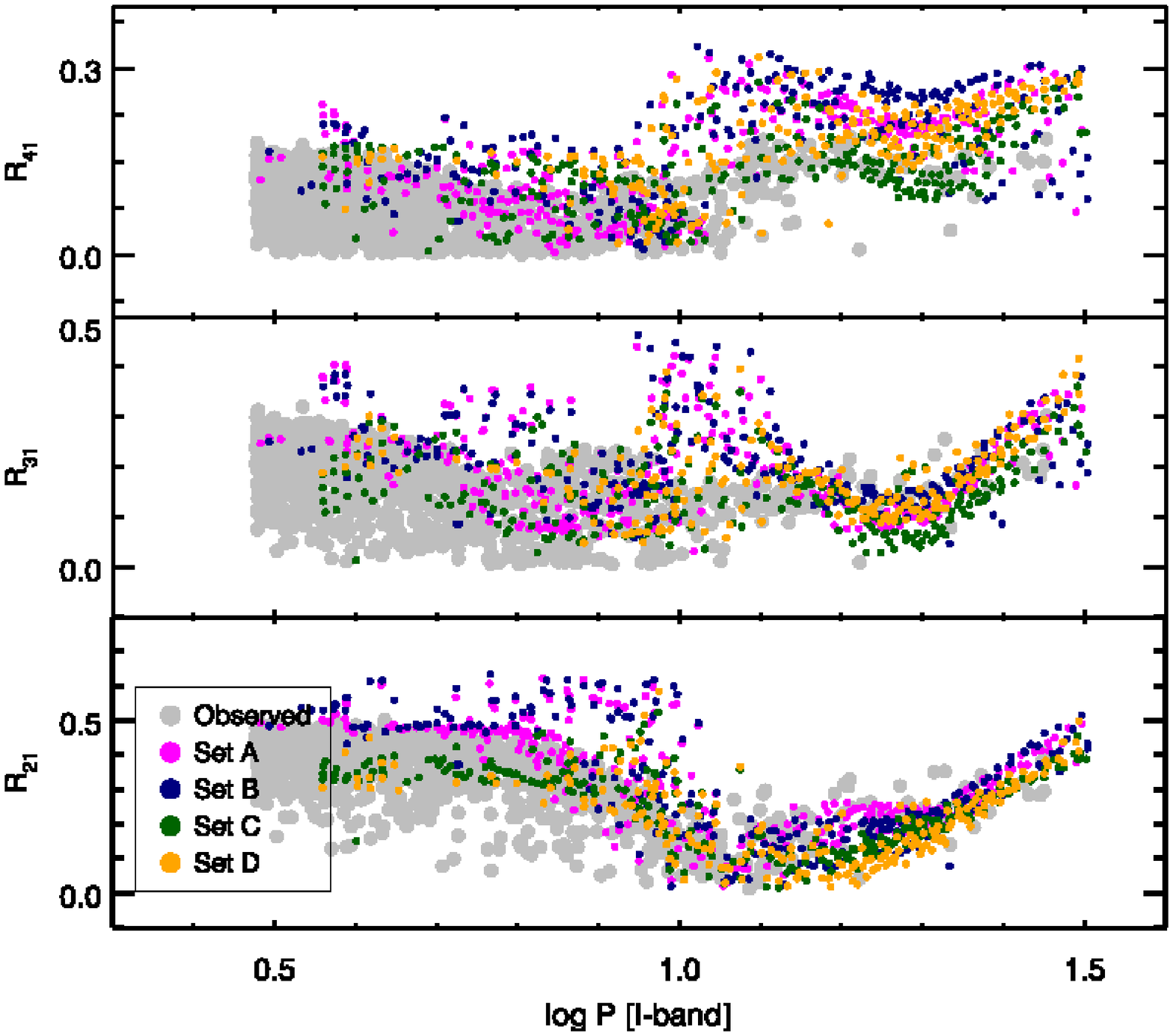}}\\
\resizebox{0.48\linewidth}{!}{\includegraphics*{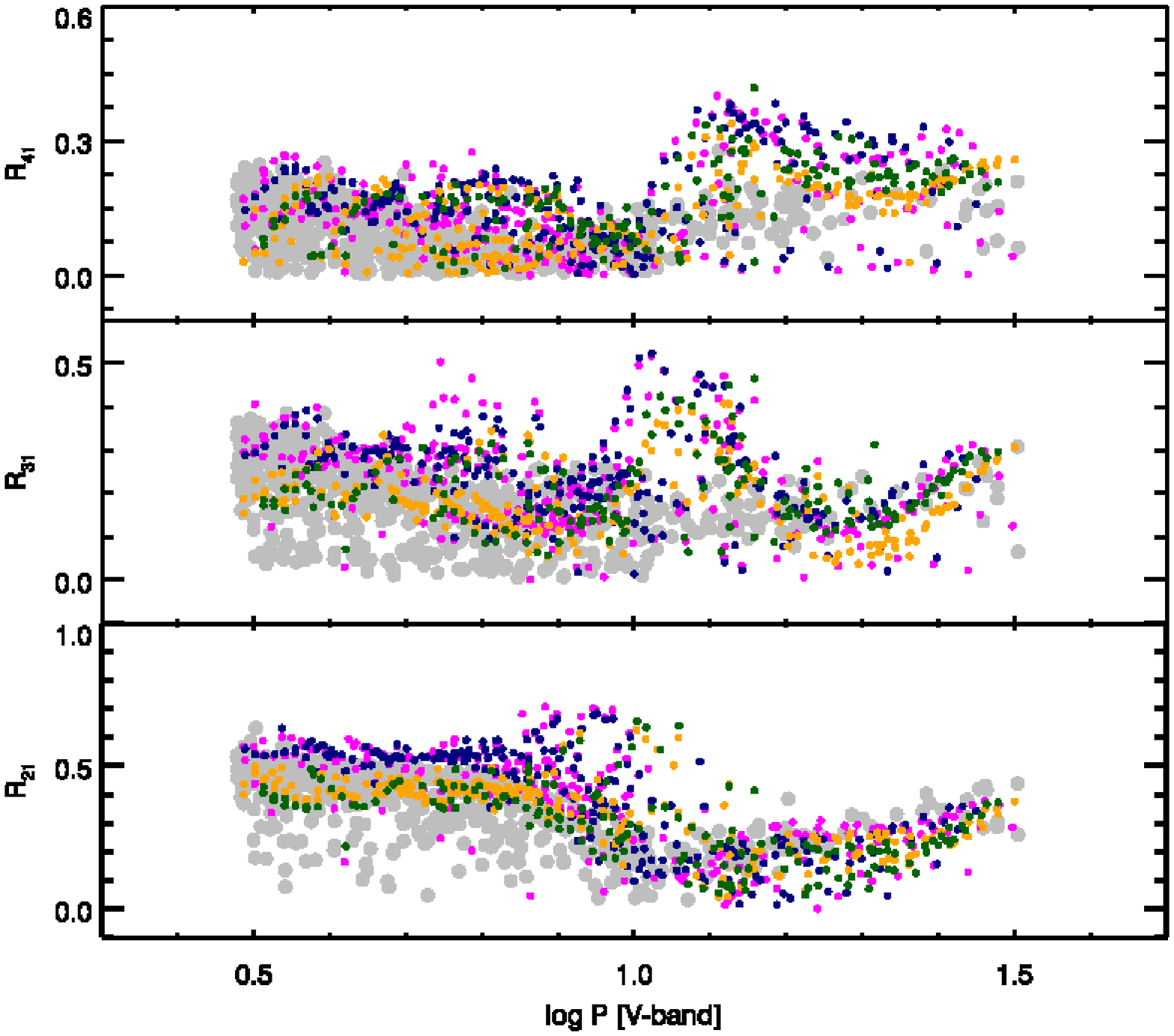}}&
\resizebox{0.48\linewidth}{!}{\includegraphics*{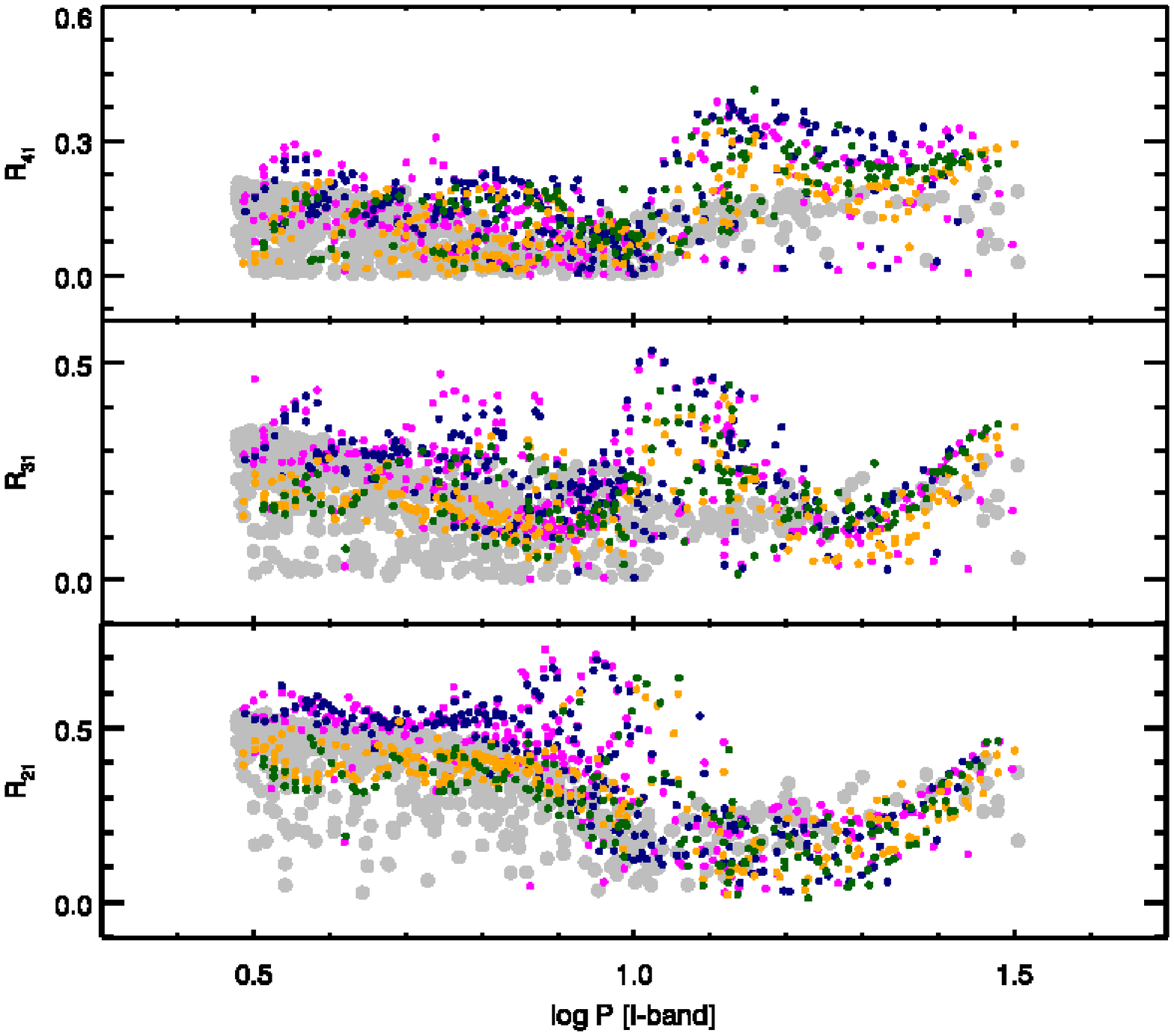}}\\
\end{tabular}
\caption{Comparison of Fourier amplitude ratios of theoretical Cepheid light curves obtained using  convection sets A, B, C, D with the observed ones for the LMC (upper panel) and SMC (lower panel) in the $VI$-bands, respectively.}
\label{fig:fou_r_theo_lmcsmc}
\end{figure*}

\begin{figure*}
%\scalebox{0.95}{
\begin{tabular}{cccc}
\resizebox{0.48\linewidth}{!}{\includegraphics*{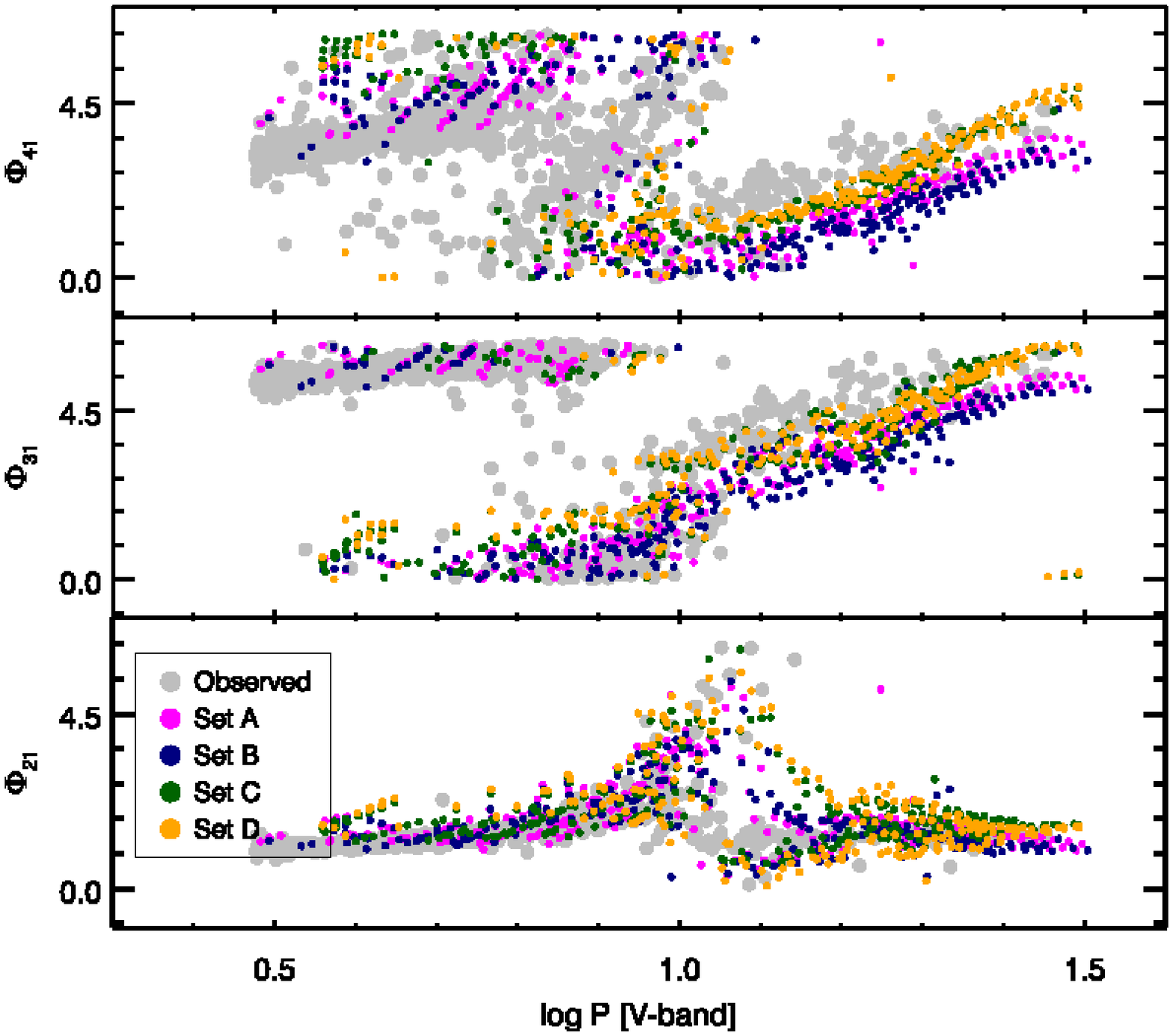}}&
\resizebox{0.48\linewidth}{!}{\includegraphics*{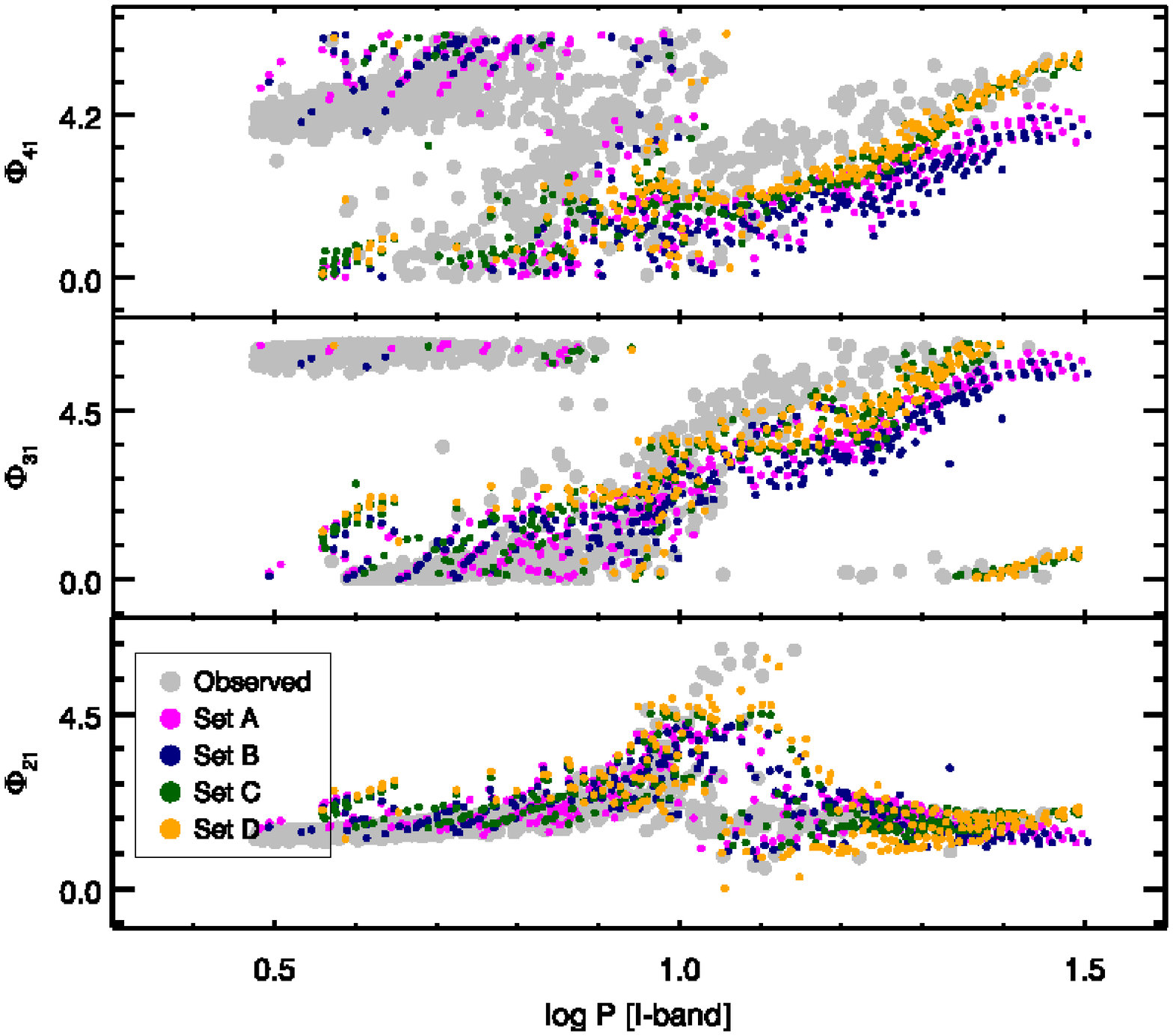}}\\
\resizebox{0.48\linewidth}{!}{\includegraphics*{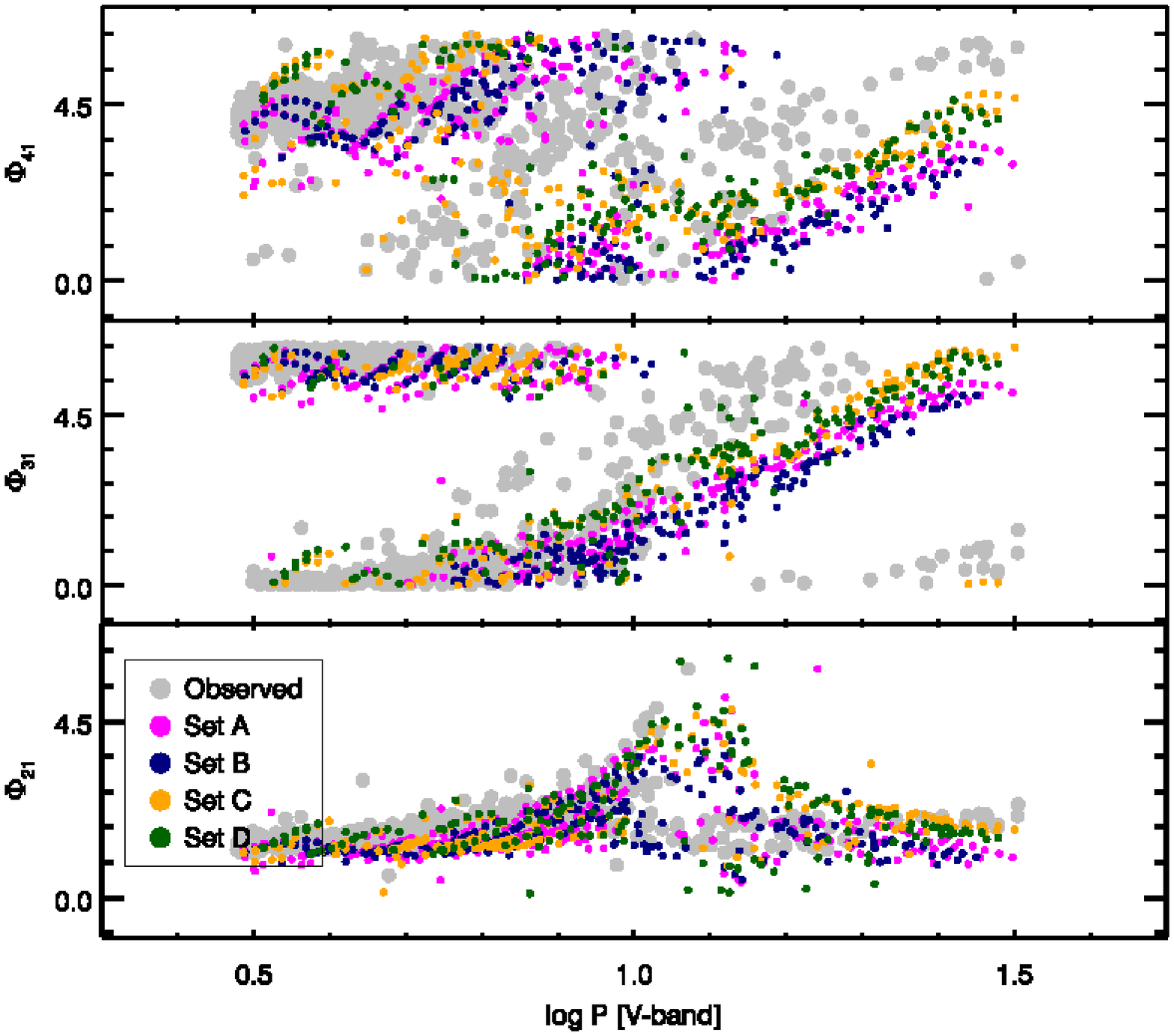}}&
\resizebox{0.48\linewidth}{!}{\includegraphics*{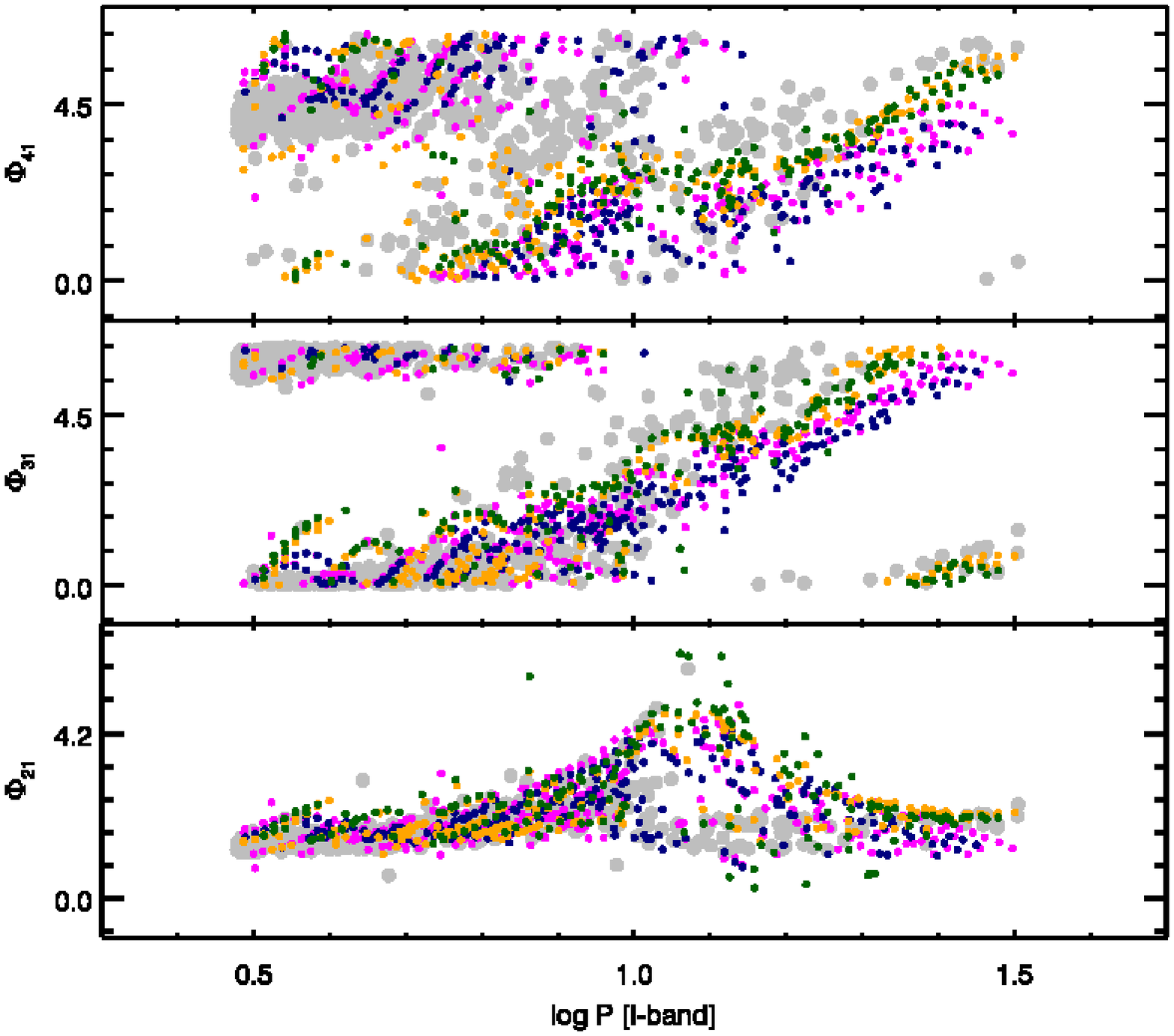}}\\
\end{tabular}
\caption{Same as Fig.~\ref{fig:fou_r_theo_lmcsmc} but for Fourier phase parameters.}
\label{fig:fou_phi_theo_lmcsmc}
\end{figure*}

\begin{figure*}
%\scalebox{0.95}{
\begin{tabular}{cccc}
\resizebox{0.48\linewidth}{!}{\includegraphics*{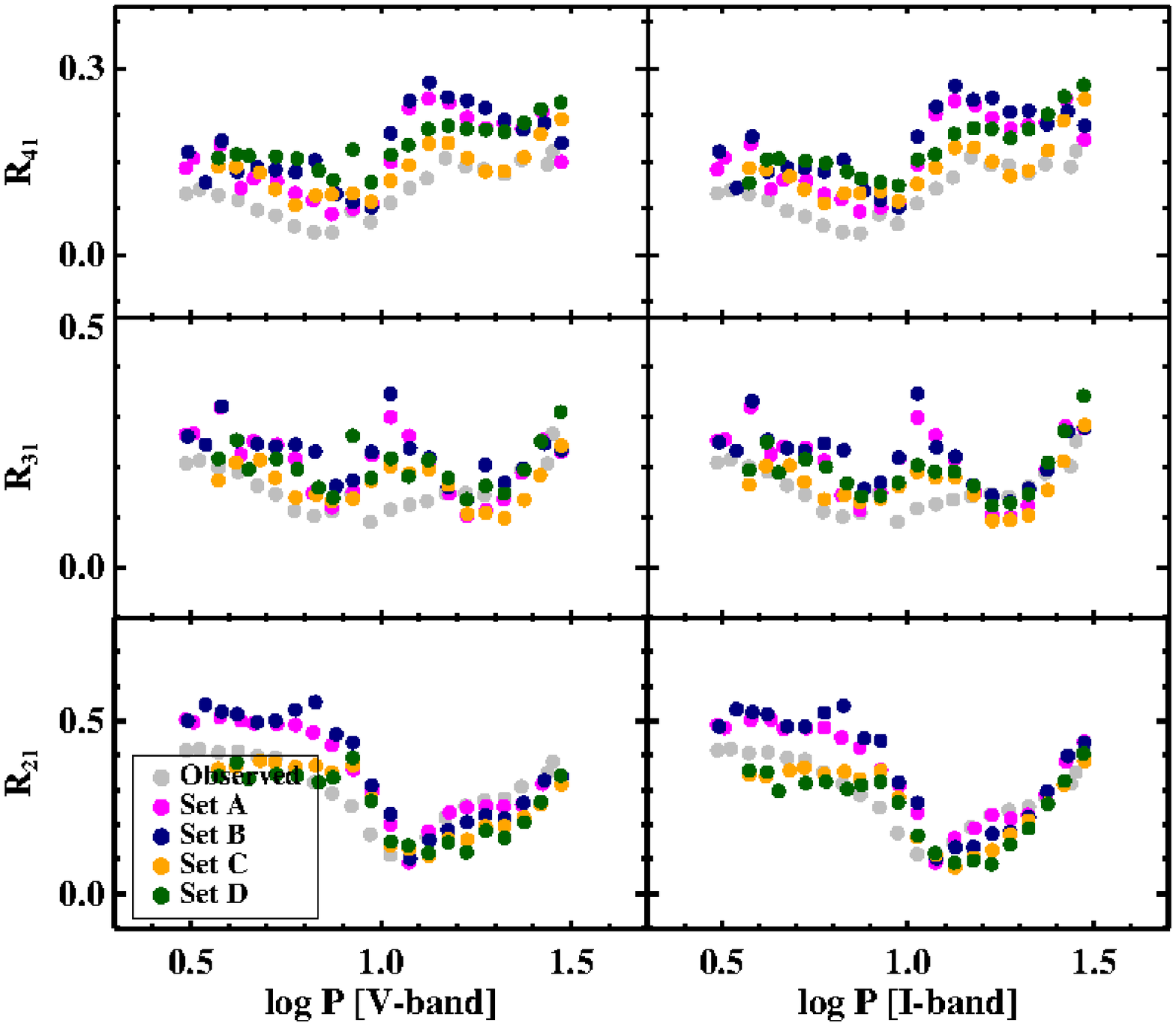}}&
\resizebox{0.48\linewidth}{!}{\includegraphics*{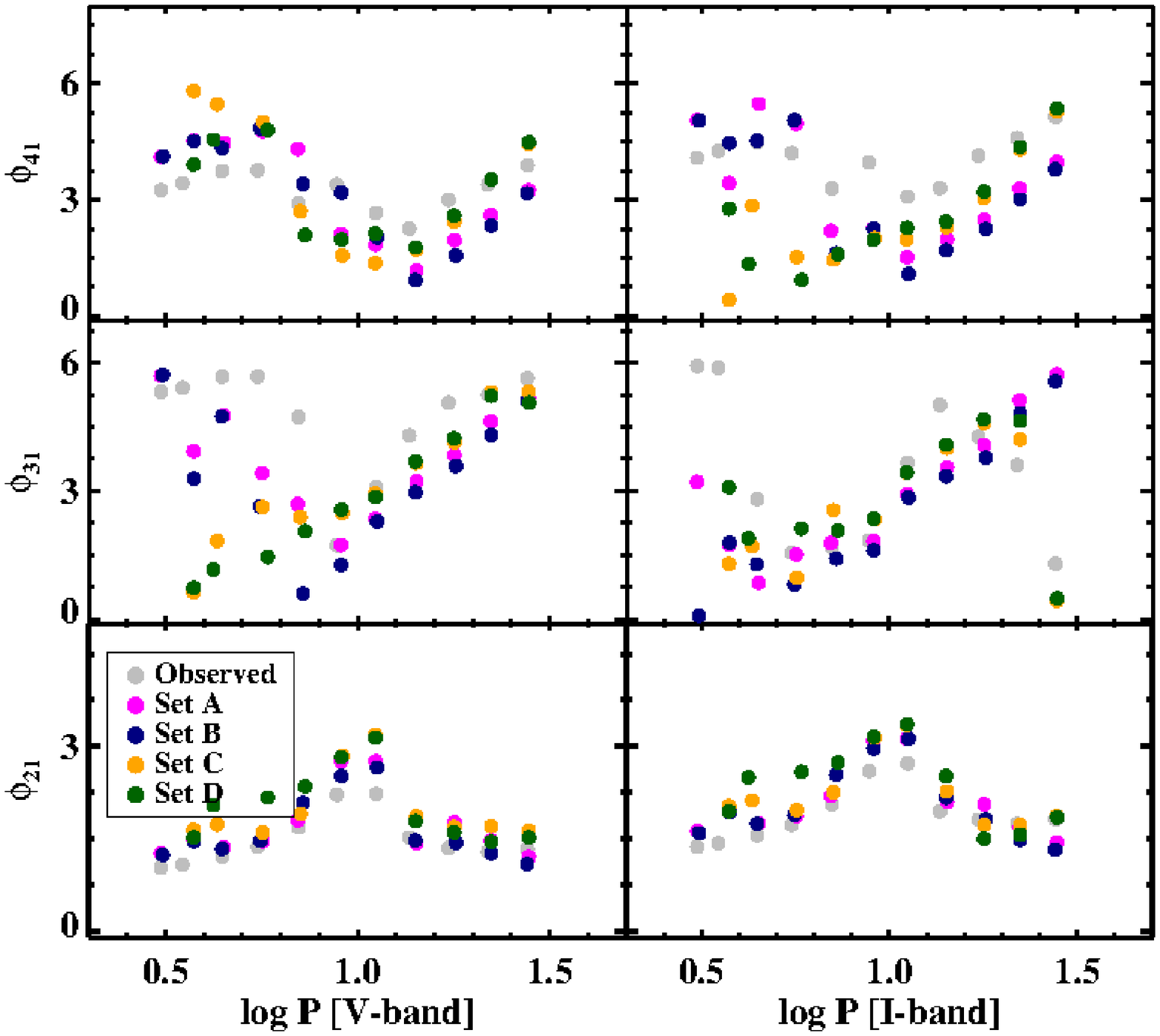}}\\
\resizebox{0.48\linewidth}{!}{\includegraphics*{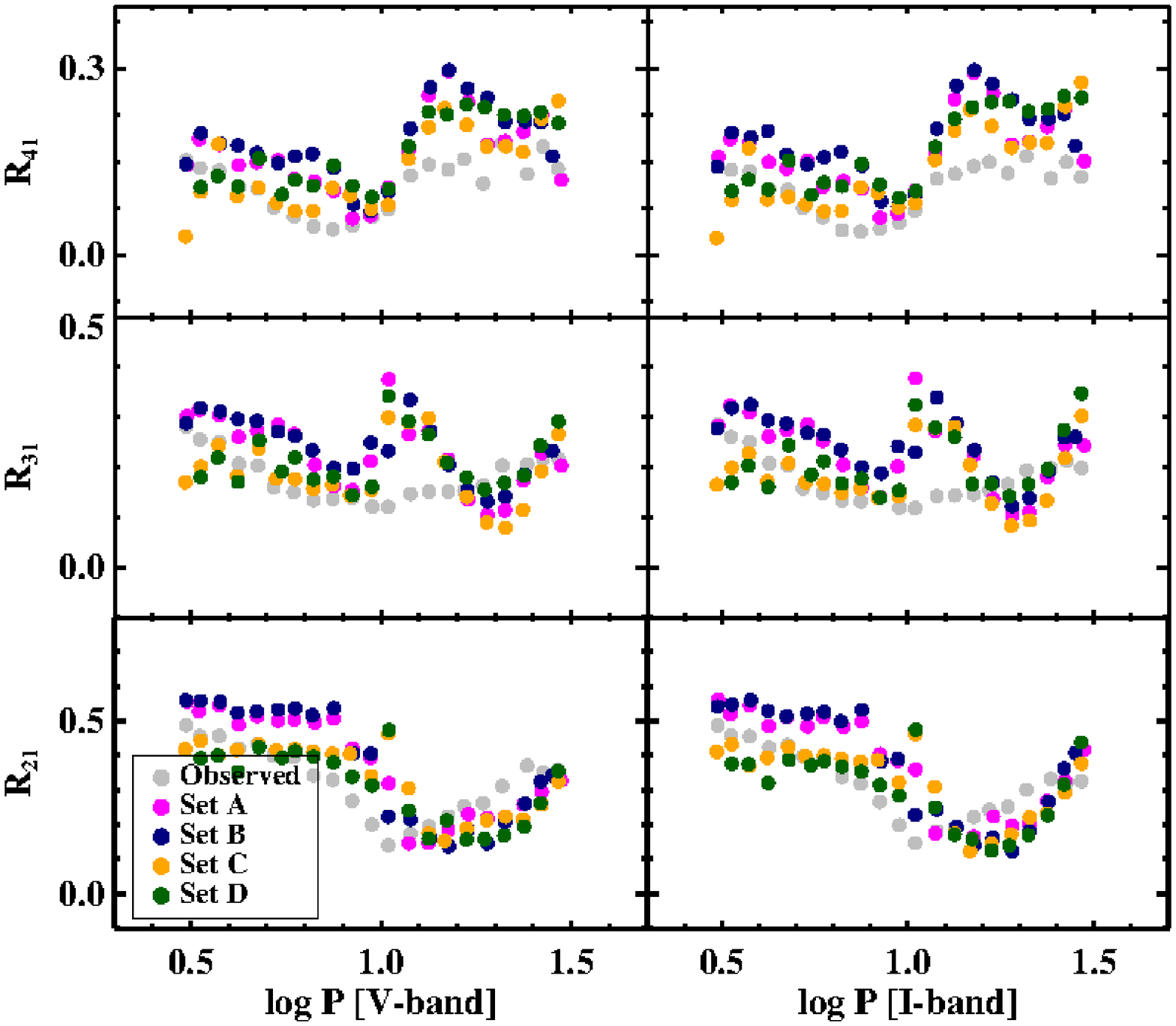}}&
\resizebox{0.48\linewidth}{!}{\includegraphics*{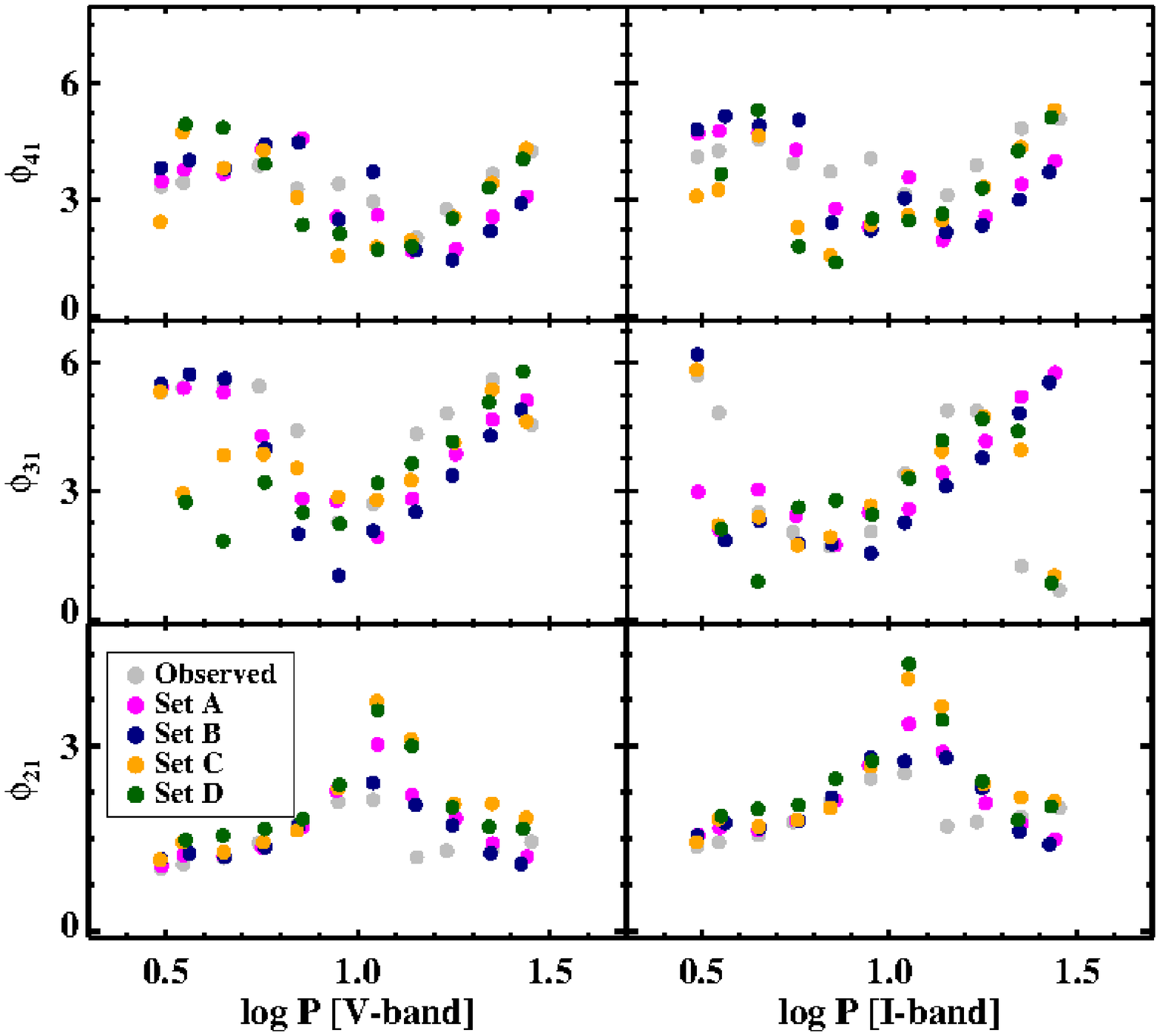}}\\
\end{tabular}
\caption{A comparison of the theoretical and observed $VI$-bands Fourier parameters for FU Cepheids in the LMC (upper panel) and SMC (lower panel). The plots are obtained by binning the entire period
range in steps of $\log{P}=0.05$ dex.}
\label{fig:fou_iv_bin}
\end{figure*}
Figs.~\ref{fig:fou_r_theo_lmcsmc} and~\ref{fig:fou_phi_theo_lmcsmc} displays the Fourier amplitudes and  phase parameters as a function of period for the Cepheid light curves obtained from the models in all four convection sets and from observations for the LMC (upper panel) and SMC (lower panel) in $(V,I)$-bands. From Fig.~\ref{fig:fou_r_theo_lmcsmc}, the following points may be noted for the amplitude parameters:
\begin{enumerate}
\item $R_{21}$ values obtained from the models using the convection sets (A, B, C, D) are found to be consistent with the observed ones for long periods in both the bands. Some of the $R_{21}$ values obtained from the four sets are found to be higher than their observed values for short periods in $0.8<\log{P}<1.0$ for both LMC and SMC.  
\item For both LMC and SMC, there is an agreement in the $R_{31}$ values obtained using convection sets A, B, C, D with the observations for long periods. For short periods between $0.7<\log{P}<1.15$, some of the $R_{31}$ obtained using the four sets are found to be higher than the  observations.
\item $R_{41}$ values obtained from the $4$ sets are consistent with the observations in the short period. However, a discrepancy of the $R_{41}$ values is observed for long periods in both LMC and SMC. 
\end{enumerate}

In all the aforementioned cases, Fourier amplitude parameters obtained  from the models using all four sets of convection parameters which seem to have higher values as compared to observations for the same short period range ($0.7<\log{P}<1.1$) are selected for both $(V,I)$-bands. The difference in mean values of these parameters as obtained from the sets and observations using the same period range is calculated. The resulting values are given in Table~\ref{tab:r_lmc}. It is evident from the table that the difference in amplitude parameters as obtained from the four convection sets and observations lies in $(10-30)\%$ range. Discrepancy is found to be minimum for $R_{41}$ as compared to $R_{21}$ and $R_{31}$ parameters. Thus, the amplitude parameters are found to be systematically larger than the observed ones towards the short period end. This is consistent with the results as obtained by \citet{bhar17a}. Full-amplitude, non-linear, convective Cepheid models were used in \citet{bhar17a} to generate the theoretical Cepheid light curves in the mass range $5.4-6.8~M_{\odot}$ employing the code, physical and numerical assumptions as in \citet[and reference therein]{marc13}.

Similarly, from Fig.~\ref{fig:fou_phi_theo_lmcsmc}, the following points may be noted for the phase parameters:
\begin{enumerate}
\item $\phi_{21}$ values obtained from the four convection sets are found to be consistent with the observations for both the LMC and SMC short/long period Cepheids.
\item For short periods, the $\phi_{31}$ values as well as $\phi_{41}$ values obtained from the models using  sets A, B, C, D were found to agree very closely with the observed ones. However, for long periods, the $\phi_{31}$  and $\phi_{41}$ values obtained using sets C and D agree more closely with observations than the other two sets (A and B) for the LMC/SMC in both the two bands. 
\end{enumerate}

\begin{figure*}
%\scalebox{0.95}{
\begin{tabular}{cc}
\resizebox{0.5\linewidth}{!}{\includegraphics*{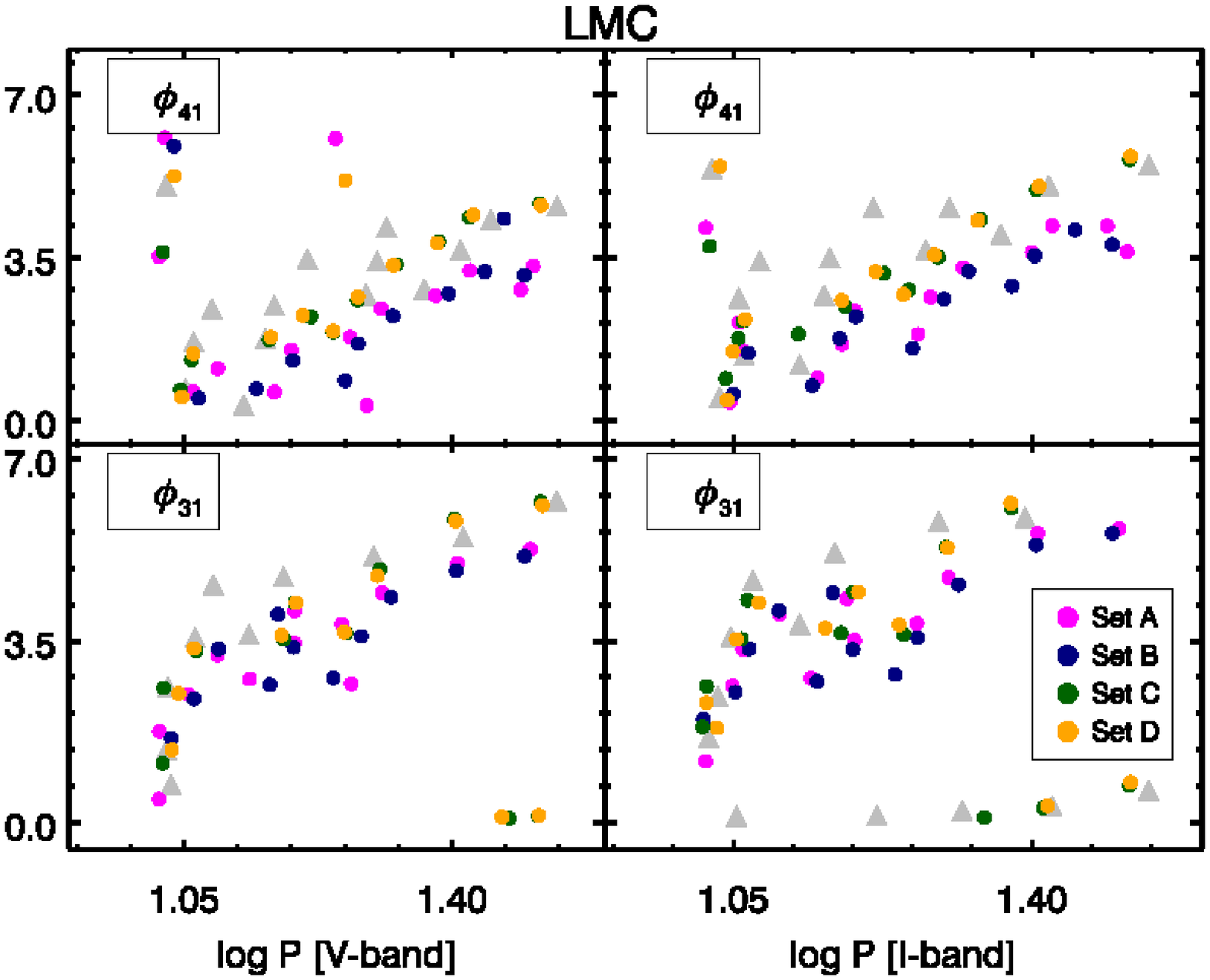}}& 
\resizebox{0.5\linewidth}{!}{\includegraphics*{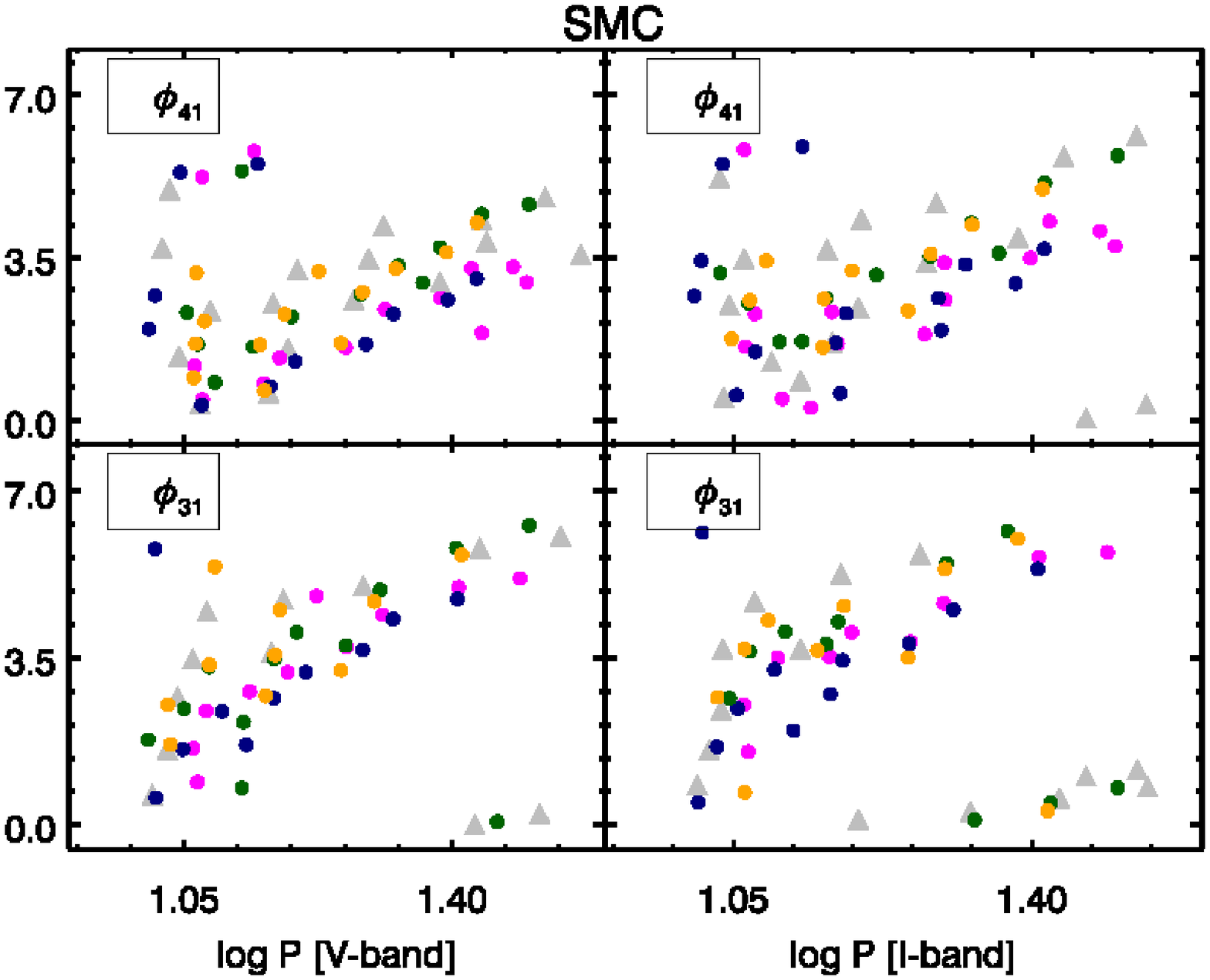}}\\
\end{tabular}
\caption{Plots depicting the average theoretical/observed $\phi_{31}$  and $\phi_{41}$ values for the LMC (left panel) and SMC
(right panel). The observed parameters are represented in grey for both LMC and SMC.}
\label{fig:phi_lmcsmc}
\end{figure*}

In order to check how closely the phase parameters $\phi_{31}$ and $\phi_{41}$ obtained as a function of $\log{P}$ using sets C and D match with the observations, the parameters are binned in ($5\times5$) coordinate grids in the $\log{P}-\phi_{31}/\phi_{41}$ plane. The average binned values are calculated and are plotted in Fig.~\ref{fig:phi_lmcsmc}. It is evident from the figure that $\phi_{31}$, $\phi_{41}$ parameters obtained using sets C and D match more closely with the observed ones. 
\begin{table}
\caption{Summary of the difference in mean values of the theoretical amplitudes ratios with respect to observations.}
%\scalebox{0.95}{
\begin{tabular}{c c c c c c c c c} \\ \hline \hline
 && Bands &  Set A &  Set B & Set C &  Set D \\ \hline
$R_{21}$ &LMC &   $V$  & $0.241$ & $0.242$ &  $0.149$ & $0.175$ \\
         & &      $I$  &  $0.265$ & $0.253$ & $0.131$ & $0.153$\\
         &SMC &   $V$  & $0.280$ & $0.322$ & $0.362$  & $0.223$ \\
         & &      $I$  & $0.304$ & $0.319$ & $0.234$ & $0.206$ \\ \hline
$R_{31}$ &LMC &   $V$  & $0.218$ & $0.225$ & $0.176$ & $0.207$ \\
         &&       $I$  &  $0.214$ & $0.223$ & $0.173$ & $0.189$\\
         &SMC &   $V$  & $0.257$  & $0.221$ & $0.216$ & $0.240$ \\ 
         & &      $I$  & $0.255$  & $0.227$ & $0.179$ & $0.210$ \\ \hline
$R_{41}$  &LMC &  $V$  & $0.109$ & $0.145$ & $0.105$ & $0.123$ \\
          & &     $I$  & $0.128$ & $0.146$ & $0.124$ & $0.136$\\ 
          &SMC &  $V$  &  $0.131$ & $0.172$ & $0.125$ & $0.141$ \\  
          & &     $I$  & $0.144$ & $0.166$ & $0.151$ & $0.150$\\  \hline
\end{tabular}
\label{tab:r_lmc}
\end{table}
 
\begin{figure*}
\begin{tabular}{cc}
\resizebox{0.5\linewidth}{!}{\includegraphics*{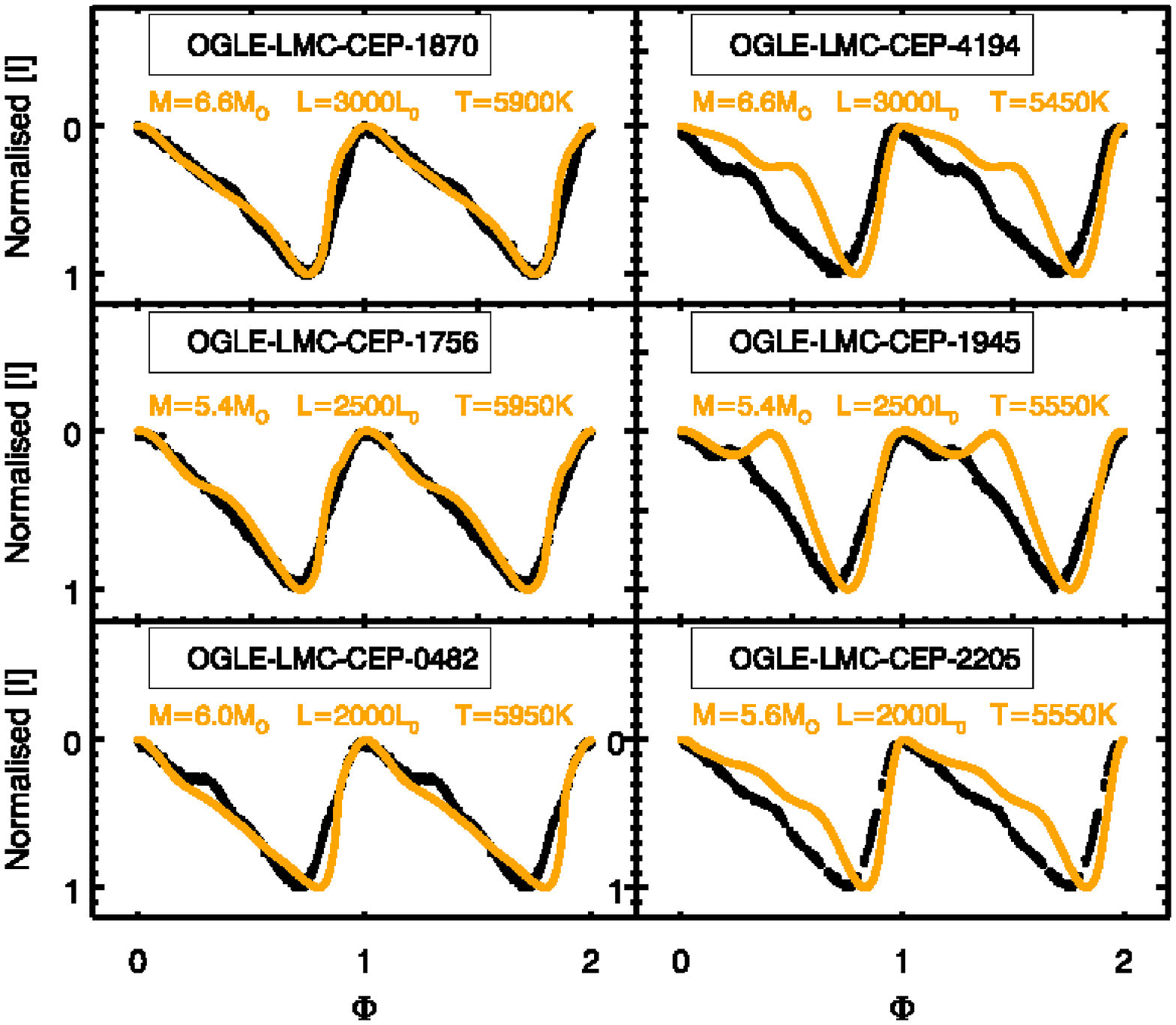}}&
\resizebox{0.5\linewidth}{!}{\includegraphics*{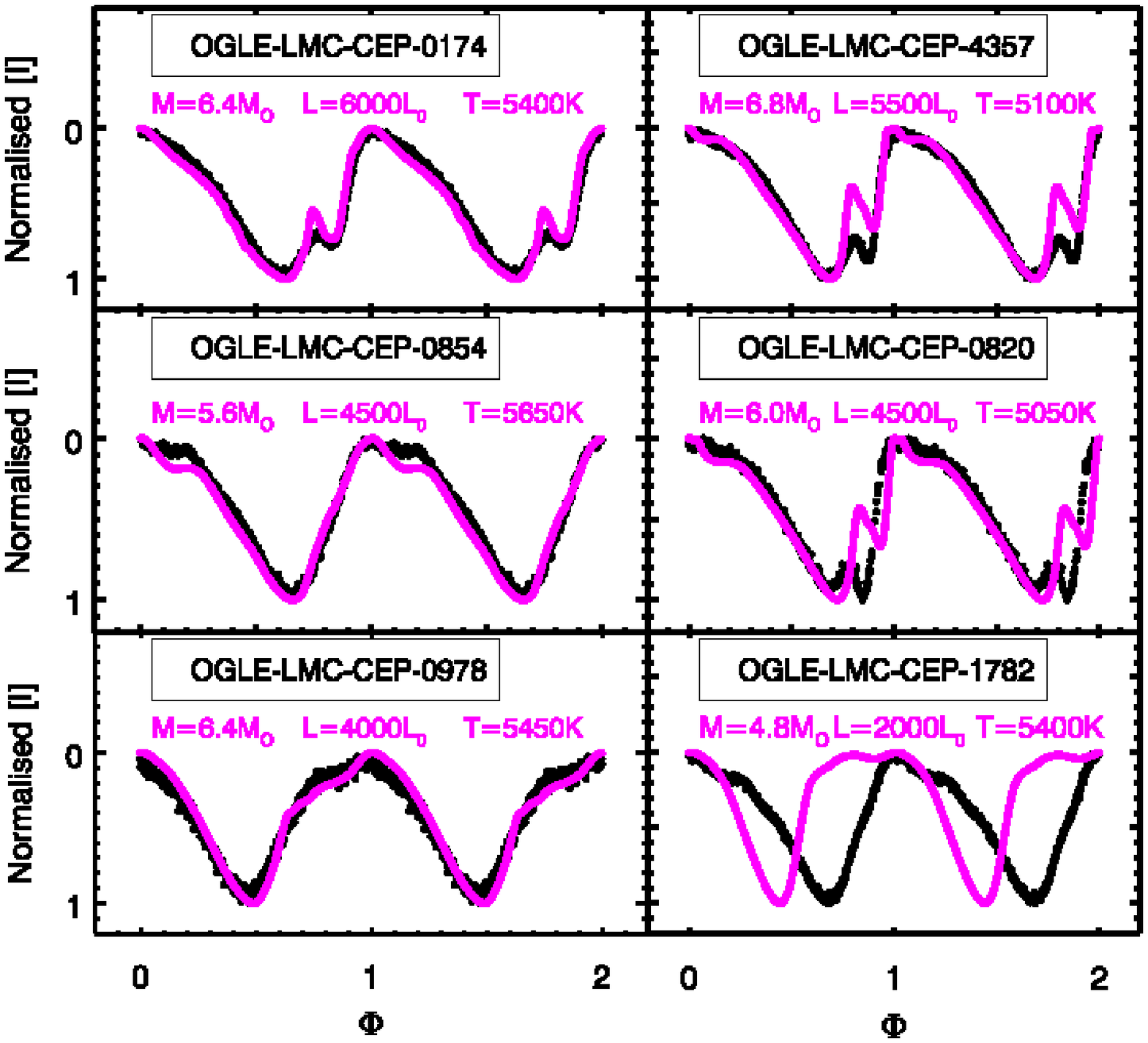}}\\
\end{tabular}
\caption{Normalized theoretical/observed light curves using set A (orange, \textit{left panel}) and using set C (magenta, \textit{right panel}) in the LMC with nearly similar period (correct up to $1$-$2$ decimal places) in $I$-band. The observed light curves are shown using black data points. The first column in each of the plots displays the light curves where the FP parameters are in agreement, whereas the second column shows the light curves where the discrepancy between the FP parameters is observed.}
\label{fig:light_curve_ot_lmc}
\end{figure*}

 The discrepancy in the FPs may be attributed to the difference in the structure of some theoretical light curves with respect to the observed ones. The same has also been reported in the analysis carried out by \citet{bhar17a}.  Normalized theoretical light curves of the models using sets A and C and observed light curves having nearly similar periods for the LMC are plotted in Fig.~\ref{fig:light_curve_ot_lmc}. From the second column of both the panels in the figure, it can be clearly seen that  the theoretical light curves appear to have higher/lower amplitudes in the ascending/descending branch of the Cepheid progression. The periods and FPs of theoretical/observed light curves of Fig.~\ref{fig:light_curve_ot_lmc} are given in Table \ref{tab:par_abcd_lmc}. The table also shows that dissimilar theoretical light curves have different FP values with a large offset from the observed ones. The theoretical light curves considered in this study have been obtained from a generalized grid consisting of limited mass ranges, as mentioned in the previous section (Section~\ref{sec:theoretical_models}). Nonetheless, the models are able to reproduce the observed light curves, but not perfectly for all individual light curves.  For some models, there occur a few discrepancies at those phases containing bumps. Since \textsc{mesa-rsp} follows \citet{smol08} in its treatment of stellar pulsation, results obtained by \citet{smol12} indicate that \textsc{mesa-rsp} can reliably model the observed light curve of a particular star by varying the values of convective parameters and as well as the input parameters $M, L, T$. This is further supported by \citet{paxt19} where light curves of models and observations are shown to match for a broad spectrum of pulsating variable star types (figures 16, 19 and 20 of \citet{paxt19}. Since a more generalized grid of model parameters has been used for modeling the observed Cepheid light curves in an ensemble fashion in the present study, individual modeling of the light curves has not been taken into account. However, any individual light curve as given in Fig.~\ref{fig:light_curve_ot_lmc} may be chosen and can be modelled well by adjusting the $M, L, T$ parameters even before one tries anything with the turbulent convection parameters.

 It is to be noted that while comparing models with observations on the basis of more and more parameters, the comparison space increases significantly when adding a comparison quantity. It is thereby a challenging task to have models that agree with observations on the basis of increasing parameters \citep{kova98}.  

\begin{figure*}
\scalebox{0.95}{=
\begin{tabular}{ccccc}
\resizebox{0.5\linewidth}{!}{\includegraphics*{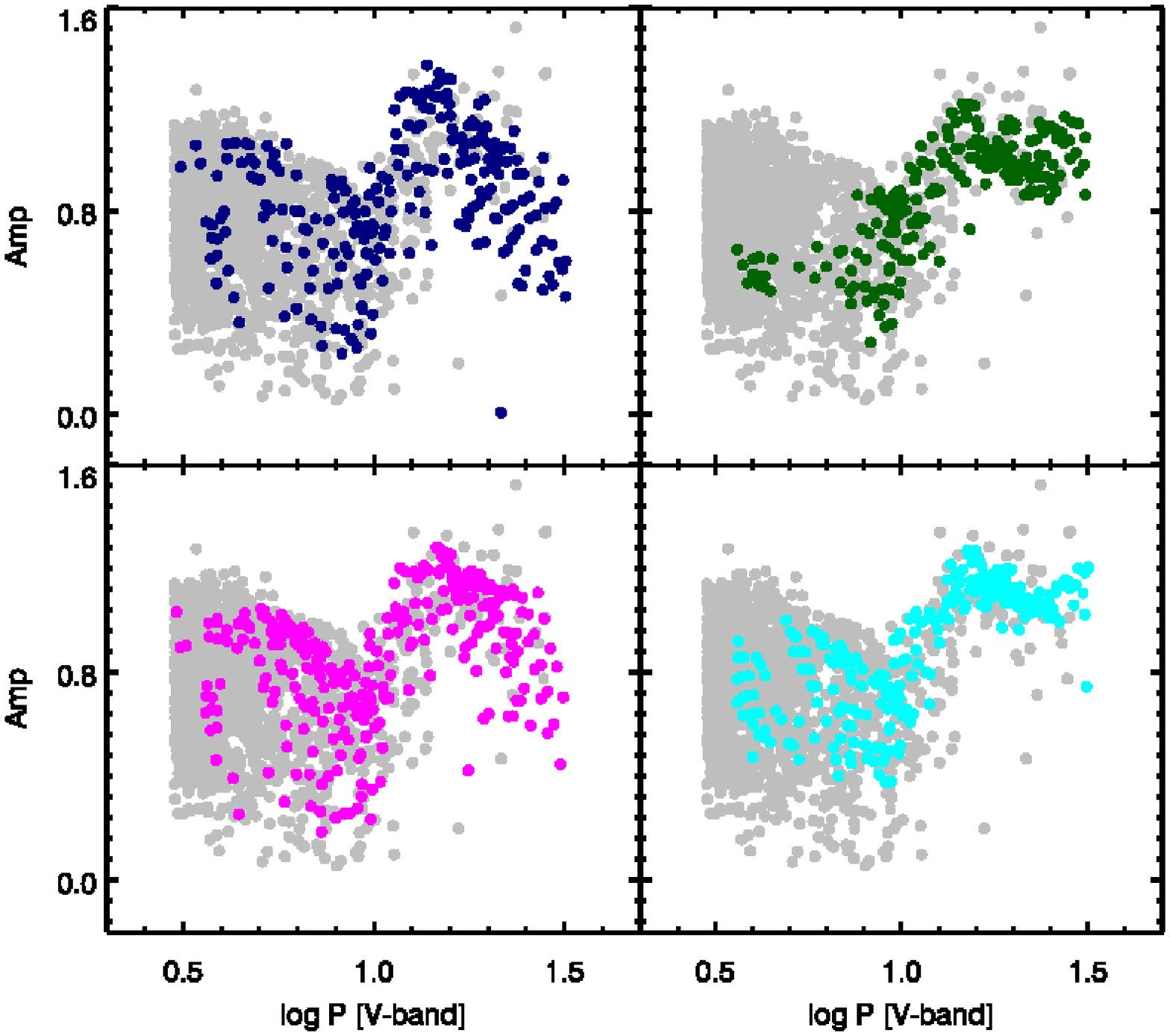}}& 
\resizebox{0.5\linewidth}{!}{\includegraphics*{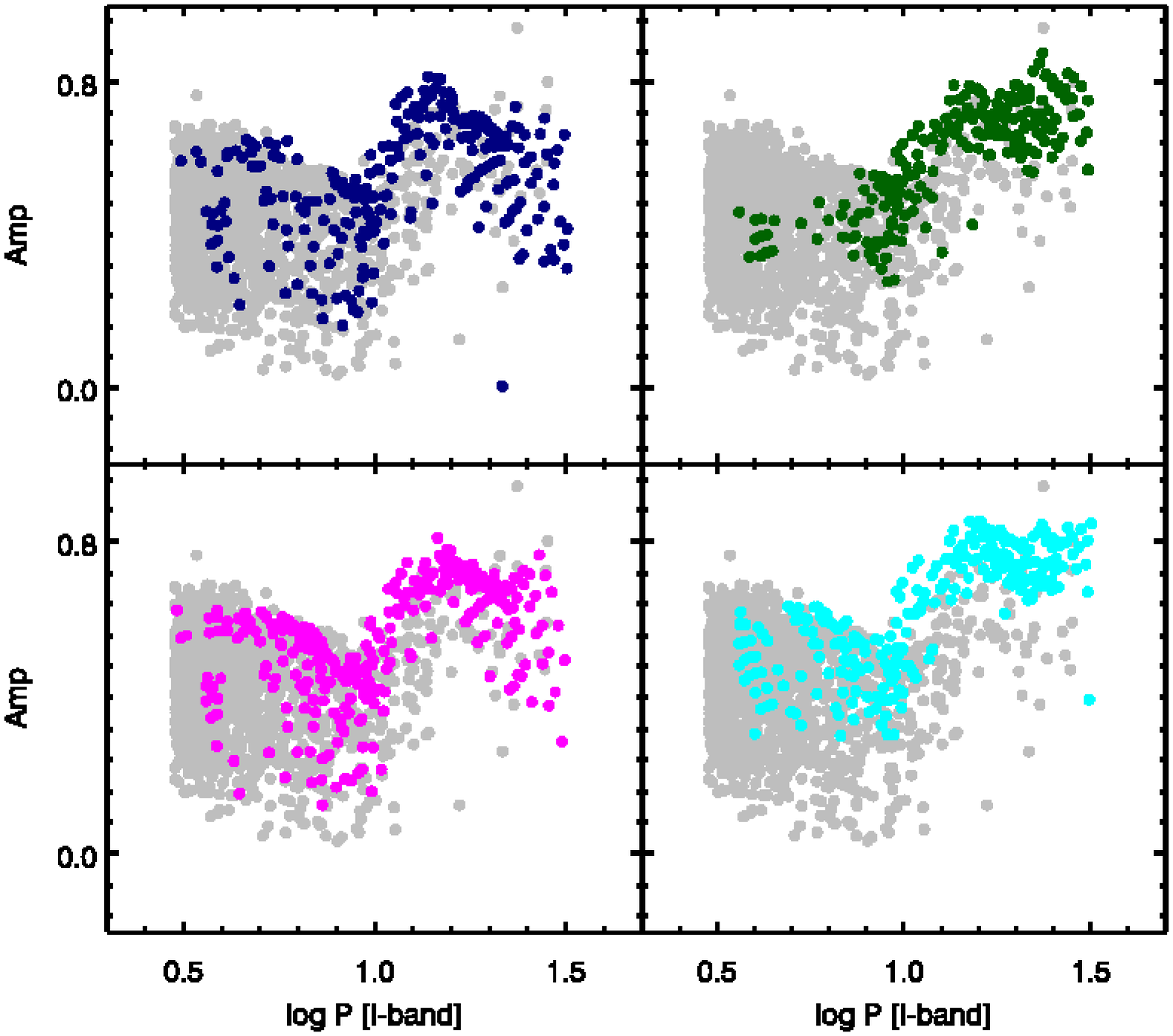}}\\
\resizebox{0.5\linewidth}{!}{\includegraphics*{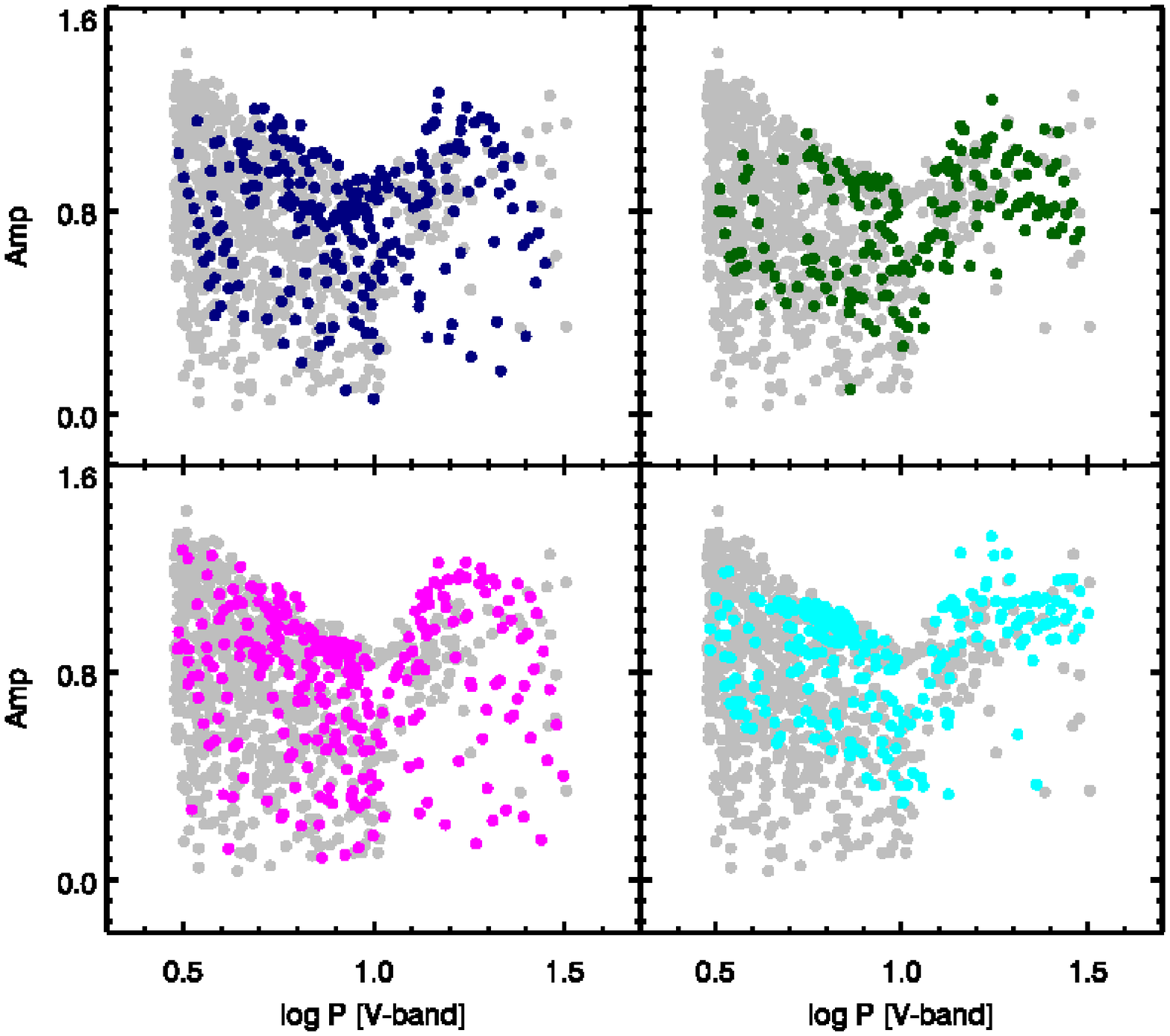}}& 
\resizebox{0.5\linewidth}{!}{\includegraphics*{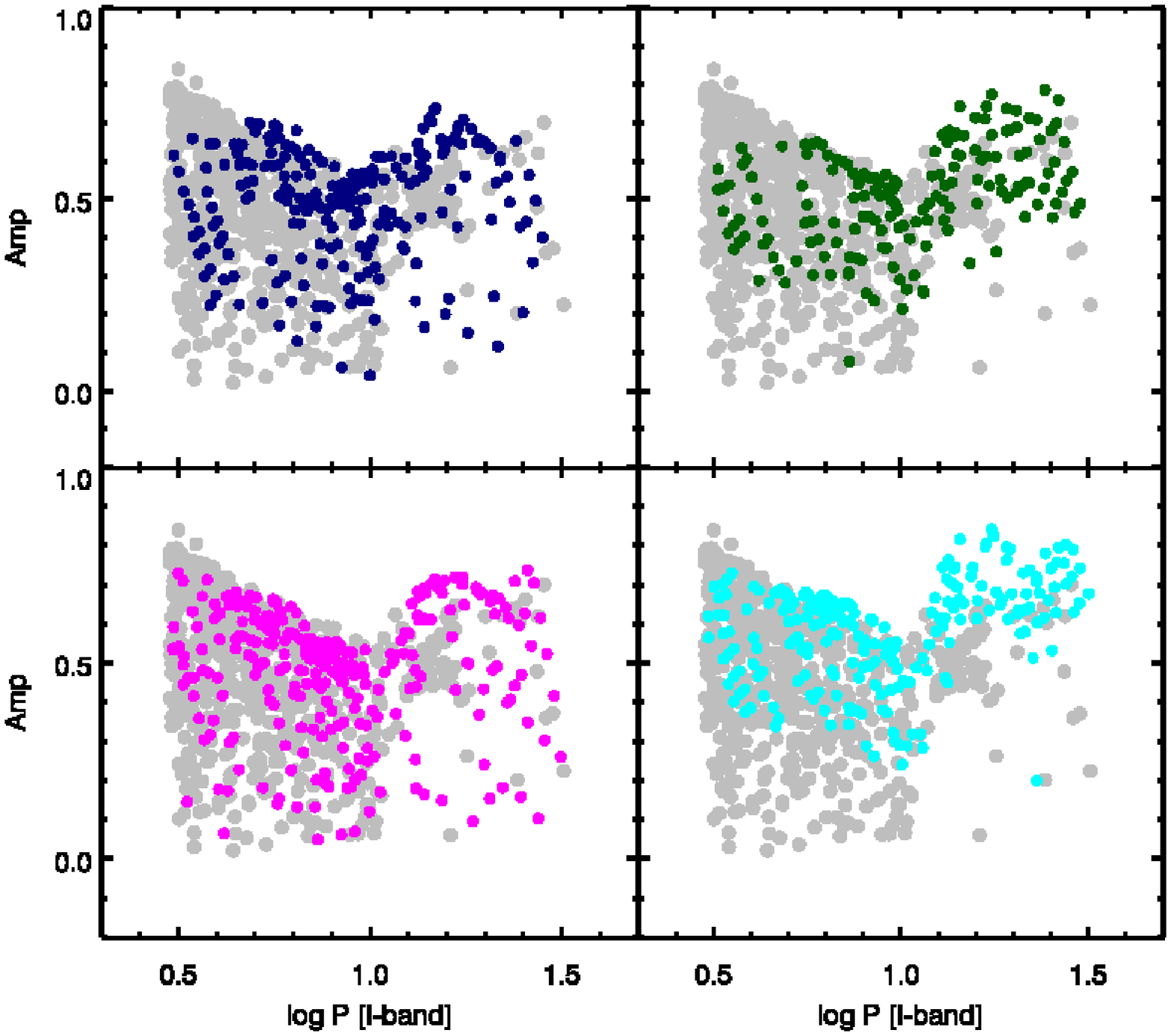}}\\
\end{tabular}}
\caption{Comparison of amplitude (Amp) variation in $VI$-bands as a function of period for the models using  sets A (magenta), B (navy blue), C (cyan) and D (dark green) for the LMC  (upper panel) and SMC (lower panel). Grey colour denote amplitudes from observations for both 
LMC and SMC.}
\label{fig:amp_lmcsmc}
\end{figure*}

Fig.~\ref{fig:amp_lmcsmc} display the theoretical amplitudes of the models and observed  amplitudes in $VI$- bands for the LMC (upper panel) and SMC (lower panel). The theoretical amplitudes obtained from sets A, B, C, D are consistent with observations at short periods in both $VI$- bands. We note that all the four convection sets detect the “turnover” in amplitudes seen in the observations at around $\log{P} \sim 1.2/1.3$. For long periods, in case of the LMC, some of the theoretical amplitudes display higher values than the observed ones in $I$- band but are found to be consistent in $V$-band. However, for SMC, some of the models display higher amplitudes values in both the bands.

\begin{table}
\caption{Summary of the number of theoretical and observed light curves used in the analysis for the LMC and SMC. Here $N$:total number of light curves, $N_{s}$: number of short period light curves and $N_{l}$: number of long period light curves.}
\begin{tabular}{c c c c c c c} \\ \hline \hline
       & & Set A & Set B & Set C & Set D & Observations \\ \hline
LMC & $N$        & $307$&   $237$& $267$ & $213$ & $1530$ \\
    & $N_{s}$    & $169$&   $108$ & $115$  & $66$ &$1407$  \\
    & $N_{l}$    & $138$&   $129$& $142$ & $147$ & $123$ \\ \hline
SMC & $N$       & $293$ & $234$& $240$ &$171$ &  $703$\\ 
    & $N_{s}$   & $203$ &$149$& $155$ & $87$ & $612$ \\ 
    & $N_{l}$   & $90$ & $85$ & $85$ & $84$ & $91$ \\ \hline
\end{tabular}
\label{tab:model_number_lmcsmc}
\end{table}

\begin{table*}
%\scalebox{0.90}{
\caption{FPs of theoretical/observed light curves related to Fig.~\ref{fig:light_curve_ot_lmc} for the LMC.}
\begin{tabular}{c c c c c c c c r r r r} \\ \hline \hline
& $P$ & $R_{21}$ & $R_{31}$ & $R_{41}$ & $\phi_{21}$ & $\phi_{31}$ & $\phi_{41}$ & Amp  \\ \hline
Left panel (first column)&          &            &             &  \\ \hline
OGLE-LMC-CEP-0482  & $4.416$  & $0.456$   & $0.232$ & $0.129$  & $1.480$  & $6.175$    &$4.459$ & $0.581$ \\
$M=\rm 6.0~M_{\odot}$, $L=\rm 2000~L_{\odot}$, $T_{\rm eff}=5950$~K & $4.410$   &$0.481$   &$0.254$ & $0.125$ & $1.587$ & $6.155$ & $4.735$ & $0.604$\\
OGLE-LMC-CEP-1756 & $5.738$  & $0.322$  & $0.081$  & $0.015$ & $1.791$ & $0.460$ & $6.116$ & $0.327$ \\
$M=\rm5.4~M_{\odot}$, $L=\rm2500~L_{\odot}$, $T_{\rm eff}=5950$~K   & $5.719$ & $0.460$ & $0.115$ & $0.086$ & $1.891$ & $0.749$ & $6.153$ & $0.575$ \\ 
OGLE-LMC-CEP-1870 & $6.021$ & $0.408$ & $0.140$ & $0.070$ & $1.863$ & $0.286$ & $5.350$ & $0.516$ \\
$ M=\rm 6.6~M_{\odot}$, $L=\rm 3000~L_{\odot}$, $T_{\rm eff}=5900$~K & $6.021$ & $0.474$ & $0.207$ & $0.089$ & $1.741$ & $0.489$ & $5.698$ & $0.595$\\ \hline
 Left panel (second column)&          &            &             &  \\ \hline
OGLE-LMC-CEP-2205 &  $5.993$ & $0.424$ &  $0.161$ &   $0.096$ & $1.788$ & $0.072$ & $5.078$ & $0.546$  \\
$M=\rm 5.6M_{\odot}$, $L=\rm 2000~L_{\odot}$, $T_{\rm eff}=5550$~K & $5.922$ & $0.535$ & $0.350$ & $0.136$ & $2.108$ & $1.103$ & $6.133$ & $0.330$ \\
OGLE-LMC-CEP-1945 & $7.685$ &  $0.303$ & $0.088$  & $0.054$ & $2.381$  & $0.958$  & $3.373$ & $0.358$ \\
$M=\rm 5.4~M_{\odot}$, $L=\rm2500~L_{\odot}$, $T_{\rm eff}=5500$~K & $7.604$ & $0.619$ & $0.110$ & $0.100$ & $2.993$ & $1.748$ & $0.985$ & $0.255$ \\
OGLE-LMC-CEP-4194 & $8.024$  & $0.337$  & $0.196$  & $0.017$ & $1.886$ & $6.194$ & $4.232$ & $0.551$ \\
$M=\rm6.6~M_{\odot}$, $L=\rm 3000~L_{\odot}$, $T_{\rm eff}=5450$~K & $8.015$ & $0.592$ & $0.223$ & $0.110$ & $2.321$ & $1.749$ & $6.256$ & $0.278$ \\ \hline
Right panel (first column)&          &            &             &  \\ \hline
 OGLE-LMC-CEP-0978 &$10.520$   & $0.206$ & $0.062$ & $0.013$ & $3.079$ & $2.295$ & $4.944$ & $0.251$  \\
$M=\rm 6.4~M_{\odot}$, $L=\rm 4000~L_{\odot}$, $T_{\rm eff}=5450$~K & $10.548$ & $0.254$ & $0.138$ & $0.034$ & $3.558$ & $2.522$  & $1.824$ & $0.465$ \\ 
 OGLE-LMC-CEP-0854 & $9.049$   & $0.225$   & $0.069$   & $0.040$  & $2.379$ &      $0.974$ &      $4.035$ & $0.488$ \\
 $M=\rm5.6~M_{\odot}$, $L=\rm 4500~L_{\odot}$, $T_{\rm eff}=5650$~K & $9.017$ & $0.267$ & $0.052$ & $0.075$ & $2.293$ & $0.338$ & $2.140$ & $0.524$\\
OGLE-LMC-CEP-0174  &$15.863$   &$0.124$   &$0.144$   &$0.127$   &$1.487$      &$4.929$      &$3.369$& $0.466$ \\
$M=\rm6.4~M_{\odot}$, $L=\rm 6000~L_{\odot}$, $T_{\rm eff}=5400$~K & $15.864$ & $0.115$ & $0.151$ & $0.145$ & $1.610$ & $4.475$ & $2.572$ & $0.853$ \\ \hline
Right panel (second column) &          &            &         &  \\ \hline
OGLE-LMC-CEP-1782 &$7.393$   &$0.311$   &$0.065$   &$0.042$  &$2.285$      &$0.857$      &$2.834$ & $0.333$ \\
$M=\rm4.8~M_{\odot}$, $L=\rm 2000~L_{\odot}$, $T_{\rm eff}=5400$~K & $7.328$  & $0.440$ & $0.136$ & $0.089$ & $3.072$ & $2.354$ & $1.707$ & $0.344$ \\
OGLE-LMC-CEP-0820& $16.836$   &$0.258$   &$0.160$   &$0.145$  &$1.975$      &$0.062$      &$4.837$& $0.585$ \\ 
$M=\rm 6.0~M_{\odot}$, $L=\rm 4500~L_{\odot}$, $T_{\rm eff}=5050$~K &  $16.808$   &$0.171$ & $0.057$ & $0.193$  & $2.299$ & $3.813$ & $2.695$ & $0.710$\\ 
OGLE-LMC-CEP-4357& $17.527$   &$0.198$   &$0.123$   &$0.145$ &$1.839$&$5.602$      &$3.978$ & $0.552$\\
$M=\rm 6.8~M_{\odot}$, $L=\rm5500~L_{\odot}$, $T_{\rm eff}=5100$~K &$17.567$ & $0.109$ & $0.098$ & $0.190$ & $2.181$ & $3.505$ & $2.492$& $0.737$  \\ \hline
\end{tabular}
\label{tab:par_abcd_lmc}
\end{table*}

\subsection{Multiphase relations in the LMC/SMC}
 Studies in the literature have reported the break in the PL relation of fundamental mode Cepheids  in the SMC at a period of $3$ d in optical bands \citep{subr15,bhar16} and also near infrared bands (\citep{ngeo10,ripe16}). In order to avoid any effect in the PL relation because of the break at the ${P}\sim 3$ d in the SMC, the light curves above $P\sim 3$ d are selected. For a consistency in the period range between models and observations, the light curves in the period range $3<P(\rm d)<32$ are considered in the analysis of PL, PC and AC relations as a function of phase for the theoretical and observed light curves for the SMC. The same has been done for the LMC, although there is no such break at $P\sim 3$ d reported in the literature. The same period range is maintained for both LMC and SMC throughout the analysis.
 
 To obtain the magnitudes of $50$ pulsation phases from both theoretical and observed light curves, equation \ref{equa:n01} is interpolated in $\Phi$ from $0$ to $1$ in steps of $0.02$ on the Fourier-fitted data points. Magnitudes corresponding to these $50$ pulsation phases are extracted from the interpolated light curves. A linear regression is then fitted to the data points thus obtained for each of the $50$ 
 phases to get the multiphase PL/PC/AC relations for both LMC/SMC. Multiphase Period-Luminosity/Period-Colour/Amplitude-Colour (MPPL/MPPC/MPAC) relations are obtained for short periods ($\log{P}<1$), long periods ($\log{P}>1.0$) and all periods. The number of theoretical and observed light curves used for the analysis for both LMC and SMC are summarized in Table~\ref{tab:model_number_lmcsmc}. 

\subsubsection{Multiphase Period-Luminosity (MPPL) relations} 
\begin{figure*}
\scalebox{0.98}{
\begin{tabular}{c c c}
\resizebox{0.5\linewidth}{!}{\includegraphics*{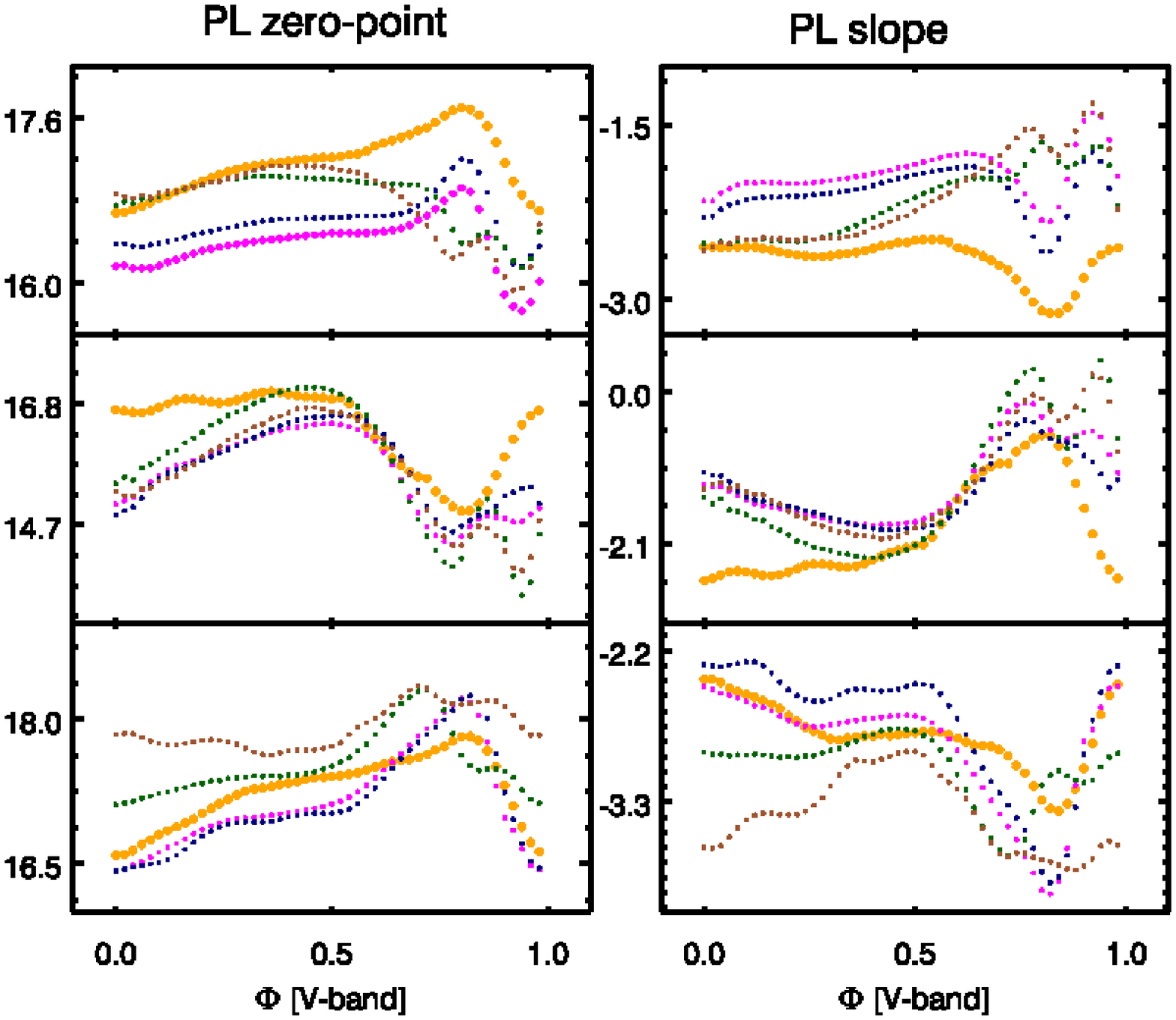}}&
\resizebox{0.5\linewidth}{!}{\includegraphics*{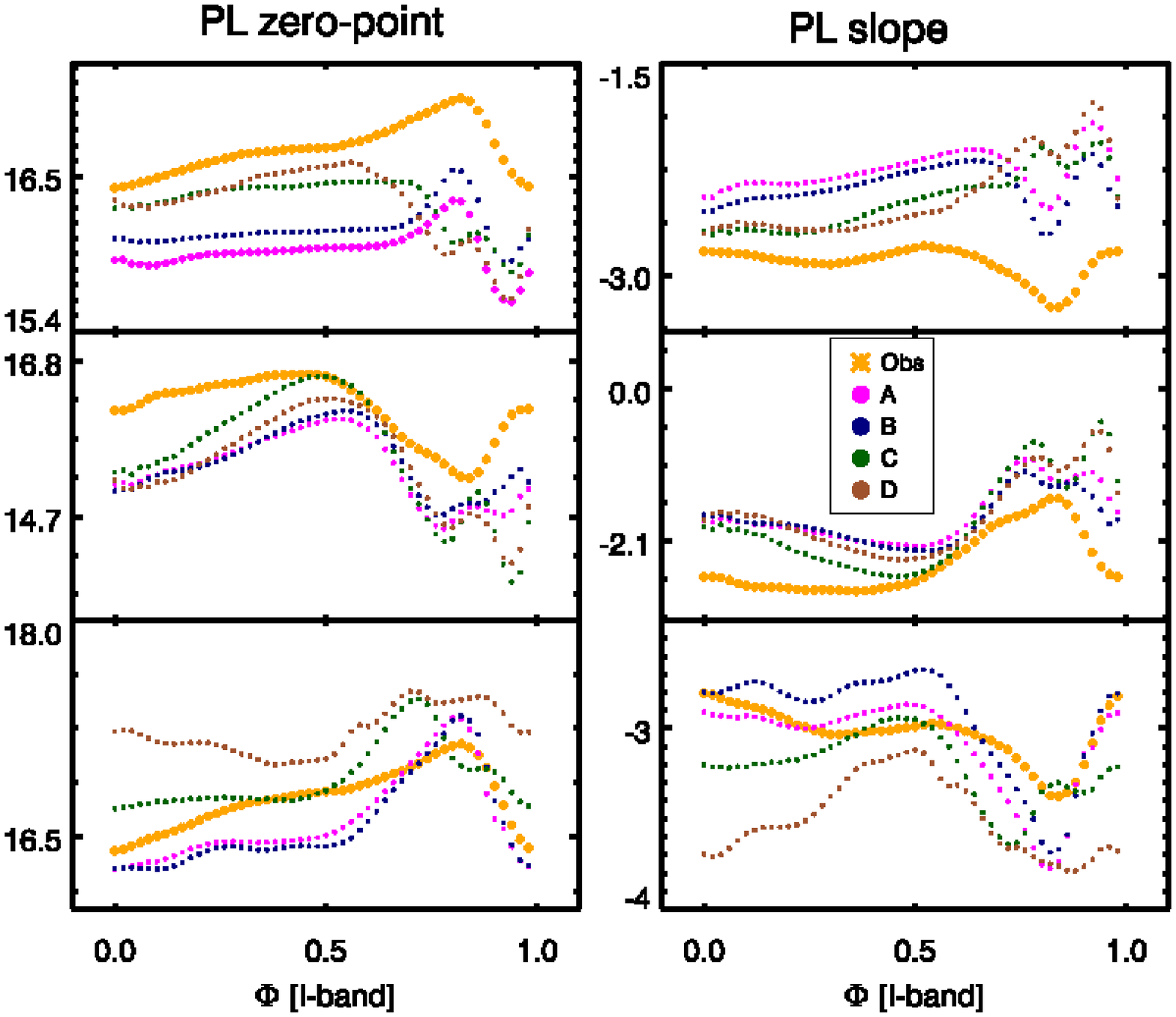}}\\ 
\resizebox{0.5\linewidth}{!}{\includegraphics*{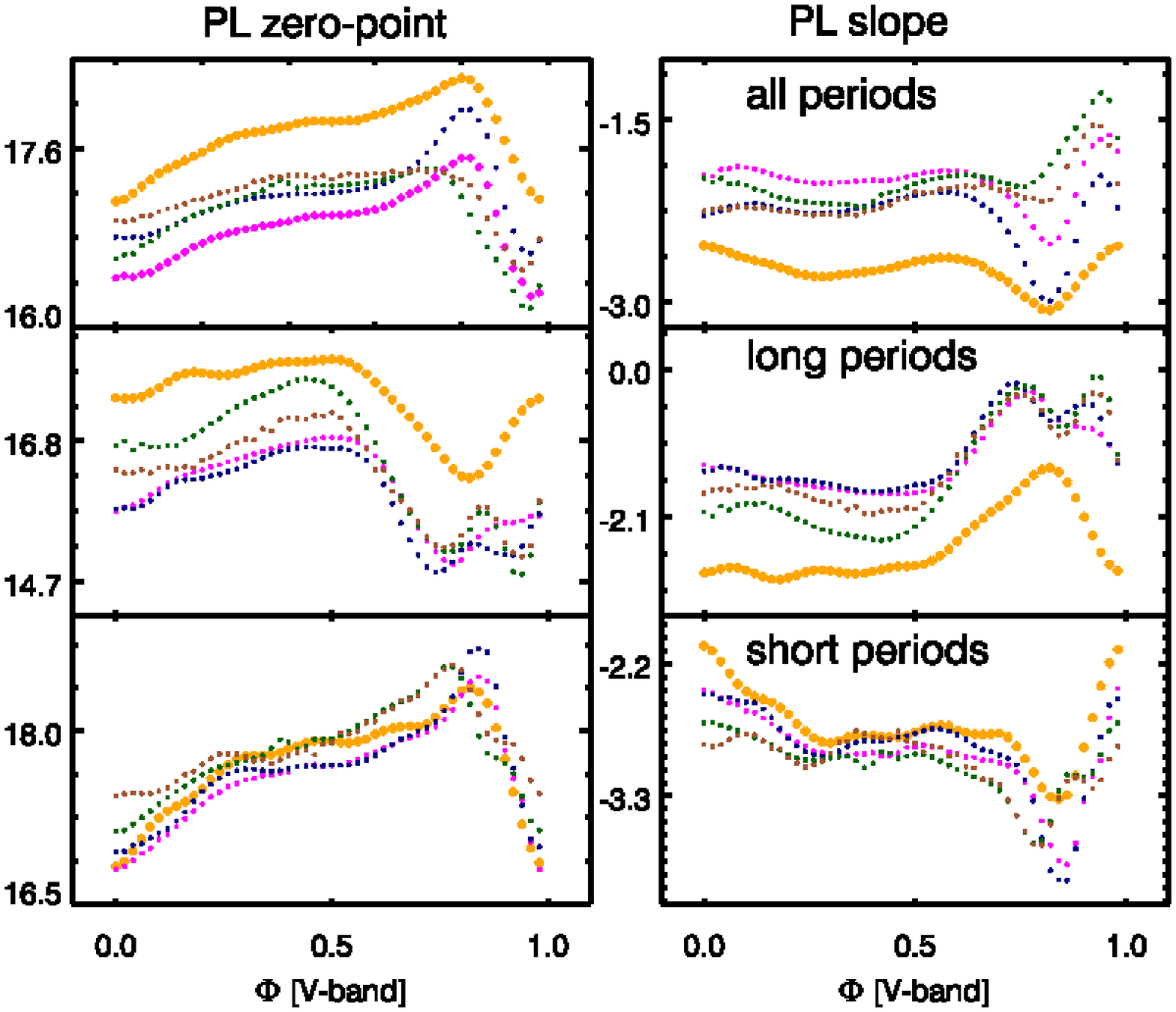}}&
\resizebox{0.5\linewidth}{!}{\includegraphics*{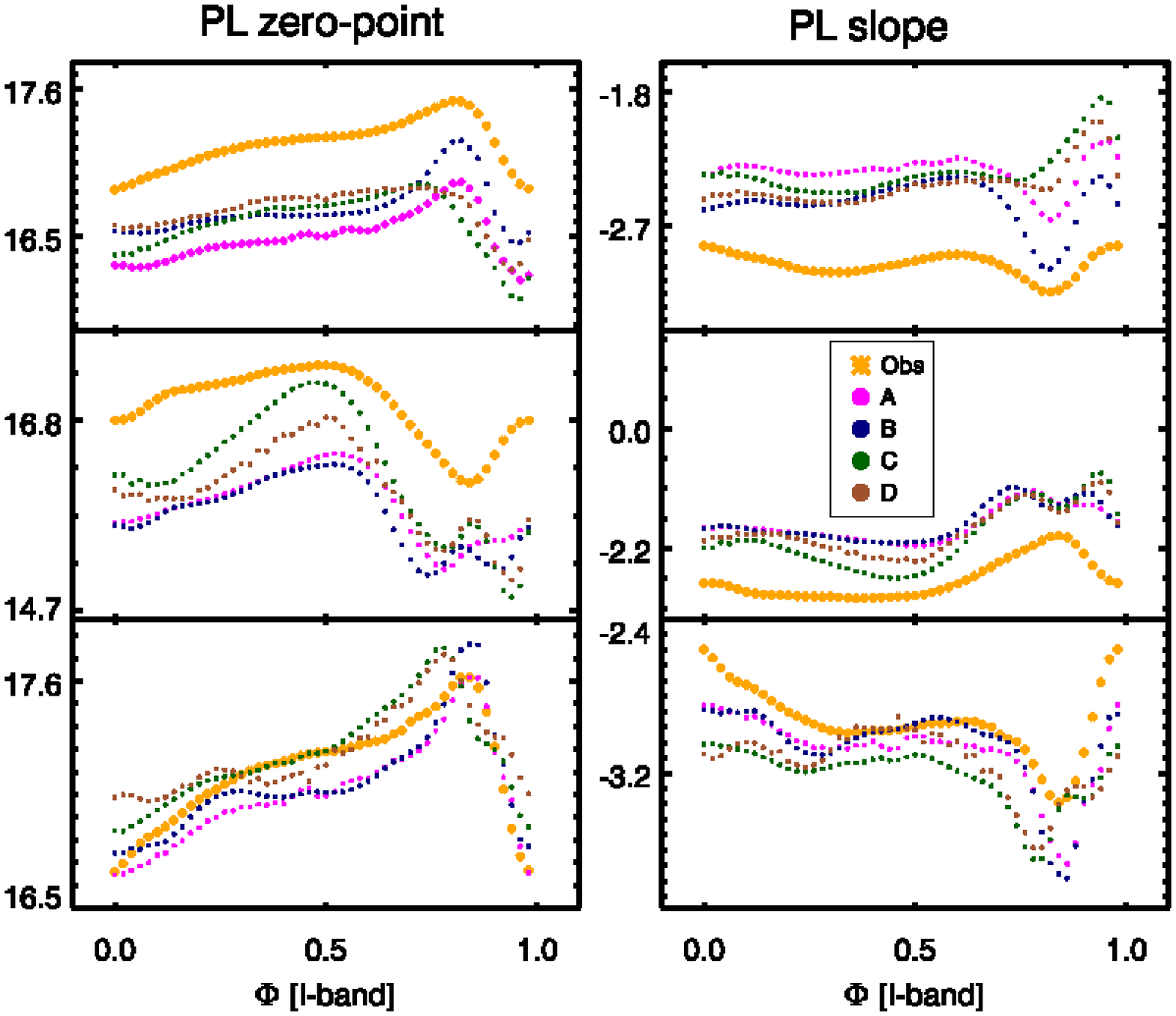}}\\ 
\end{tabular}}
\caption{Plots of the MPPL relations obtained from Cepheid models and observations for the  LMC (upper panel) and SMC (lower panel). Magenta, navy blue, dark green and brown denote convection sets A, B, C and D, respectively. Orange represent observed MPPL relations in $V$- band  and $I$-band. In each plot, lower/middle/upper portions correspond to short/long/all periods, respectively.}
\label{fig:plvi_thob_lmcsmc}
\end{figure*}

 The coefficients of the observed MPPL relations (orange) using SM map for extinction correction are depicted in Fig.~\ref{fig:plvi_thob_lmcsmc}. Distance modulus of $\rm \mu_{LMC}=18.49\pm0.008(\rm statistical)\pm0.047(\rm systematic)$ mag for the LMC as quoted in the literature \citep{piet19} is added to the intercept of the theoretical MPPL relation to be compared with the observed relations. The same has also been done for the SMC considering $\rm \mu_{SMC}=18.97\pm0.016\pm0.028$ mag \citep{grac20}. Also, the MPPL relations of the models obtained from all the four convection sets are plotted and shown in the same figure.

  The upper/lower panel of Fig.~\ref{fig:plvi_thob_lmcsmc} shows the MPPL relations obtained for the LMC/SMC. In each of the plots, the lower, middle and upper panel represents short, long, and all periods (short and long combined), respectively. The plots show that the coefficients of the MPPL relations obtained from  observations vary as follows: 
\begin{enumerate}
\item Empirical relations:
\begin{enumerate}
\item For the LMC, the MPPL slopes display minimum values at phase $\Phi \sim 0.85$ (the same phase where the PL zero-point exhibit maximum value) for short and all periods. At the same phase, the slopes and zero-points for long periods are found to be maximum and minimum, 
respectively in both $VI$-bands.  
\item For the SMC, the minimum slopes and maximum zero-points of the MPPL relations are found at $\Phi \sim~0.85$ for short periods
and all periods. However, for long period, the maximum slopes and minimum zero points occur at $\Phi \sim~0.8$ in both $VI$-bands. 
\end{enumerate}
 For both LMC and SMC, it is seen that short periods and all periods exhibit similar pattern of variations. This may be because the data set is mainly dominated by short period Cepheids (approximately $700$ short periods as opposed to nearly $100$ long periods).
\item Theoretical Relations:
\begin{enumerate}
\item LMC: For short periods, the minimum slopes as well as maximum zero-points of the MPPL relations occur at phase
$\Phi \sim 0.8$ for sets A \& B and at $\Phi \sim 0.7$ using sets C \& D. For long periods, all the four convection sets exhibit maximum slopes and minimum zero-points at $\Phi \sim 0.8$. For all periods, the slopes and zero-points are found to be minimum and maximum at $\Phi \sim 0.8$ for sets A \& B; and at $\Phi \sim 0.75$ for sets C \& D. It seems that higher number of short/long period theoretical light curves as opposed to one another plays a significant role on the nature of the MPPL relations particularly at $\Phi \sim 0.75-0.85$ as seen in all the convection sets  (model numbers are given in Table~\ref{tab:model_number_lmcsmc}). 
\item SMC: For short periods, the MPPL slopes and zero-points obtained using sets A and B display minimum and maximum values at phase, $\Phi \sim 0.85$  while those using convection sets C and D at $\Phi \sim 0.8$ in both the bands. For long/all periods, the slopes and zero-points are maximum and minimum at phase $\Phi \sim 0.85$ in both $VI$-bands for all convection sets. 
\end{enumerate}
\end{enumerate}

In case of the LMC, the occurrence of minimum/maximum theoretical PL slopes/zero-points using sets A and B matches more closely with the observations than the other two sets (C and D) for short and all periods. For long periods, the maximum/minimum  of the observed PL slopes/zero-points are seen at, $\Phi \sim 0.85$ whereas those obtained from the theoretical MPPL relations are seen at $\Phi \sim 0.8$. For short periods and long periods, a consistency of the theoretical MPPL slopes with the observed ones has been found for most of the phases in both the bands. For all periods, the theoretical MPPL relations obtained using sets C and D at phase $\Phi \sim 0.6-1.0$ are found to be in contrast with the observations. Higher number of long periods as compared to short periods in these two sets seems to have a significant effect on the nature of the MPPL relations for all periods. However, the nature of the theoretical MPPL relations obtained using sets A \& B are found to be consistent with the observations. 

In SMC, the theoretical MPPL relations obtained using all four convection sets are found to be consistent with the observed MPPL relations for short periods; in both $VI$- bands. Contrary to the LMC, a discrepancy between theoretical and observed MPPL relations is observed for long periods in the SMC; this may be due to lesser number of long period theoretical light curves as compared to the observed ones (see Table~\ref{tab:model_number_lmcsmc}). It is to be noted that for all periods 
which is dominated by more number of short period Cepheids, the nature of theoretical MPPL relations obtained from models is quite similar to the observed ones for SMC. Although the models and the observations display good agreement for short periods, the large offset between the models and observations in the long period range seems to have contributed to the discrepancy arising for all periods.

Overall, the MPPL relations obtained from the four convection sets and from observations displays dynamic variability as a function of $\Phi$ in all the three cases (short/long/all periods) for both bands. It is found that the theoretical MPPL relations from set A are consistent with set B, whereas the relations obtained using set C are consistent with set D; both for the LMC and SMC. It is interesting to note that the theoretical MPPL slopes and zero-points obtained from models for both LMC and SMC at long periods display two peaks of maxima and minima across the four sets, the first peak occurring at  $\Phi \sim 0.75$, while the second one at $\Phi \sim 0.95$. The second peak in sets C and D is more pronounced than that obtained using sets A and B in both $VI$-bands. The occurrence of these two peaks in the theoretical MPPL relations seem to have an impact on the nature of the MPPL relations for all periods, with the effect more pronounced for sets C and D. It can also be seen that short and long period Cepheids exhibit contrasting behaviour in both $VI$-bands, for both the models and observations. Moreover, at phases $\Phi \sim 0.5-1.0$, the relations are found to be shallower for short periods and steeper for long periods. For short periods, an offset of the MPPL relations  between models and observations is observed at $\Phi \sim 0.8$, for both LMC and SMC. 
 
It is interesting to note that there is a clear distinction among different sets in the multiphase plane for most of the phases; while in the FP plane, there is an overlap of parameters among these sets. A small discrepancy in the FP plane can lead to a large discrepancy in the PL plane as observed in Figs.~\ref{fig:fou_r_theo_lmcsmc}, \ref{fig:fou_phi_theo_lmcsmc}, \ref{fig:plvi_thob_lmcsmc}. Thus, the multiphase comparison between models and observations in the PL plane for both LMC and SMC proves to be a more stringent way to constrain the theory of pulsations as compared to that done in the FP plane. Difference between the theoretical and observed light curve structures plays a significant role for the discrepancy arising between models and observations. Different number of light curves between the models and observations, particularly for long periods in the SMC, is likely to have contributed to the offset between the two in the multiphase PL plane.
\subsubsection{Multiphase Period-Colour/Amplitude-Colour (MPPC/MPAC) relations}

\begin{figure*}
\scalebox{0.95}{
\begin{tabular}{cccc}
\resizebox{0.5\linewidth}{!}{\includegraphics*{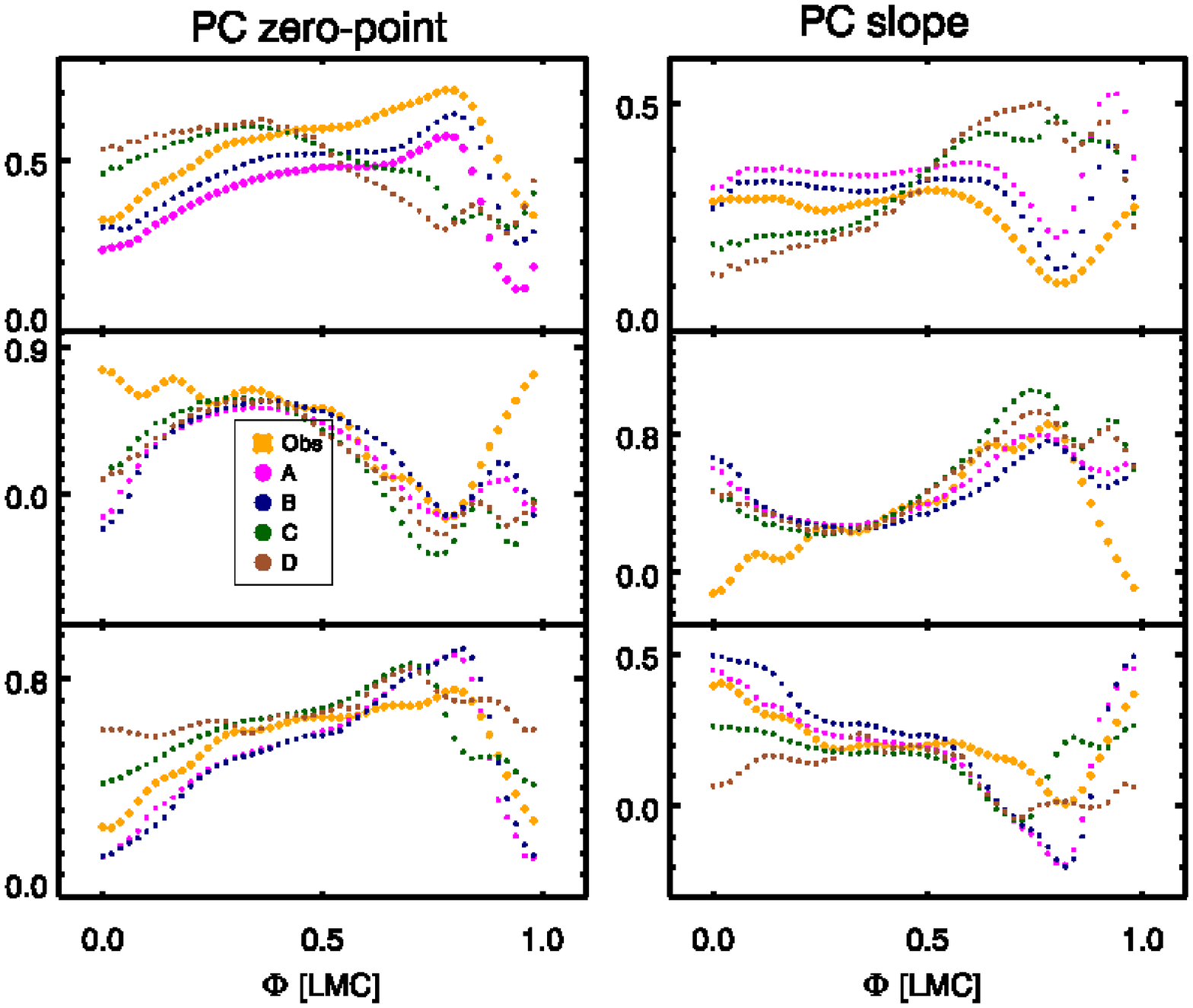}}&
\resizebox{0.5\linewidth}{!}{\includegraphics*{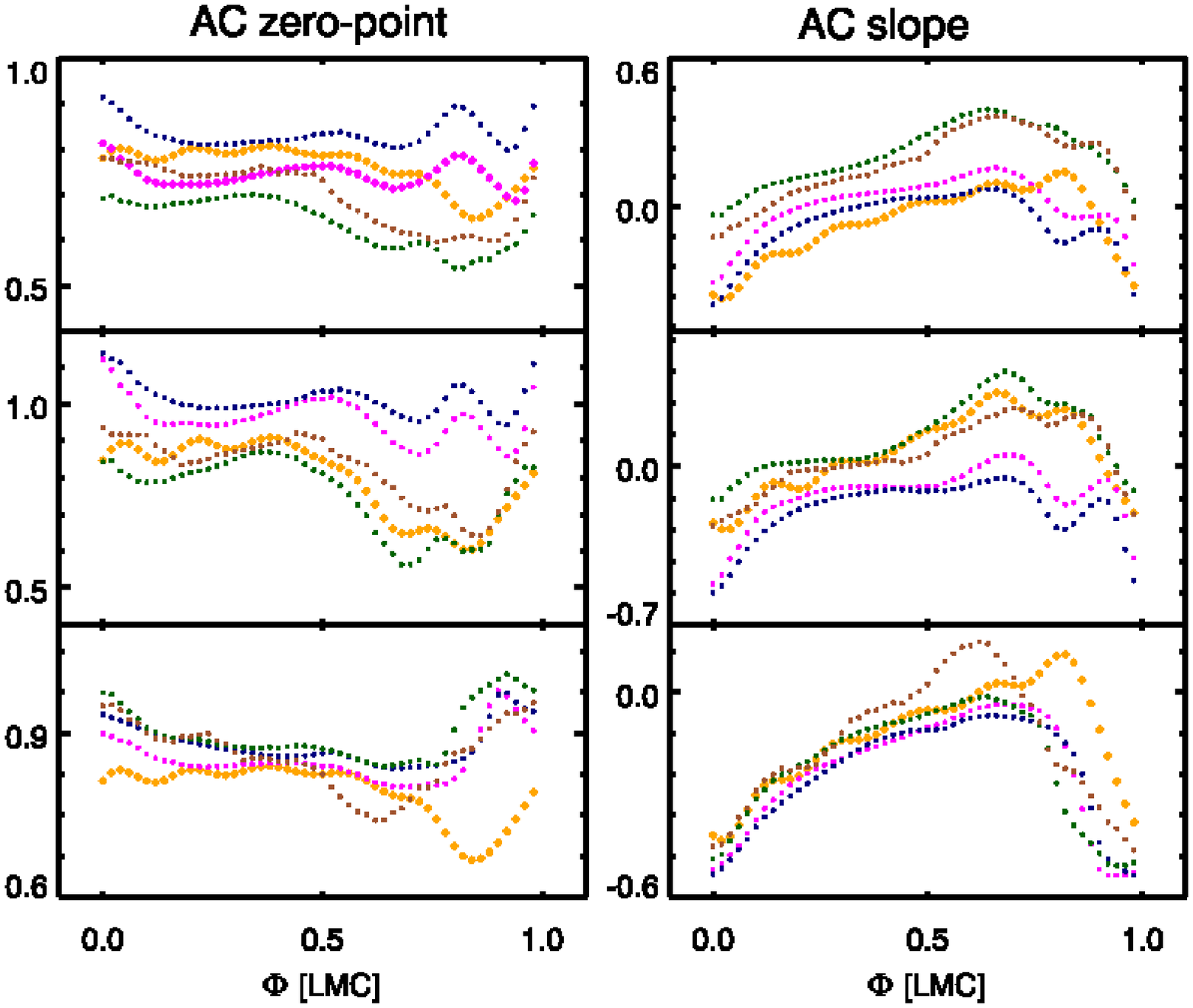}}\\
\resizebox{0.5\linewidth}{!}{\includegraphics*{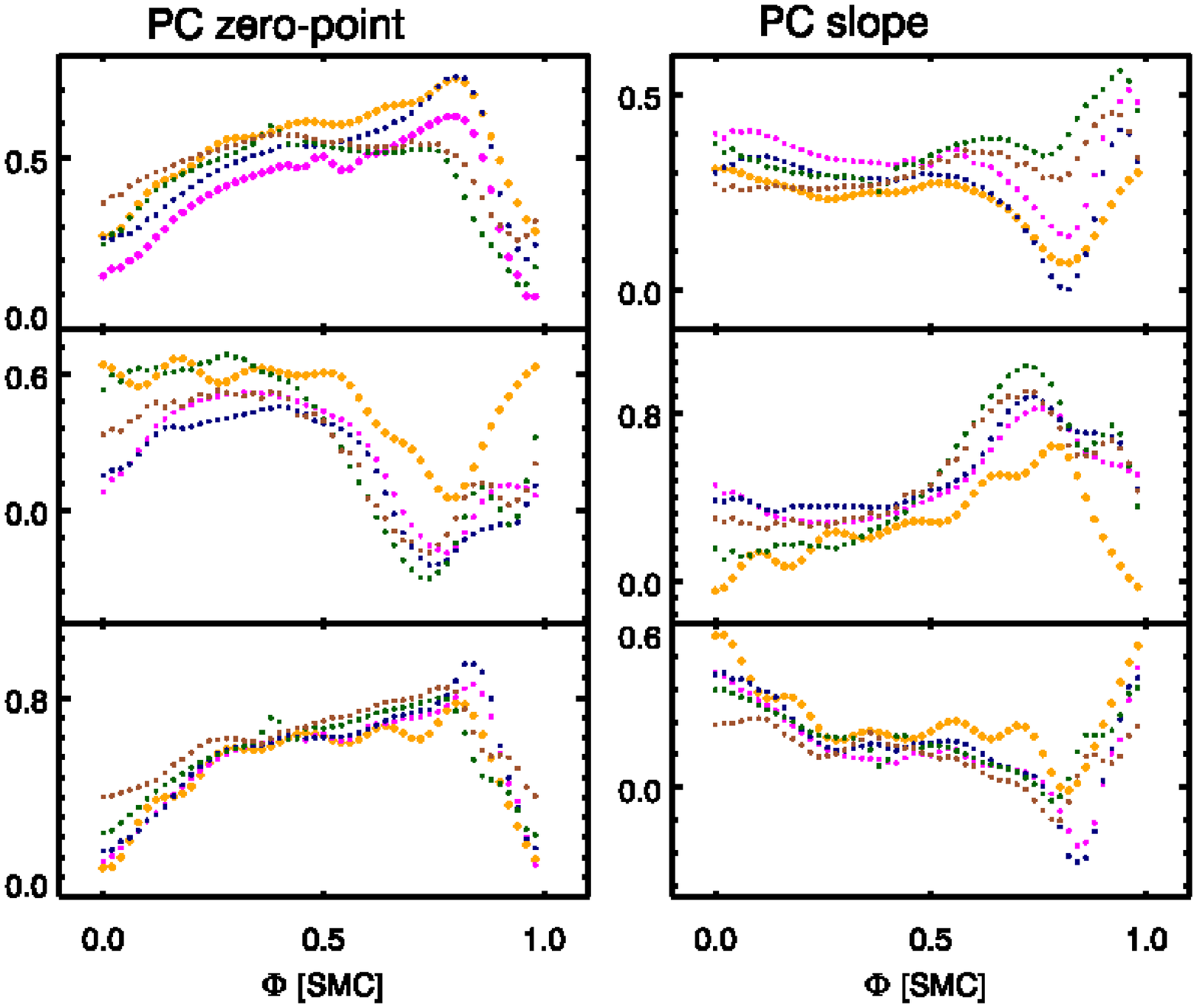}}&
\resizebox{0.5\linewidth}{!}{\includegraphics*{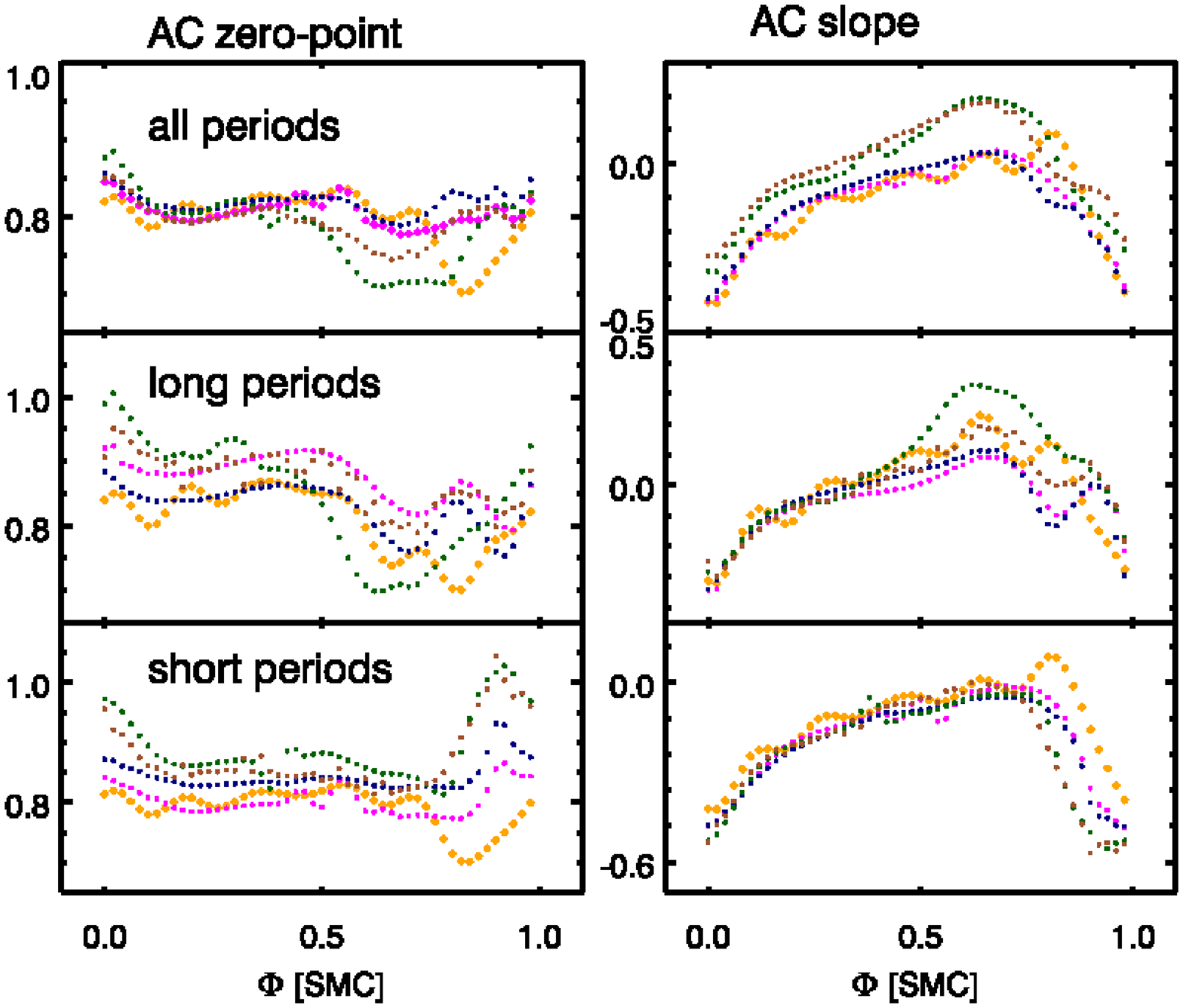}}\\
\end{tabular}}
\caption{Same as Fig.~\ref{fig:plvi_thob_lmcsmc} but for MPPC and MPAC relations.}
\label{fig:pc_thob_lmc}
\end{figure*}

Fig.~\ref{fig:pc_thob_lmc} displays the comparative plots showing the multiphase relations obtained from the models in all four convection sets and observations in the PC and AC planes for the LMC (upper panel) and SMC (lower panel), respectively.  

\begin{enumerate}
\item Empirical relations: The MPPC relations obtained from observations for both LMC and SMC are found to be highly dynamic with pulsational phase. Similar to the MPPL relations, MPAC relations between short and long periods are in contrast. As in the case of the MPPL relations, short and all periods also display similar nature of slopes and zero-points.
\item Theoretical relations: 
\begin{itemize}
\item [(a)] LMC: It has been observed that long period Cepheids display maximum slope and minimum zero-point  at $\Phi \sim 0.8$ in all the sets. For short periods, the MPPC relations obtained  using sets A \& B and sets C \& D exhibit minimum/maximum slope/zero-point at $\Phi \sim 0.85$. 
\item [(b)] SMC:  For short periods, the MPPC relations display maximum slope and minimum zero-point  at $\Phi \sim 0.85$ using sets A \& B and using sets C and D, respectively. All the sets exhibit maximum/minimum slopes/zero-points at $\Phi \sim 0.75$. For all periods, maximum/minimum slopes/zero-points are seen at $\Phi \sim 0.8$ in all the sets.
\end{itemize}
\end{enumerate}
For the LMC, it has been observed that long period Cepheids display maximum slope and minimum zero-point  at $\Phi \sim 0.8$ for both the models and observations. It can be seen from the plots in Fig.~\ref{fig:pc_thob_lmc} that at the phase corresponding to maximum light, the MPPC slopes obtained from observations display slopes nearly equal to zero, whereas those obtained from models have non-zero slopes. For short periods, the MPPC relations obtained from observations and using sets A \& B exhibit minimum slope at $\Phi \sim 0.85$. However, the minimum slope using sets C \& D occurs at $\Phi \sim 0.7$. In both the two phases, the zero-point is maximum. A large discrepancy between theoretical and observed MPPC relation at $\Phi \sim 0$ is seen only for set D. For all periods, discrepancy in MPPC relation using sets C and D with observed relations is  more prominent in the phase range $\Phi \sim 0.5-1.0$, while the relations obtained using set A and B are consistent with the observations for the same range. Significantly, the higher number of long period light curves in sets C and D seems to take over the nature of the MPPC relations at all periods.
In case of the SMC, the MPPC relations of the Cepheid models obtained using all four convection sets are found to be in consistent with the observed ones for most of the phases in short/long/all periods. There is a large dip seen in the theoretical MPPC relations as compared to observations, particularly at phases $\Phi \sim 0.7-0.95$ for short periods. A steep rise in the MPPC slopes at phase $\Phi \sim 0.65-0.95$ is observed for long periods for both the models and observations. 

The MPPC relations obtained from theoretical light curves of the models and as well as observed light curves are dynamic with pulsational phase, highly non-linear at $\Phi \sim 0.7-0.85$. Theoretical and observed MPPC relations at short and long periods also display contrasting behaviour. Similar to the MPPL relations, MPPC slopes obtained using short and long period Cepheids display shallowest and steepest slopes at phase $\Phi \sim 0.65-0.95$. For long periods, at phases close to maximum light, the MPPC plots obtained from observations display slopes close to zero 
for the LMC and a non-zero slope for the SMC. For both LMC and SMC, the observed short period Cepheids  display minimum slopes close to zero at $\Phi \sim 0.8$. These results are consistent with the results as reported in \citet{ngeo06}. At short periods, the minimum MPPC slopes obtained from models in all the four sets are closer to zero at $\Phi \sim 0.8$ for the LMC; whereas a large dip has been observed for the SMC. It is interesting to note that although the discrepancies between models and observations are seen in the PL plane, the two seem to agree better in the PC plane for most of the phases in all the three cases (short/long/all periods). 

The left panel of Fig.\ref{fig:pc_thob_lmc} display the MPAC relations for both models and observations. From the figure, the comparative study of the MPAC obtained from observations and the models are summarized. It can be seen from the plots that for short periods, most of the slopes obtained from both models and observations are negative. For long period, however, the slopes become positive at $\Phi \sim 0.5-0.8$ for both models and observations. The nature of the MPAC relations for short and long periods are similar, both for models and as well as observations. This is consistent with the results obtained by \citet{ngeo06b}. For LMC, the MPAC relations using convection sets C and D are found to be more consistent with the observations for long periods. On the other hand, the MPAC slopes obtained using sets A \& B are found to be significantly smaller as compared to observations. However, the MPAC relations obtained using sets A, B and D are found to be more consistent with observations for the SMC. For short periods, the agreement between theoretical and observed MPAC relations has been observed for both LMC and SMC. It can be seen for all periods, sets A \& B are more consistent with the observations as compared to the other two sets (C and D) for both LMC and SMC. The MPAC slopes obtained using sets C \& D are found to be significantly higher than the observations for the LMC/SMC. 

\subsection{ PL, PC, AC relations (models and observations) at mean light for the LMC/SMC}
\label{sec:plpcac_mean}
\begin{figure}      
\begin{tabular}{c}
\resizebox{1.0\linewidth}{!}{\includegraphics*{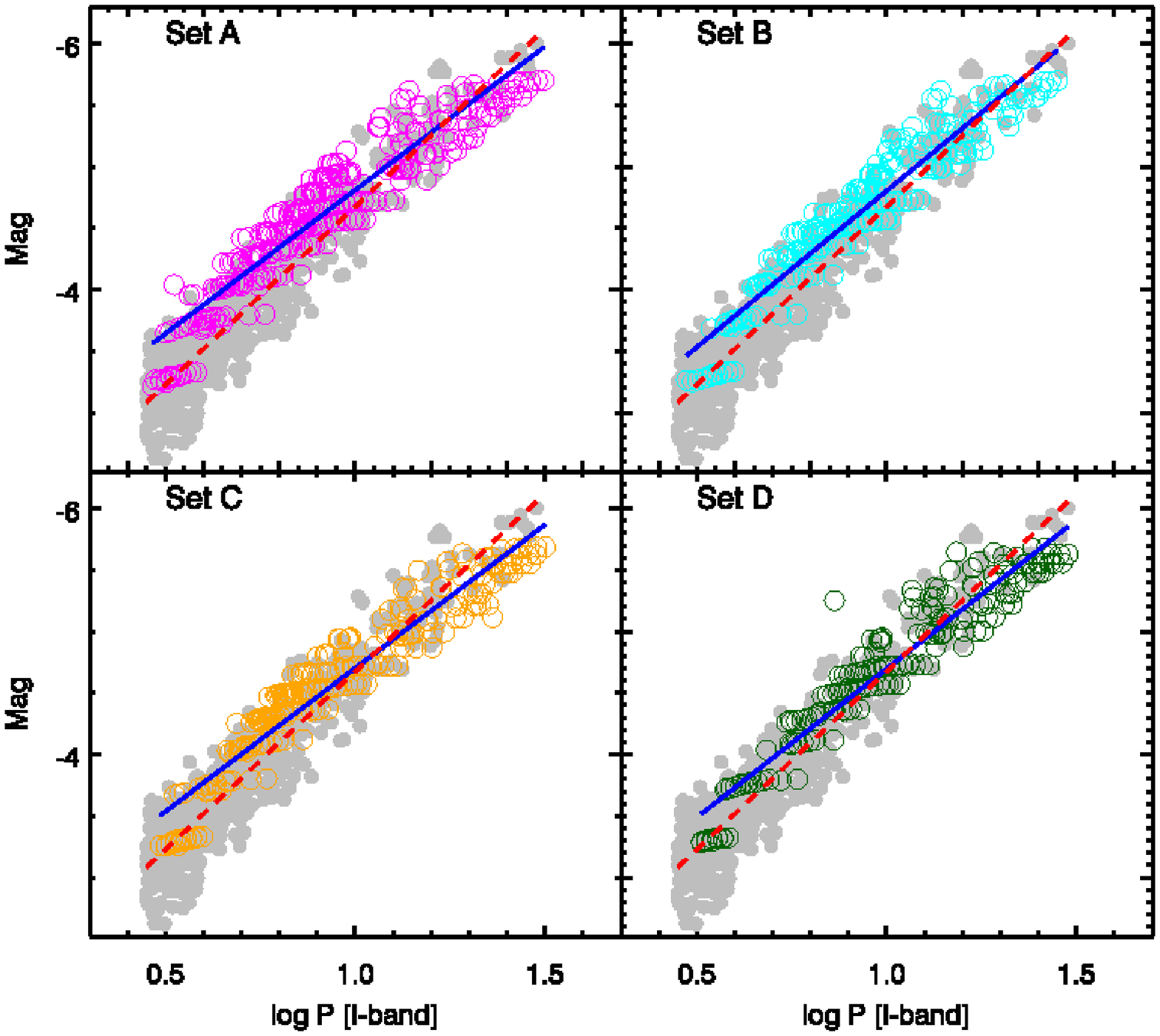}}
\end{tabular}
\caption{Comparative plots of the Cepheid PL relation between the models using convection sets A (magenta), B (cyan), C (orange), D (dark green) and  observations (grey) at mean light in $I$-band for the SMC. Solid blue and dotted red lines  represent linear regression fits to the PL relation obtained from models and observations, respectively.}
\label{fig:pl_plot}
\end{figure}

The PL/PC/AC relations at mean light are carried out using a term $A_{0}$ of the Fourier decomposition for both the models and observations within the same period range. The results of PL relations for the LMC and SMC at mean light are given in Tables~\ref{tab:pl_lmc} and \ref{tab:pl_smc}. From the tables, it is evident that the PL relation obtained from the models in all the four sets are in close agreement with the observations, particularly for short periods in both bands for the LMC and SMC,  within the obtained error bars. However, large errors on the coefficients of the theoretical relations obtained at mean light as compared to observations makes it difficult to choose the best model describing the observed dataset. For the LMC, the PL relation of the models obtained using set D are comparatively higher than the observed one at mean light  in both $VI$- bands. It is evident from the table that, for long periods, the model PL slopes with convection sets A, B, C, D do not agree well with the observations. Theoretical magnitudes of the models do not cover some of the observed range for long periods, as seen in Fig.~\ref{fig:pl_plot}. This results in the discrepancy of the PL relations between the models and observations for long periods.  

\begin{table*}
\caption{PL slopes and intercepts of the form $Y=a+b\log{P}$  at mean light; for short ($a_{s}, b_{s}$), long ($a_{l}, b_{l}$), all period ($a_{all}, b_{all}$) obtained from observations and models using sets A, B, C, D in the LMC. Bold-faced entries in the table  represent PL coefficients obtained from models  which agree well with the  observations.}
\scalebox{0.95}{
\begin{tabular}{c c c c c c c c} \\ \hline \hline
& $b_{s}$ & $a_{s}$ &  $b_{l}$ &$a_{l}$  &$b_{all}$ &$a_{all}$ \\ \hline
&&&LMC [$V$- band] &&&\\ \hline

Obs &$-2.981\pm0.035$ & $16.658\pm0.022$   & $-2.083\pm0.194$  & $16.600\pm0.226$  &$-2.637\pm0.032$&$17.239\pm0.023$   \\
Set A &$\mathbf{-2.905}\pm\mathbf{0.148}$ & $\mathbf{17.152}\pm\mathbf{0.121}$ & $-1.201\pm0.127$  & $15.529\pm0.158$ &$-1.873\pm0.057$ &  $16.341\pm0.059$ \\
Set  B  &$\mathbf{-2.729}\pm\mathbf{0.165}$   &$\mathbf{17.125}\pm\mathbf{0.136}$ &  $-1.325\pm0.133$  &$15.679\pm0.165$& $-2.047\pm0.061$&$16.571\pm0.066$ \\
Set C & $\mathbf{-3.047}\pm\mathbf{0.146}$ & $\mathbf{16.873}\pm\mathbf{0.171}$  &$-1.249\pm0.122$   &$15.693\pm0.154$ &  $-2.095\pm0.056$ &$16.756\pm0.061$ \\
Set D &$\mathbf{-3.389}\pm\mathbf{0.184}$   &$\mathbf{17.908}\pm\mathbf{0.162}$ & $-1.206\pm0.126$   &$15.617\pm0.158$  & $-2.127\pm0.075$&$16.784\pm0.086$ \\ \hline
&&&LMC [$I$- band] &&& \\ \hline

Obs &$-2.999\pm0.035$&$16.765\pm0.023$ & $-2.435\pm0.153$ & $16.169\pm0.177$ & $-2.888\pm0.023$ & $16.694\pm0.017$ \\
Set A &$\mathbf{-3.089}\pm\mathbf{0.103}$ & $\mathbf{16.620}\pm\mathbf{0.084}$ & $-1.737\pm0.112$ & $15.353\pm0.138$ & $-2.224\pm0.046$ & $15.941\pm0.047$ \\
Set B &$\mathbf{-2.945}\pm\mathbf{0.135}$ &$\mathbf{16.588}\pm\mathbf{0.118}$ & $-1.807\pm0.116$  & $15.439\pm0.143$ & $-2.348\pm0.050$ & $16.106\pm0.053$ \\
Set C & $\mathbf{-3.210}\pm\mathbf{0.124}$  & $\mathbf{16.905}\pm\mathbf{0.140}$ & $-1.845\pm0.110$ & $15.550\pm0.137$ & $-2.417\pm0.045$ & $16.268\pm0.049$ \\
Set D & $-3.494\pm0.335$ & $17.229\pm0.309$ & $-1.777\pm0.111$  &$15.454\pm0.138$ & $-2.442\pm0.060$ & $16.298\pm0.069$ \\ \hline
\end{tabular}}
\label{tab:pl_lmc}
\end{table*}

The results of PC/AC relations obtained at mean light for the LMC and SMC are given in Tables~\ref{tab:pc_lmcsmc} and \ref{tab:ac_lmcsmc}. It is evident from the Table~\ref{tab:pc_lmcsmc} that the  all four sets of model PC slopes show better agreement with the observed ones at mean light for short/long/all periods for both LMC and SMC. It can be seen from Table~\ref{tab:ac_lmcsmc}, that for both LMC and SMC, models using convection sets A, B, C, D  are consistent with the observations for short periods.  On the other hand, a discrepancy between models and observations is observed for long periods for both LMC and SMC. 
\section{Distances to the LMC and SMC}
\label{sec:distance}
There is evidence in the literature that there exists a break in the mean light Cepheid PL relation for the LMC at $\log{P}\sim1$  \citep{tamm03,sand04,kanb04,kanb06,kanb07b,sand09,ngeo08,ngeo09,kodc15,bhar16}.
Also, there are both observational and theoretical evidences of non-linearity at $\log{P}\sim 1$ 
in both the LMC and SMC multiphase Cepheid PL relations \citep{ngeo06, kanb10}. Furthermore, Fig.~\ref{fig:plvi_thob_lmcsmc} provides compelling evidence that the MPPL relations for both observations and models are highly dynamic with pulsational phase both for LMC/SMC. The figure suggests  that the non-linearity at some phases is extreme and at some phases, reduced. However, these are smoothed over at mean light and this has a greater effect on the SMC \citep{ngeo06}.  Hence, there is no such break reported at $\log{P}\sim 1$ for the SMC as quoted in the literature \citep{kanb04,kanb06,kanb10,subr15,ngeo15a,ripe16,jacy16,scow16,ripe17}. To avoid the effect of the  mean light Cepheid PL relation at $P \sim 10$ d in the LMC, Cepheids with period $\log{P}>1.0$  are considered to calculate the distance to the LMC. The same has also been done for the SMC, although there is no such break at mean light. The other reason for using long period Cepheids is that they are brighter and hence are useful for accurate and precise distance determination to the target galaxy.

Simon and his collaborators have established the correlation between 
various Fourier parameters and other physical parameters for RR Lyrae stars \citep{simo81,simo82,simo83,simo85,simo88,simo93}. The direct advantage of using Fourier parameters in the distance determination is that they are  reddening-free. Cepheids follow a period-luminosity-colour relation following the Stephan-Boltzmann law. If the Fourier parameter $R_{21}$ is added replacing the colour term, it is found that the error sum of squares is  reduced significantly. However, the effect of adding other Fourier parameters or their combinations in the PL relation is beyond the scope of the present study and will be done in a future project.

Theoretical magnitudes obtained from the models in all the four sets are fitted using two separate regressions, $Y_{r}=g+h\log{P}$ and $Y_{f}=\alpha+\beta \log{P}+\gamma R_{21}$. Here $Y_{r}$ and $Y_{f}$ are the dependent variables representing the reduced model and full model, respectively. It is worth mentioning here that this is the first approach to associate a Fourier parameter in addition to the PL relation to determine the Cepheid distances. To check the significance of the addition of $R_{21}$ to the regression, statistical $F$-test as described in \citet{kanb04b} is used. The result studied at $95\%$ confidence level is considered to be statistically significant if the probability of the $F$-test, $p(F)<0.05$. The coefficients of the reduced model and full model along with their rms value, the $F$- statistics and the corresponding probability of the $F$- statistics ($p(F)$) are given in Table~\ref{tab:ftest_lmc}. For the LMC,  $F$- statistics provides evidence that the reduced model can be rejected for
sets C and D in $V$- band,  and not for sets A and B. However, in $I$- band, it can be seen that the reduced model can be rejected only for convection set A. For SMC, the $F$-statistics as displayed in the Table~\ref{tab:ftest_lmc} for convection sets C and D give the evidence that the reduced model can be rejected in both $VI$- bands.

\begin{table*}
\caption{A summary of $F$- test in sets A, B, C, D. Here $g$ and $h$ are the coefficients of the reduced model ($Y_{r}=g+h\log{P}$);  $\alpha, \beta, \gamma$ are the coefficients of the full model ($Y_{f}=\alpha+\beta \log{P}+\gamma R_{21}$).
$D_{f}$ and $D_{r}$ denote distances obtained using the coefficients of the full model ($Y_{f}$) and reduced model ($Y_{r}$).}
\centering
\scalebox{0.85}{
\begin{tabular}{c c c c c c c c c c c c} \hline \hline
& $g$ & $h$ &$\sigma_{{rms}}$& $\alpha$ & $\beta$ & $\gamma$ &$\sigma_{{rms}}$ & $D_{f}$ & $D_{r}$ & $F$ & $p(F)$\\ \hline
&&&&&LMC [$V$-band] &&&&&&\\ \hline
Set A & $-3.246\pm0.197$ & $-0.984\pm0.153$ & $\mathbf{0.194}$ & $-3.342\pm0.207$ & $-0.807\pm0.193$ & $-0.523\pm0.349$ & $\mathbf{ 0.136}$  &  $49.848\pm0.417$ & $49.841\pm0.446$
& $2.348$ & $0.111$ \\
Set B & $-3.244\pm0.212$ & $-0.995\pm0.165$ &$\mathbf{0.191}$ & $-3.613\pm0.233$ & $-0.534\pm0.213$  &$-0.982\pm0.305$ & $\mathbf{0.134}$& $50.949\pm0.434$
& $50.103\pm0.449$ & $10.393$ & $0.002$ \\
Set C & $-3.164\pm0.198$ & $-0.971\pm0.152$ &$\textbf{0.376}$& $-4.215\pm0.269$ & $0.152\pm0.249$ & $-2.056\pm0.380$ &$\textbf{0.265}$& $51.274\pm0.436$
 & $48.623\pm0.428$ & $29.204$ & $0.000$  \\ 
Set D& $-2.872\pm0.158$ & $-1.206\pm0.125$ &$\mathbf{0.194}$& $-3.0161\pm0.142$ & $-1.009\pm0.142$ & $-0.582\pm0.213$ &$\mathbf{0.133}$ & $49.410\pm0.436$ & $48.824\pm0.431$ & $7.386$ & $0.008$ \\ \hline
&&&&&LMC [$I$-band]&&&&&&\\ \hline
Set A & $-3.187\pm0.124$ & $-1.694\pm0.099$ &$\mathbf{0.153}$& $-2.893\pm0.207$ & $-2.084\pm0.122$ & $0.814\pm0.168$ &$\mathbf{0.099}$& $49.883\pm0.445$ & $50.004\pm0.439$ & $24.742$ & $0.000$   \\
Set B & $-3.079\pm0.131$ & $-1.783\pm0.105$ &$\mathbf{0.152}$& $-2.861\pm0.2598$ & $-2.060\pm0.122$ & $0.590\pm0.151$ & $\mathbf{0.101}$& $49.680\pm0.442$ & $50.216\pm0.440$ & $15.281$ & $0.000$ \\
Set C& $-2.988\pm0.118$ & $-1.504\pm0.093$ &$\mathbf{0.371}$& $-2.724\pm0.416$ & $-2.092\pm0.140$ & $0.523\pm0.192$ & $\mathbf{0.251}$&$48.929\pm0.435$ & $48.387\pm0.426$ & $10.844$ & $0.001$  \\ 
Set D& $-3.081\pm0.122$ & $-1.738\pm0.097$ & $\mathbf{0.150}$&$-2.986\pm0.277$ & $-1.842\pm0.129$ & $0.192\pm0.160$ &$\mathbf{0.105}$& $48.940\pm0.448$  & $50.366\pm0.430$ & $1.444$ & $0.216$ \\ \hline
&&&&&SMC [$V$-band]&&&&& \\ \hline
Set A & $-2.856\pm0.246$ & $-1.298\pm0.199$ & $\mathbf{0.245}$ & $-2.804\pm0.213$ & $-1.426\pm0.210$ & $0.482\pm0.274$ & $\mathbf{0.169}$ & $61.253\pm0.449$ & $61.469\pm0.442$ & $3.080$ & $0.218$   \\
Set B & $-3.016\pm0.249$ & $-1.217\pm0.209$ & $\mathbf{0.240}$ & $-3.036\pm0.242$ & $-1.270\pm0.210$ & $0.422\pm0.270$ & $\mathbf{0.166}$ & $62.599\pm0.470$ & $62.858\pm0.460$ & $2.440$ & $0.316$ \\
Set C& $-2.427\pm0.205$ & $-1.527\pm0.164$ & $\mathbf{0.207}$ & $-2.692\pm0.173$ & $-1.457\pm0.150$ & $0.746\pm0.178$& $\mathbf{0.132}$ & $57.658\pm0.435$ & $57.012\pm0.421$ & $17.443$ & $0.000$  \\ 
Set D& $-2.691\pm0.247$  &$-1.363\pm0.200$ & $\mathbf{0.237}$ &  $-2.998\pm0.200$&$-1.255\pm0.184$&$0.805\pm0.192$& $\mathbf{0.151}$ & $59.254\pm0.439$&$58.985\pm0.426$&$17.563$ &$0.000$ \\ \hline
&&&&&SMC [$I$- band] &&&&&& \\ \hline
Set A & $-3.108\pm0.193$ &$-1.771\pm0.156$ & $\mathbf{0.192}$ & $-3.067\pm0.216$&$-1.901\pm0.191$&$0.523\pm0.224$& $\mathbf{0.130}$ & $60.796\pm0.548$ & $60.772\pm0.451$& $7.174$ &$0.008$   \\
Set B &$-3.204\pm0.192$&$-1.732\pm0.161$ &   $\mathbf{0.184}$ & $-3.265  \pm0.247$&$-1.772\pm0.155$&$0.535\pm0.191$ & $\mathbf{0.123}$ &$62.187\pm0.556$ &$62.037\pm0.471$ & $7.847$ & $0.006$ \\
Set C& $-2.727\pm0.160$&$-2.004\pm0.127$ & $\mathbf{0.162}$ & $-2.873\pm0.112$ &$-2.003\pm0.119$&$0.606\pm0.100$ & $\mathbf{0.099}$   &$59.069\pm0.485$ &$58.626\pm0.435$&$25.761$ & $0.000$  \\ 
Set D& $-2.951\pm0.185$ &$-1.858\pm0.156$ & $\mathbf{0.186}$ & $-3.116\pm0.142$&$-1.847\pm0.135$&$0.685\pm0.129$ & $\mathbf{0.112}$ &$59.411\pm0.489$ & $59.213\pm0.489$ & $28.111$ &$0.000$ \\\hline
\end{tabular}}
\label{tab:ftest_lmc}
\end{table*}

In any case, however, the coefficients generated by both the reduced and full models from all the convection sets are used to calculate the absolute magnitudes of the observed light curves. These coefficients are used in the relation $M_{\lambda_{1}}=\alpha+\beta\log{P}^o+ \gamma R_{21}^o$ and $M_{\lambda_{2}}=g+h\log{P}^o$, where $\lambda \equiv \left(V,I\right)$. $\log_{P}^o$ and $R_{21}^{o}$ in the relations are taken from observations. The  distance modulus $\mu_{0}$ is calculated using the relation, $\mu_{0}= A_{0}-M_{\lambda_{1,2}}$  where $A_{0}$ are the extinction corrected magnitudes in $VI$- band, where the extinction corrected has been done using SM map. Individual distances are then  calculated using the relation, $D=10^{(0.2\mu_{0}+1.0)}$.

The average distance to the LMC and SMC along with their errors are obtained using Monte Carlo simulations as described in \citet{deb15}. The simulations are carried out for $10^5$ iterations. The $\mu$ and $\sigma$ values are obtained from the Gaussian fit to the distance distribution obtained using sets A, B, C and D. The average distances to the LMC and SMC obtained using the coefficients from the reduced and full model are given in columns 7 and 8 of Table.~\ref{tab:ftest_lmc}. The calculated distances are in good agreement with the published literature values of $D_{\textsc{lmc}}=49.59\pm0.09$ kpc and $D_{\textsc{smc}}=62.44\pm0.47$ kpc \citep{piet19, grac20}. From the table, it can be seen that the addition of $R_{21}$ plays a significant role in accurate  distance determination. This is quite evident involving cases where the $F$- test is significant. For cases where the $F$- test is not significant, the values of distance determined for the LMC using the coefficients obtained from  the reduced model and full model are similar in both the cases (sets A \& B). From the table, it is clear that the distances to the SMC obtained using sets A and B are close to the literature values, but are  shorter by $4-5$ percent while using sets C and D. However, it is important to note that various studies in the literature found that  SMC is heavily elongated along the line of sight up to $\sim 20$ kpc \citep{jacy16, scow16, ripe17}. The mean distance is therefore dependent on the dataset used to calculate it.  

\section{Conclusion}
\label{sec:diss}
A detailed analysis of theoretical light curves of Cepheid models generated with the help of \textsc{mesa-rsp} using four convection sets for metallicities appropriate for the LMC and SMC has been carried out. The observed Cepheid light curves are taken from OGLE-IV database. Comparison of the light curves obtained from Cepheid models and those obtained from observations has been done in terms of different light curve parameters and various derived relations using them.  Parameters in the FP plane as a function of period and  PL/PC/AC relations as a function of phase are investigated. 

For short periods, it has been observed that the theoretical amplitudes are in consistent with the observations in all
the four convection sets (A, B, C, D); however for long periods, some of the theoretical amplitudes are systematically higher than the observed ones. The same has been reported in the study of Cepheid light curves by \citet{bhar17a}, particularly at optical wavelengths. In the multiphase relations, it has been observed that higher numbers of short/long period in the dataset as opposed to one another seems to have played a significant role on the nature of the relations for all periods, clearly seen at  $\Phi \sim 0.5-0.9$. It is interesting to see that the MPPL/MPPC slopes are opposite for short/long period Cepheids obtained from observations especially at $\Phi \sim 0.8$ for both LMC and SMC, and the fact that this behaviour is supported by models as well. It is worth mentioning here that the multiphase PC/PL relations can be an additional and tighter constraint on models over just the FPs. In the multiphase planes, the slopes of the PL/PC relations look at how models are interrelated to each other; the ensemble of models must match the ensemble of observations. Discrepancy in the multiphase planes may be attributed to significantly small number of theoretical light curves as compared to observations. The results of the present study are summarized below:
 \begin{enumerate}
\item Theoretical Fourier amplitudes obtained from the models in all four convection parameter sets are found to be consistent with the observations for long periods. Offset between Fourier amplitude parameters for short periods are seen ($0.7<\log {P}<1.1$) for both the LMC and SMC, which is consistent with \citet{bhar17a}. 
\item Fourier phase parameters obtained in all the four convection sets were found to be consistent with the observed ones for short periods. However, for long periods, the phase values obtained from the models using the four sets are comparatively smaller than the observations. It has been found that the Fourier phase parameters from sets C and D are closer to the observations for long periods than the other two sets (A \& B) for both LMC and SMC. 
\item The MPPL relations of the observed light curves corrected for extinctions using SM map in the LMC and SMC are found to be highly dynamic in nature, with the effect  more pronounced at  $\Phi \sim 0.8-0.85$ for both $VI$-bands. The general form of the PL relations using observations is similar for both LMC and SMC. A contrasting behaviour is observed between short and long periods. This behaviour of the MPPL relation is consistent with the findings of \citet{ngeo06} and \citet{ngeo12a}. 
\item It has been found that the theoretical PL relations obtained in all the four convection sets are highly dynamic as a function of phase. At phase $\Phi \sim 0.6-0.9$, it is found that the theoretical MPPL relations of Cepheid models overlap with the  observed relations for both LMC and SMC. Contrasting behaviour of the theoretical/observed MPPL relations between short and long periods was also observed. For long periods, theoretical MPPL relations display two peaks of maxima/minima at phases $\Phi \sim 0.75$ and $\Phi \sim 0.95$; more pronounced in sets C \& D than sets A \& B for both LMC and SMC.  
\item Theoretical MPPC relations obtained from the Cepheid models in all the four convection sets  are found to agree well with the observations in the PC plane for most of the phases in both LMC and SMC. Offset between the MPPC slopes for models and observations has been observed at phase $\Phi \sim 0.8$.    
\item The theoretical PL/AC relations at mean light obtained from the models using the four sets agree better with the observed ones in case of short periods for both LMC/SMC. However, for long periods, models do not match well with the observations for most of the sets. On the other hand, the model PC slopes obtained from all the convection sets are found to be consistent with the observations for all period ranges (short/long/all periods) for both LMC and SMC. 
\item Statistical $F$- test shows that the addition of the Fourier parameter $R_{21}$ to the PL relation in finding distances to the LMC and SMC is significant. For some sets, this improves the LMC/SMC distance and is comparable to the literature values. The addition of $R_{21}$ reduces the error sum of squares significantly for cases where the $F$- test is significant.

\end{enumerate}

 It has been found that the parameters from different sets overlap in the FP plane, but clearly distinct in the multiphase plane. The “opposite” behaviours of short and long period Cepheids on the multiphase PC/PL planes, especially at $\Phi \sim 0.8$ may be related to the Hertzsprung progression, which is a subject for future work. It has been observed that the PC relations for theories works well, indicating that the diffusion approximation is satisfactory and supports the work of \citet{das20}. We also make a point for the need to compute models with period $P <3$ d and also models of higher mass. We noted here that the parameters in convection sets A, B, C, D have no pre-determined values and this is a preliminary study using the set of values as given in \citet{smol08}. A more detailed investigation involving changing the parameters in the convection sets that can discriminate between different theories will be in the future work. Such studies can, in principle, lend insight into the role of turbulent convection in stellar pulsation, in particular, the contrasting behaviour of short and long period multiphase slopes and may lead to optimizing the parameterizing of 3d convection into 1d prescriptions.  
 
 \section*{Acknowledgements}
\addcontentsline{toc}{section}{Acknowledgements}
The authors thank an anonymous referee for making various helpful comments and providing useful suggestions which have significantly improved the manuscript. KK thanks the Council of Scientific and Industrial Research (CSIR), Govt. of India for the Senior Research Fellowship (SRF). SKM thanks State University of New York, Oswego, NY 23126, USA and Cotton University, Guwahati, Assam for the support. SD thanks Council of Scientific and Industrial Research (CSIR), Govt. of India, New Delhi for a financial support through the research grant ``03(1425)/18/EMR-II''. MD thanks CSIR for Junior Research Fellowship provided through CSIR-NET under the project. S. Das acknowledges the support of the KKP-137523 `SeismoLab' Élvonal grant of the Hungarian Research, Development and Innovation Office (NKFIH).  AB acknowledges funding from the European Union's Horizon 2020 research and innovation programme under the Marie Skłodowska-Curie grant agreement No.886298. The authors acknowledge IUCAA, Pune for the use of High Performance Computing facility Pegasus.  The authors also acknowledge the use of MESA-r15140 software in  this project \citep{paxt11, paxt13, paxt15, paxt18, paxt19}. 

\section*{Data Availability}
The \citet{skow21} map is available from \url{http://ogle.astrouw.edu.pl/cgi-ogle/get_ms_ext.py}. The OGLE-IV data is downloaded from \url{http://ftp.astrouw.edu.pl/ogle/ogle4/OCVS/}. The theoretical models will be made available on reasonable request to the corresponding authors. 

\bibliographystyle{mnras}
\bibliography{multiphase_pl} % if your bibtex file is called example.bib

\appendix
\clearpage
\section{PL/PC/AC relations tables}
The results of the coefficients of the PL/PC/AC relations for theoretical/observed light curves of Cepheids in the LMC/SMC at mean light are summarized in tables~\ref{tab:pl_smc}, \ref{tab:pc_lmcsmc}, \ref{tab:ac_lmcsmc} respectively. The results in the mentioned tables are discussed in Section~\ref{sec:plpcac_mean}. 
 
\begin{table*}
\caption{ Same as Table.~\ref{tab:pl_lmc} but for SMC.}
\scalebox{0.95}{
\begin{tabular}{c c c c c c c c} \\ \hline \hline
	& $b_{s}$ & $a_{s}$ &  $b_{l}$ &$a_{l}$  &$b_{all}$ &$a_{all}$ \\ \hline
	
		 & &&SMC [$V$- band] &&\\ \hline
Obs &$-2.941\pm0.071$&$17.756\pm0.061$ & $-2.510\pm0.242$ & $17.501\pm0.289$ & $-2.734\pm0.053$ & $17.784\pm0.040$ \\
Set A & $\mathbf{-2.948}\pm\mathbf{0.117}$  &$\mathbf{17.611}\pm\mathbf{0.093}$ & $-1.298\pm0.200$ & $16.113\pm0.246$  & $-1.987\pm0.064$  &$16.889\pm0.061$ \\
Set B   & $\mathbf{-2.913}\pm\mathbf{0.117}$ & $\mathbf{17.687}\pm\mathbf{0.094}$ & $-1.217\pm0.210$ & $15.954\pm0.249$  & $-2.275\pm0.067$  &$17.191\pm0.065$\\
Set C   & $\mathbf{-3.040}\pm\mathbf{0.111}$ & $\mathbf{17.823}\pm\mathbf{0.087}$  & $-1.527\pm0.164$ & $16.543\pm0.206$ & $-1.970\pm0.057$   &$17.037\pm0.055$ \\
Set D & $\mathbf{-3.028}\pm\mathbf{0.149}$ & $\mathbf{17.888}\pm\mathbf{0.120}$  & $-1.363\pm0.200$ & $16.278\pm0.247$ & $-2.107\pm0.075$   &$17.174\pm0.078$ \\ \hline
&&&SMC [$I$- band] &&& \\ \hline
Obs &$-2.932\pm0.071$ & $17.212\pm0.048$ & $-2.782\pm0.180$ & $17.002\pm0.215$ & $-2.964\pm0.041$ & $17.231\pm0.032$\\
Set A & $\mathbf{-3.076}\pm\mathbf{0.091}$  & $\mathbf{17.048}\pm\mathbf{0.073}$ &  $-1.772\pm0.156$ & $15.862\pm0.193$ & $-2.319\pm0.050$ & $16.479\pm0.048$ \\
Set B& $\mathbf{-3.059}\pm\mathbf{0.094}$ & $\mathbf{17.108}\pm\mathbf{0.075}$ & $-1.732\pm0.161$ & $15.766\pm0.192$ &  $-2.536\pm0.053$ & $16.702\pm0.051$ \\
Set C& $\mathbf{-3.212}\pm\mathbf{0.091}$  & $\mathbf{17.246}\pm\mathbf{0.070}$ & $-2.004\pm0.128$ & $16.242\pm0.160$  & $-2.331\pm0.046$   &$16.600\pm0.045$ \\
Set D & $\mathbf{-3.145}\pm\mathbf{0.119}$ & $\mathbf{17.253}\pm\mathbf{0.096}$ & $-1.858\pm0.156$ & $16.019\pm0.193$  &$-2.418\pm0.059$  & $16.691\pm0.061$ \\ \hline
\end{tabular}}
\label{tab:pl_smc}
\end{table*}

\begin{table*}
\centering
\caption{PC slopes and intercepts of the form $Y=a+b\log{P}$  at mean light; for short ($a_{s}, b_{s}$), long ($a_{l}, b_{l}$), all period ($a_{all}, b_{all}$) obtained from observations and models using sets A, B, C, D in the LMC/SMC. Bold-faced entries in the table  represent PC coefficients obtained from models  which agree well with the  observations.}
\scalebox{0.95}{	
\begin{tabular}{c c c c c c c} \\ \hline \hline
  & $b_{s}$ & $a_{s}$ & $b_{l}$ & $a_{l}$	 & $b_{all}$  & $a_{all}$ \\ \hline
  &&&LMC&&& \\ \hline
  Obs   &$0.210\pm0.018$  & $0.570\pm0.021$  & $0.359\pm0.074$ & $0.430\pm0.087$ &$0.250\pm0.012$ & $0.545\pm0.008$\\
Set A  &$\mathbf{0.185}\pm\mathbf{0.033}$  & $\mathbf{0.532}\pm\mathbf{0.027}$ & $\mathbf{0.493}\pm\mathbf{0.028}$ & $\mathbf{0.226}\pm\mathbf{0.035}$ &$\mathbf{0.351}\pm\mathbf{0.012}$ & $\mathbf{0.400}\pm\mathbf{0.013}$\\
Set B  &$\mathbf{0.216}\pm\mathbf{0.034}$  & $\mathbf{0.537}\pm\mathbf{0.028}$ & $\mathbf{0.458}\pm\mathbf{0.029}$ & $\mathbf{0.268}\pm\mathbf{0.036}$ &$\mathbf{0.334}\pm\mathbf{0.013}$& $\mathbf{0.465}\pm\mathbf{0.014}$\\
Set C  &$\mathbf{0.163}\pm\mathbf{0.028}$  & $\mathbf{0.621}\pm\mathbf{0.036}$ & $0.555\pm0.029$ & $0.191\pm0.037$ &$\mathbf{0.322}\pm\mathbf{0.012}$ & $\mathbf{0.488}\pm\mathbf{0.013}$\\
Set D  &$\mathbf{0.105}\pm\mathbf{0.036}$  & $\mathbf{0.679}\pm\mathbf{0.071}$ & $0.533\pm0.030$ & $0.209\pm0.037$ &$\mathbf{0.316}\pm\mathbf{0.016}$ & $\mathbf{0.485}\pm\mathbf{0.019}$\\
 \hline 
&&&SMC&&& \\ \hline
Obs & $0.247\pm0.024$  &$0.544\pm0.016$  &$0.273\pm0.086$ &$0.449\pm0.102$ &$0.231\pm0.015$   &$0.553\pm0.011$\\
Set A  & $\mathbf{0.128}\pm\mathbf{0.031}$  & $\mathbf{0.560}\pm\mathbf{0.024}$ &$0.474\pm0.045$ &$0.251\pm0.055$ &$\mathbf{0.332}\pm\mathbf{0.016}$   &$\mathbf{0.410}\pm\mathbf{0.015}$\\
Set B  & $\mathbf{0.147}\pm\mathbf{0.028}$  & $\mathbf{0.579}\pm\mathbf{0.023}$ &$0.515\pm0.050$ &$0.188\pm0.060$ &$\mathbf{0.261}\pm\mathbf{0.016}$   &$\mathbf{0.489}\pm\mathbf{0.015}$\\
Set C  & $\mathbf{0.172}\pm\mathbf{0.026}$  & $\mathbf{0.570}\pm\mathbf{0.020}$ &$0.477\pm0.038$ &$0.301\pm0.047$ &$\mathbf{0.361}\pm\mathbf{0.012}$   &$\mathbf{0.437}\pm\mathbf{0.012}$\\
Set D  & $\mathbf{0.116}\pm\mathbf{0.037}$  & $\mathbf{0.634}\pm\mathbf{0.030}$ &$0.495\pm0.045$ &$0.259\pm0.056$ &$\mathbf{0.311}\pm\mathbf{0.017}$    &$\mathbf{0.483}\pm\mathbf{0.018}$\\ \hline 
\end{tabular}}
\label{tab:pc_lmcsmc}
\end{table*}

\begin{table*}
\centering
\caption{Same as Table~\ref{tab:pc_lmcsmc}  but for AC relations.}
\scalebox{0.95}{	
\begin{tabular}{c c c c c c c} \\ \hline \hline
  & $b_{s}$ & $a_{s}$ & $b_{l}$ & $a_{l}$	 & $b_{all}$  & $a_{all}$ \\ \hline
  &&&LMC&&& \\ \hline

Obs  &  $-0.126\pm0.011$ & $0.795\pm0.008$ & $0.056\pm0.036$ & $0.796\pm0.034$ & $-0.065\pm0.011$ & $0.763\pm0.009$ \\
Set A  & $\mathbf{-0.229}\pm\mathbf{0.012}$ & $0.853\pm0.010$ & $-0.126\pm0.031$ & $0.961\pm0.031$ & $0.008\pm0.025$ & $0.744\pm0.022$ \\ 
Set B  & $\mathbf{-0.233}\pm\mathbf{0.012}$ & $0.880\pm0.009$ & $-0.193\pm0.021$ & $1.016\pm0.020$ & $\mathbf{-0.067}\pm\mathbf{0.023}$ & $\mathbf{0.835}\pm\mathbf{0.020}$ \\
Set C  & $\mathbf{-0.210}\pm\mathbf{0.019}$ & $0.900\pm0.013$ & $0.127\pm0.057$ & $0.755\pm0.061$  & $0.211\pm0.023$ & $0.638\pm0.022$ \\
Set D  & $\mathbf{-0.143}\pm\mathbf{0.030}$ & $0.83\pm0.019$ & $\mathbf{0.046}\pm\mathbf{0.054}$ & $\mathbf{0.830}\pm\mathbf{0.054}$  & $0.167\pm0.025$ & $0.697\pm0.023$ \\
 \hline 
&&&SMC&&& \\ \hline
Obs &    $-0.196\pm0.007$ & $0.844\pm0.005$ & $-0.039\pm0.034$ & $0.829\pm0.029$ & $-0.112\pm0.020$ & $0.819\pm0.016$ \\
Set A  &   $\mathbf{-0.217}\pm\mathbf{0.011}$ & $\mathbf{0.890}\pm\mathbf{0.009}$ & $0.053\pm0.038$ &  $0.846\pm0.036$ & $-0.013\pm0.033$ & $0.786\pm0.030$ \\ 
Set B &    $\mathbf{-0.207}\pm\mathbf{0.016}$ & $\mathbf{0.870}\pm\mathbf{0.012}$ & $-0.009\pm0.046$ &  $0.875\pm0.040$ & $0.001\pm0.035$ & $0.795\pm0.028$ \\
Set C &    $\mathbf{-0.124}\pm\mathbf{0.011}$ & $\mathbf{0.805}\pm\mathbf{0.009}$ & $-0.007\pm0.058$ &   $0.826\pm0.048$ & $\mathbf{-0.113}\pm\mathbf{0.012}$ & $\mathbf{0.811}\pm\mathbf{0.013}$ \\ 
Set D &    $\mathbf{-0.124}\pm\mathbf{0.012}$ & $\mathbf{0.744}\pm\mathbf{0.010}$ & $0.001\pm0.063$ & $0.875\pm0.040$ &  $\mathbf{-0.112}\pm\mathbf{0.013}$ & $\mathbf{0.750}\pm\mathbf{0.011}$ \\ \hline 

\end{tabular}}
\label{tab:ac_lmcsmc}
\end{table*}

\section{Error in the MPPL relations for the LMC/SMC}
\label{app:error_mppl}
\begin{figure*}
%\scalebox{0.95}{
\begin{tabular}{cc}
\resizebox{0.48\linewidth}{!}{\includegraphics*{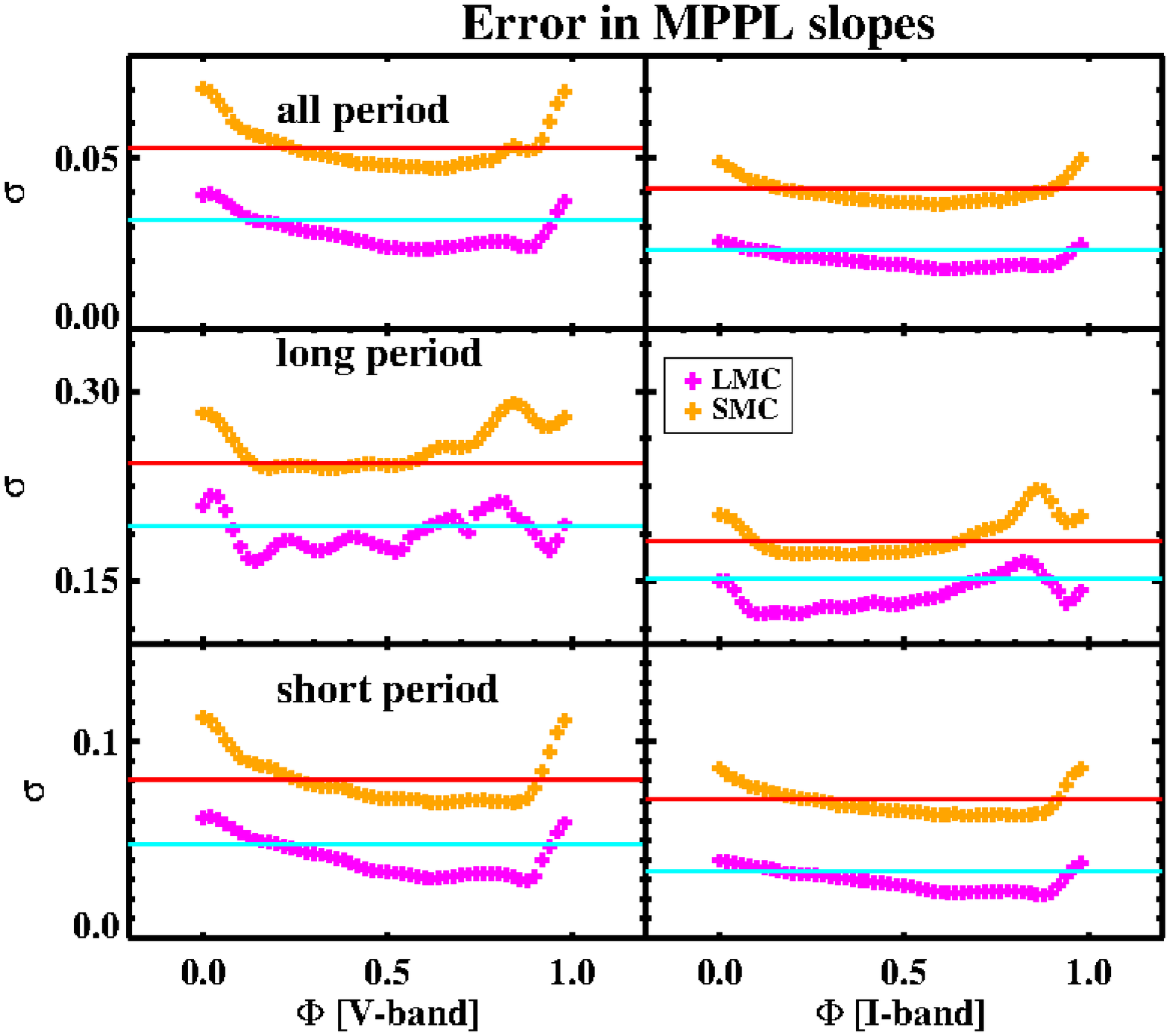}}& 
\resizebox{0.48\linewidth}{!}{\includegraphics*{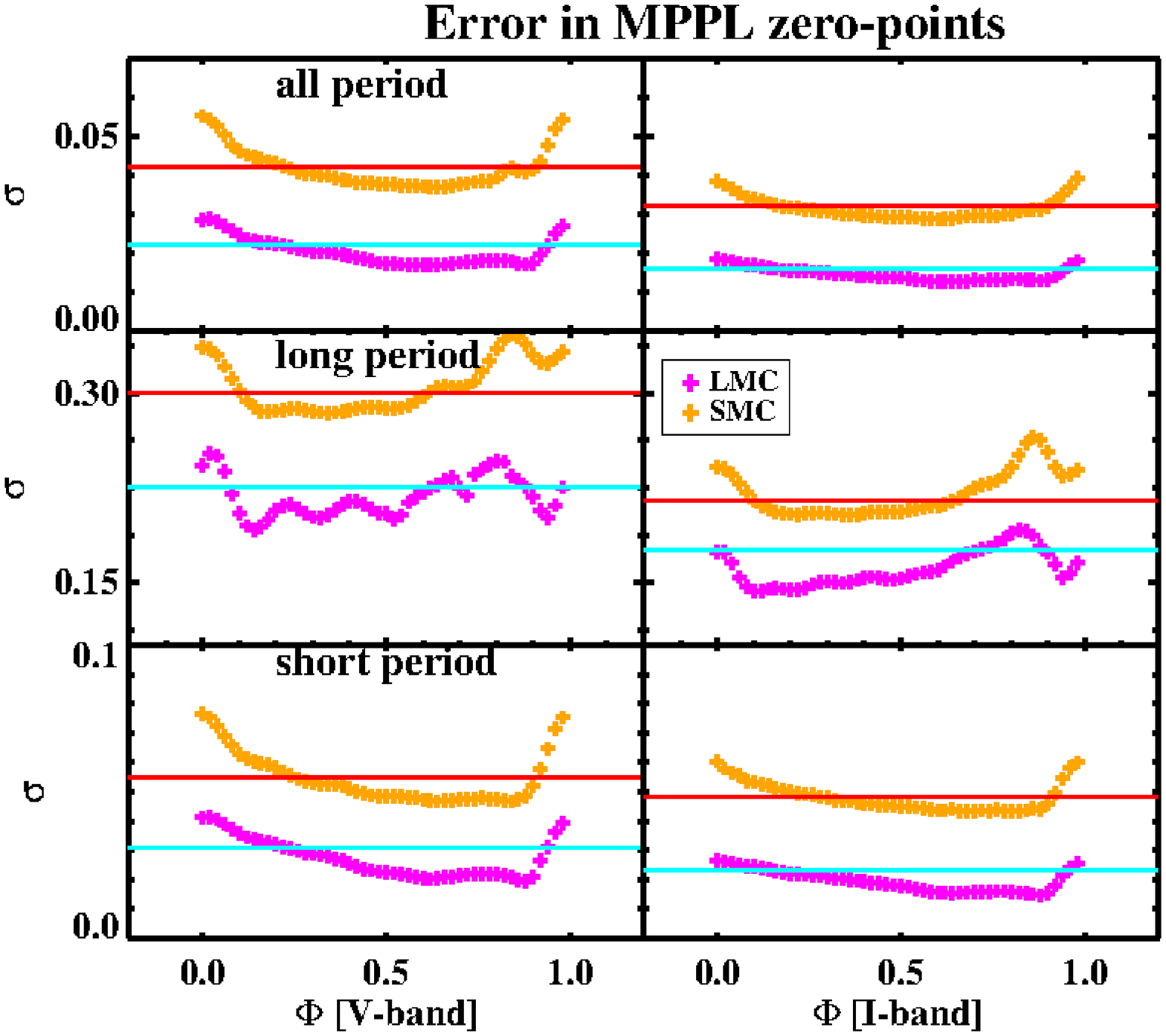}}\\
\end{tabular}
\caption{\textit{Left panel}: Errors in the MPPL slopes for the LMC/SMC. \textit{Right panel}: Same as left panel, but for errors in the MPPL zero-points. Cyan and 
red lines represent the errors in PL slopes/zero-points obtained at mean light for 
the LMC and SMC, respectively. }
\label{fig:error_lmcsmc}
\end{figure*}

The errors in the MPPL relations obtained using the observational data for the LMC/SMC are shown in Fig.~\ref{fig:error_lmcsmc}.
\section{MESA INLIST}
\label{app:inlist}
{\fontsize{6.2}{6pt}\selectfont
\begin{verbatim}
 &star_job

      show_log_description_at_start = .false.

      create_RSP_model = .true.

      save_model_when_terminate = .true.
      save_model_filename = 'final.mod'

      initial_zfracs = 6

      color_num_files=2
      color_file_names(2)='blackbody_johnson.dat'
      color_num_colors(2)=5

      set_initial_age = .true.
      initial_age = 0

      set_initial_model_number = .true.
      initial_model_number = 0

      profile_starting_model = .true.
       set_initial_cumulative_energy_error = .true.
      new_cumulative_energy_error = 0d0

/ ! end of star_job namelist
&eos
      use_FreeEOS = .true.
/

&kap
      kap_file_prefix = 'a09'
      kap_lowT_prefix = 'lowT_fa05_a09p'
      kap_CO_prefix = 'a09_co'
      Zbase = 0.008

! opacity controls
      cubic_interpolation_in_X = .false.
      cubic_interpolation_in_Z = .false.
      include_electron_conduction = .true.
      use_Zbase_for_Type1 = .true.
      use_Type2_opacities = .true.
      kap_Type2_full_off_X  =  0.71d0
      kap_Type2_full_on_X  =  0.70d0
      kap_Type2_full_off_dZ  =  1d-3
      kap_Type2_full_on_dZ  =  1d-2

/
&controls
 ! must set mass, Teff, L, X, and Z.
         RSP_mass = 1.6 ! (Msun)
         RSP_Teff = 6900 ! (K)
         RSP_L = 25 ! (Lsun)
         RSP_X = 0.736 ! hydrogen mass fraction
         RSP_Z = 0.008 ! metals mass fraction
        
         RSP_alfa = 1.2d0   ! mixing length; alfa = 0 gives a purely radiative model.
         RSP_alfac = 1.0d0  ! convective flux; Lc ~ RSP_alfac
         RSP_alfas = 1.0d0  ! turbulent source; Lc ~ 1/ALFAS; PII ~ RSP_alfas
         RSP_alfad = 1.0d0  ! turbulent dissipation; damp ~ RSP_alfad
         RSP_alfap = 0.0d0  ! turbulent pressure; Pt ~ alfap
         RSP_alfap = 0.0d0  ! turbulent pressure; Pt ~ alfap
         RSP_alfat = 0.0d0  ! turbulent flux; Lt ~ RSP_alfat; overshooting.
         RSP_alfam = 0.25d0 ! eddy viscosity; Chi & Eq ~ RSP_alfam
         RSP_gammar = 0.0d0 ! radiative losses; dampR ~ RSP_gammar    
        
         RSP_theta = 0.5d0  ! Pgas and Prad
         RSP_thetat = 0.5d0 ! Pturb
         RSP_thetae = 0.5d0 ! erad in terms using f_Edd
         RSP_thetaq = 1.0d0 ! avQ
         RSP_thetau = 1.0d0 ! Eq and Uq
         RSP_wtr = 0.6667d0 ! Lr
         RSP_wtc = 0.6667d0 ! Lc
         RSP_wtt = 0.6667d0 ! Lt
         RSP_gam = 1.0d0    ! Et src_snk
         
         ! controls for building the initial model
         RSP_nz = 200 ! total number of zones in initial model
         RSP_nz_outer = 60 ! number of zones in outer region of initial model
         RSP_T_anchor = 11d3 ! approx temperature at base of outer region
         RSP_T_inner = 2d6 ! T at inner boundary of initial model

         RSP_max_outer_dm_tries = 100 ! give up if fail to find outer dm in this many attempts
         RSP_max_inner_scale_tries = 100 ! give up if fail to find inner dm scale factor in this many attempts
         RSP_T_anchor_tolerance = 1d-8
         RSP_relax_initial_model = .true.
         RSP_relax_alfap_before_alfat = .true. ! else reverse the order
         RSP_relax_max_tries = 1000
         RSP_relax_dm_tolerance = 1d-6
         use_RSP_new_start_scheme = .false.
         RSP_kick_vsurf_km_per_sec = 0.1d0 
         RSP_fraction_1st_overtone = 0.0d0
         RSP_fraction_2nd_overtone = 0d0
        
        ! random initial velocity profile.  added to any kick from eigenvector.
         RSP_Avel = 0d0 ! kms. linear in mesh points from 0 at inner boundary to this at surface
         RSP_Arnd = 0d0 ! kms. random fluctuation at each mesh point.

         ! period controls
         RSP_target_steps_per_cycle = 600
         RSP_min_PERIOD_div_PERIODLIN = 0.5d0
         RSP_mode_for_setting_PERIODLIN = 0 
         RSP_default_PERIODLIN = 34560 

        ! when to stop
         RSP_max_num_periods =2000 ! ignore if < 0

         RSP_GREKM_avg_abs_frac_new = 0.1d0 ! fraction of new for updating avg at each cycle.
         ! timestep limiting
         RSP_initial_dt_factor = 1d-2 ! set initial timestep to this times linear period/target_steps_per_cycle
         RSP_v_div_cs_threshold_for_dt_limit = 0.8d0
         RSP_max_dt_times_min_dr_div_cs = 2d0 ! limit dt by this
         RSP_max_dt = -1 ! seconds
         RSP_report_limit_dt = .false.
         RSP_cq = 4.0d0 ! viscosity parameter (viscosity pressure proportional to cq)
         RSP_zsh = 0.1d0 ! "turn-on" compression in units of sound speed.
         RSP_Qvisc_linear = 0d0
         RSP_Qvisc_quadratic = 0d0
         RSP_use_Prad_for_Psurf = .false.
         RSP_use_atm_grey_with_kap_for_Psurf = .false.
         RSP_tau_surf_for_atm_grey_with_kap = 3d-3 ! for atm_grey_with_kap
         RSP_fixed_Psurf = .true.
         RSP_Psurf = 0d0 ! ignore if < 0.  else use as surface pressure.

         ! solver controls
         RSP_tol_max_corr = 1d-8
         RSP_tol_max_resid = 1d-6
         RSP_max_iters_per_try = 100
         RSP_max_retries_per_step = 8
         RSP_report_undercorrections = .false.
         RSP_nz_div_IBOTOM = 30d0 ! set IBOTOM = RSP_nz/RSP_nz_div_IBOTOM
         RSP_min_tau_for_turbulent_flux = 2d2
         ! rsp hooks
         use_other_RSP_linear_analysis = .false.
         use_other_RSP_build_model = .false.
         
         RSP_efl0 = 1.0d2
         RSP_nmodes = 3 
         RSP_trace_RSP_build_model = .false.

! output controls

      num_trace_history_values = 3
      trace_history_value_name(1) = 'rel_E_err'
      trace_history_value_name(2) = 'log_rel_run_E_err'
      trace_history_value_name(3) = 'rsp_GREKM_avg_abs'

      photo_interval = 1000
      profile_interval = 1
      history_interval = 1
      terminal_interval = 4000
      max_num_profile_models = -1
      log_directory='LOGS'
      photo_directory='photos'
/ ! end of controls namelist
&pgstar
/ ! end of pgstar namelist
\end{verbatim}
}

% Don't change these lines
\bsp	% typesetting comment
\label{lastpage}
\end{document}